\documentclass[aps,prd,reprint,preprintnumbers,showpacs,nofootinbib,superscript address]{revtex4-2}
\usepackage[utf8]{inputenc}
\usepackage{amssymb}
\usepackage{amsmath}
\usepackage{mathtools}
\usepackage[dvipsnames]{xcolor}
\usepackage{notoccite}
\usepackage{tocbasic}
\usepackage{graphicx}
\usepackage{natbib}
\usepackage{tensor}
\usepackage{etoolbox}
\usepackage{siunitx}
\DeclareSIUnit\parsec{pc}
\DeclareSIUnit\au{AU}
\DeclareSIUnit\solarmass{M_\odot}
\DeclareSIUnit\parsecfull{parsec}
\DeclareTOCStyleEntry[numwidth=20pt,linefill=\bfseries\TOCLineLeaderFill]{tocline}{section}
\DeclareTOCStyleEntry[entryformat=\textit,numwidth=10pt,linefill=\TOCLineLeaderFill]{tocline}{subsection}
\DeclareTOCStyleEntry[entryformat=\textit,numwidth=10pt,linefill=\TOCLineLeaderFill]{tocline}{subsubsection}
\allowdisplaybreaks
\usepackage{hyperref}
\hypersetup{%
	colorlinks = true,
	linkcolor = Blue,
	citecolor = Blue,
	filecolor = Blue,
	urlcolor = Blue%
}
\usepackage[capitalize]{cleveref}
\AddToHook{cmd/appendix/before}{\crefalias{section}{appendix}}

\newrobustcmd{\KB}{\tensor{K}{_{\text{B}}}}
\newrobustcmd{\KBE}{\tensor*{K}{_{\text{B}}^{\text{eff}}}}
\newrobustcmd{\KBEN}{\tensor*{\bar{K}}{_{\text{B}}^{\text{eff}}}}
\newrobustcmd{\GN}{\tensor*{G}{_{\text{N}}}}
\newrobustcmd{\Rh}{{r_{\text{S}}}}

\makeatletter
\renewcommand{\paragraph}{%
  \@startsection{paragraph}{4}%
  {\z@}{1.21ex \@plus 1ex \@minus .2ex}{0.9em}%
  {\normalfont\normalsize\bfseries}%
}
\usepackage{titlesec}
\newrobustcmd{\pea}[1]{\emph{#1}\textbf{.\ \ \ ---}}
\titleformat{\paragraph}[runin]{\normalfont\normalsize\bfseries}{\emph\theparagraph}{1em}{\pea}
\titleformat{\section}[block]{\normalfont\bfseries\centering}{\MakeUppercase\thesection}{1em}{\MakeUppercase}
\makeatother

\begin{document}

\title{How black hole mimickers and Shapiro-free lenses signal effective dark matter}

\author{Lirui Yang}
\email{ly344@cantab.ac.uk}
\affiliation{Department of Quantum Information, Yonsei University, Incheon 21983, Republic of Korea}%
\affiliation{Kavli Institute for Cosmology, Madingley Road, Cambridge CB3 0HA, UK}%
\affiliation{Astrophysics Group, Cavendish Laboratory, JJ Thomson Avenue, Cambridge CB3 0HE, UK}%
\affiliation{Kavli Institute for Astronomy and Astrophysics, Peking University, Beijing 100871, China}%
\author{Will Barker}%
\email{barker@fzu.cz}
\affiliation{Central European Institute for Cosmolgy and Fundamental Physics, Institute of Physics of the Czech Academy of Sciences, Na Slovance 1999/2, 182 00 Prague 8, Czechia}
\affiliation{Kavli Institute for Cosmology, Madingley Road, Cambridge CB3 0HA, UK}%
\affiliation{Astrophysics Group, Cavendish Laboratory, JJ Thomson Avenue, Cambridge CB3 0HE, UK}%
\author{Tobias Mistele}
\email{txm523@case.edu}
\affiliation{Max Planck Institute for Extraterrestrial Physics, Giessenbachstrasse 1, 85748 Garching, Germany}
\affiliation{Department of Astronomy, Case Western Reserve University, 10900 Euclid Avenue, Cleveland, Ohio 44106, USA}%
\author{Amel Durakovic}
\email{amel.durakovic@pmf.unsa.ba}
\affiliation{Central European Institute for Cosmolgy and Fundamental Physics, Institute of Physics of the Czech Academy of Sciences, Na Slovance 1999/2, 182 00 Prague 8, Czechia}
\affiliation{Observatoire Astronomique de Strasbourg - UMR 7550, 11 rue de l'Université, 67000 Strasbourg, France}%
\affiliation{Faculty of Science, University of Sarajevo, Zmaja od Bosne 33-35, 71000 Sarajevo, Bosnia-Herzegovina}%

\begin{abstract}
	We report the existence of two exotic compact objects in the leading relativistic model of modified Newtonian dynamics, namely \ae{}ther-scalar-tensor theory. This model is consistent with precision cosmology and gravitational wave constraints on tensor speed. Black hole mimickers could subtly change observations: gravitational waves from their mergers might show unusual echoes or altered ringdown patterns, and images of their horizon-scale shadows might be slightly different from those of a true black hole. Shapiro-free lenses are massless objects that deflect light without any gravitational time delay, producing distinctive lensing events. These predictions connect to ongoing and future gravitational-wave searches, horizon-scale imaging, and time-domain lensing surveys.
\end{abstract}

\maketitle
\tableofcontents

\section{Introduction}\label{section: Introduction}

\paragraph*{Dark matter} The standard dark-energy/cold-dark-matter ($\Lambda$CDM) model of cosmology, based in part on general relativity (GR), has been successful on cosmological scales. On smaller galactic scales we observe a tight correlation between the visible, baryonic mass and dark matter, summarised in the baryonic Tully--Fisher relation~\cite{McGaugh2000,Mistele2024} and the radial acceleration relation~\cite{Mistele2023b,Brouwer2021,Lelli2017b}. Another explanation for these correlations is in terms of new dynamics at low accelerations, i.e., modified Newtonian dynamics (MOND)~\cite{MOND3,MOND4,MOND5}. MOND by itself, however, is concerned only with the non-relativistic limit, i.e. it is an alternative to Newtonian gravity, not an alternative to GR. Thus, various relativistic models that reduce to MOND have been proposed~\cite{Famaey:2011kh}, most notably \ae{}ther-scalar-tensor (\AE{}ST) theory~\cite{Skordis:2020eui}. \AE{}ST stands out as the first model that is consistent with precision observations of the microwave background/large scale structure, and with gravitational waves (GWs)  implying a fast tensor mode, while allowing for a MOND-like phenomenology~\cite{Mistele2023b,Mistele2023}. The phenomenology of \AE{}ST has so far been explored relatively little, however, in the strong-field regime relevant to compact objects. Previous works have considered neutron stars~\cite{Reyes:2024oha,Reyes:2025oet}, unstable solutions~\cite{AeST_BH1}, and `stealth' black holes (BHs) which are indistinguishable from BHs in GR~\cite{costas_stealth_bh}. In this work we propose two more exotic compact objects with relatively subtle observational signatures: BH mimickers are hard to \emph{distinguish} from GR BHs, and lenses without Shapiro delay are hard to see \emph{at all}. We are not obliged to put these objects forwards as dark matter candidates, since \AE{}ST is designed already to provide the necessary \emph{effective} dark matter. Rather, their distinct characteristics, if observed, would offer a smoking gun for \AE{}ST itself.

\begin{figure}[ht!]
\includegraphics[width=\linewidth]{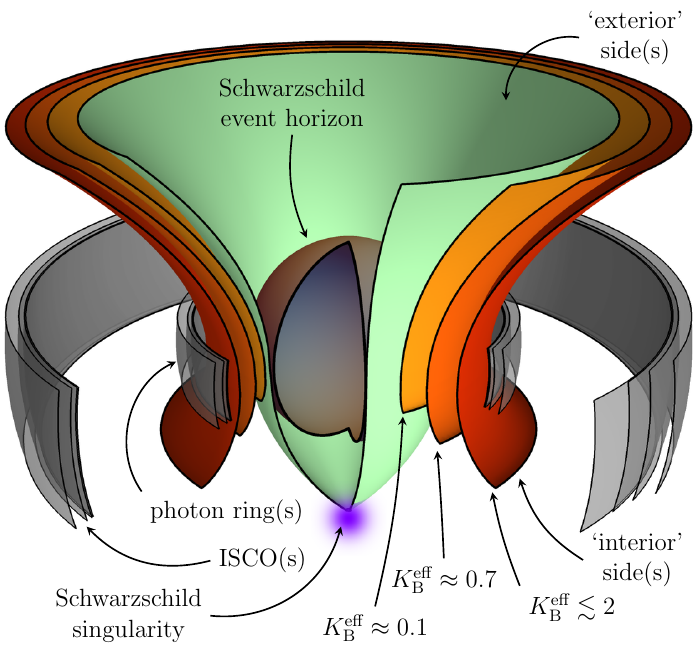}
	\caption{\label{BlackHoleMimickers} Embedding visualisation of spatial geometry. Black hole mimickers in \ae{}ther-scalar-tensor theory (orange/red, see~\cref{ElingJacobsonWormholeLineElement,eq:time_norm,RadiusInTermsOfSpaceFunc}) emulate the Schwarzschild geometry (green, see~\cref{Schwarzschild}) with the same Newtonian mass~$M$, but have slightly dilated features and a narrow wormhole throat instead of an event horizon. An alternative horizon lies on the `interior' side (here truncated).}
\end{figure}

\paragraph*{Confidence in black holes} We know that objects of BH density are very common. Doppler shifts in many X-ray binaries indicate~$\gtrsim \SI{3}{\solarmass}$ companions too heavy for neutron stars~\cite{Bolton1972,Remillard2006}; astrometry near the compact radio source Sgr~A* indicates~$\sim \SI{e6}{\solarmass}$ contained within~$\sim\SI{e2}{\au}$ of the galactic centre~\cite{Schodel2002,Ghez2008} (indeed, modern galaxy evolution is founded on accretion near such supermassive compact objects~\cite{LyndenBell1971,Kormendy2013}). Three lines of evidence suggest, moreover, that these BH candidates have horizons as predicted by GR. Firstly, they lack a visibly accreting surface, which would glow thermally or otherwise be illuminated by Type I X-ray bursts~\cite{Narayan2002,Narayan2008}. Secondly, GWs from mergers seen by LIGO/Virgo indicate ringing down of quasi-normal modes consistent with a Kerr-type horizon; the GWs appear moreover to be free from surface echos, and can be parameterised by mass and spin~\cite{Abbott2016,Cardoso2016}. Thirdly, direct VLBI imaging of Sgr~A* and M87* by the EHT reveals shadows similarly consistent with Kerr spacetime~\cite{Falcke2000,EHT2019,EHT2022}. These methods also constrain the strong-field geometry to nearly Kerr above the horizon, through the circularity of the directly imaged photon ring, inspiral phasing of merger GWs, and Fe K$\alpha$ X-ray emission near the innermost stable circular orbit (ISCO)~\cite{EHT2022,Abbott2016,Tanaka1995,Fabian2000}. Astrometric measurements of precession and gravitational redshift near Sgr~A* are less stringent, but consistent with at least the Schwarzschild exterior~\cite{GRAVITY2018,GRAVITY2020}. In summary, exotic objects which only \emph{mimick} BHs face a comprehensive battery of tests. Nonetheless, \AE{}ST contains a strong mimicker candidate (see~\cref{BlackHoleMimickers}). These may exist alongside the genuine `stealth' BHs of \AE{}ST, which themselves are less diagnostically useful.

\paragraph*{Black hole mimickers} We propose the BH mimicker of \AE{}ST to be a highly asymmetric wormhole, leading only to a BH-type object on its `interior' side. To keep our key results within in this opening section, we quote here our predicted line element on the `exterior' side, i.e. that which is most relevant to observations. For reference, in GR with Newton--Cavendish constant~$G$, the Schwarzschild BH of mass~$M$ is described by
\begin{equation}\label{Schwarzschild}
	\mathrm{d} s^2 = -\left(1-\tfrac{2GM}{r}\right)\mathrm{d} t^2 +\frac{\mathrm{d} r^2}{\left(1-\tfrac{2GM}{r}\right)}  + r^2\, \mathrm{d} \Omega^2\,\, , 
\end{equation}
where~$\mathrm{d}\Omega^2\equiv\mathrm{d}\theta^2+\sin^2(\theta)\mathrm{d}\phi^2$ is the angular measure. The `exterior' side of our mimicker generalises~\cref{Schwarzschild} to
\begin{equation}\label{ElingJacobsonWormholeLineElement}
	 \mathrm{d} s^2 = -e^{2\mathcal{N}(r)}\, \mathrm{d} t^2 + e^{2\mathcal{M}(r)}\, \mathrm{d} r^2 + r^2\, \mathrm{d} \Omega^2.
\end{equation}
Whilst~\cref{Schwarzschild} would imply~$\mathcal{N}(r)=-\mathcal{M}(r)$, the BH mimicker instead requires the cumbersome relation
\begin{equation}\label{eq:time_norm}
\begin{aligned}
	e^{2\mathcal{N}(r)}&=
	\left[ \frac{\sqrt{2}-\sqrt{2-\KBE}}{\sqrt{2}+\sqrt{2-\KBE}}\right. \\ 
	&\hspace{-30pt} \times \left.\frac{\sqrt{2+\KBE(e^{2\mathcal{M}(r)}-1)}+\sqrt{2-\KBE}}{\sqrt{2+\KBE (e^{2\mathcal{M}(r)}-1)}-\sqrt{2-\KBE}}\right]^{\sqrt{\frac{2}{2-\KBE}}}.
\end{aligned}
\end{equation}
The formula in~\cref{eq:time_norm} depends strongly on the dimensionless parameter~$\KBE{}<2$. For the moment, it suffices to know that~$\KBE{}$ depends on (i) two of the constant \AE{}ST model parameters and (ii) some dynamically acquired flux of scalar hair. The actual formula for~$\mathcal{M}(r)$ is defined as being the \emph{inverse} of an `$r\left(\mathcal{M}\right)$' formula
\begin{align}
	&\hspace{-5pt} r= \frac{\GN M\KBE e^{\mathcal{M}(r)}}{\sqrt{4+2\KBE\left(e^{2\mathcal{M}(r)}-1\right)}-2}\left[\frac{\sqrt{2}+\sqrt{2-\KBE}}{\sqrt{2}-\sqrt{2-\KBE}}\right.
	\nonumber\\
	&\hspace{-10pt} \times\left.\frac{\sqrt{2+\KBE\left(e^{2\mathcal{M}(r)}-1\right)}-\sqrt{2-\KBE}}{\sqrt{2+\KBE\left(e^{2\mathcal{M}(r)}-1\right)}+\sqrt{2-\KBE}}\right]^{\frac{1}{\sqrt{4-2\KBE}}} ,\label{RadiusInTermsOfSpaceFunc}
\end{align}
where the Newtonian~$\GN{}$ in~\cref{RadiusInTermsOfSpaceFunc} is also shifted relative to the bare~$G$ in~\cref{Schwarzschild} by \AE{}ST model parameters. The take-home point is that~\cref{ElingJacobsonWormholeLineElement,eq:time_norm,RadiusInTermsOfSpaceFunc} `mimick'~\cref{Schwarzschild} to arbitrary precision as~$\KBE{}\to 0$. This is illustrated in~\cref{BlackHoleMimickers}, which shows a faithful embedding of 2D equatorial constant-time slices in 3D space.\footnote{This is a common method of visualisation: the 3D cylindrical line element~$\mathrm{d}z^2+\mathrm{d}\rho^2+\rho^2\mathrm{d}\phi^2$ is equated with~\cref{Schwarzschild,ElingJacobsonWormholeLineElement} at~$t=\text{const.}$ and~$\theta=\pi/2$. By identifying~$r$ with~$\rho$, one gets an equation for~$\mathrm{d}z/\mathrm{d}\rho$ describing the 2D embedded surface.} Where the Schwarzschild geometry has an event horizon, however, the BH mimicker always has a slight throat, on the `interior' side of which lies a null singularity. The fact that the horizon is replaced by a throat is confirmed by the standard `throat conditions' in~\cref{ThroatConds}. We imagine BH mimickers as accounting for some clandestine fraction of the observed BH population. They would presumably evade all radio, astrometric and GW bounds on the external geometry, and all optical, X-ray and GW bounds on surface characteristics. Signals for BH mimickers from \AE{}ST may include (i) corrections to radio imaging within the photon ring from `interior'-side emissions, and (ii) non-BH ringdown.

\begin{figure}[ht!]
\includegraphics[width=\linewidth]{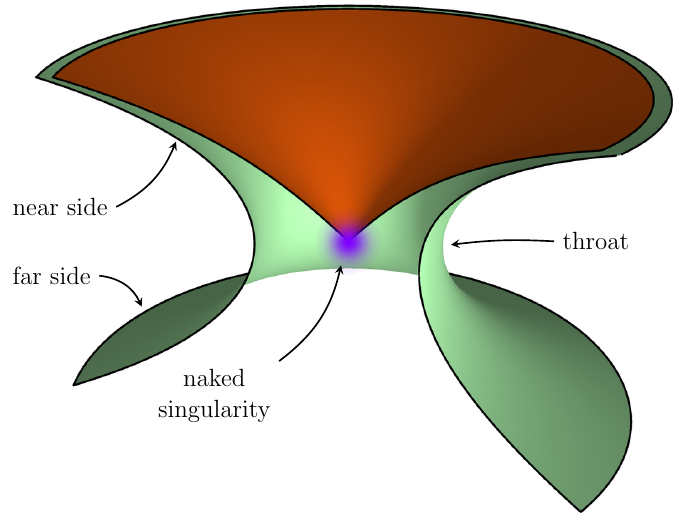}
	\caption{\label{ShapiroFreeLenses} Embedding visualisation of spatial geometry. Lenses without Shapiro delay in \ae{}ther-scalar-tensor theory (red, see `$(-)$' configuration in~\cref{EllisWormhole}) are massless regions of curved space ending in a naked singularity. They are the analytic continuation of Ellis wormholes (green, see `$(+)$' configuration in~\cref{EllisWormhole}).}
\end{figure}

\paragraph*{Confidence in lenses} We will also be interested in hard-to-find exotic objects that do \emph{not} mimick BHs, but whose subtle lensing properties would provide a smoking gun for \AE{}ST (see~\cref{ShapiroFreeLenses})~\cite{Gott1985,Huterer2003}. Statistical microlensing campaigns (MACHO, EROS, OGLE) constrain the fraction of compact lenses in galactic halos across~$\SIrange{e-7}{10}{\solarmass}$ to below a few percent of CDM~\cite{Tisserand2007,Mroz2024}, while forthcoming wide‐field surveys (LSST) will extend sensitivity to sub-lunar masses~\cite{Inoue2017,Niikura2019}. Pulsar timing constrains populations of exotic lenses which have become entrained in binary systems. Precision monitoring of binary pulsars --- most notably the Hulse--Taylor and subsequent millisecond systems --- has confirmed the Shapiro delay (lensing in time) predicted by GR to sub-percent accuracy~\cite{Taylor1989,Kramer2021}. No anomalous timing residuals attributable to scalar charges or exotic multipole structure have been observed in existing systems~\cite{Lazaridis2009,Freire2012}. Pulsar timing arrays (NANOGrav, EPTA, PPTA) likewise report a stochastic background consistent with a population of supermassive BH binaries and show no non-quadrupolar radiation or unexpected spectral features~\cite{NANOGrav2023,EPTA2023}. Notice, however, how all these constraints presume CDM-type lenses, which `weigh' and thus participate in Newtonian gravity. Recalling, therefore, that we have no need of dark matter candidates in \AE{}ST, this motivates considering an extreme case: a `weightless' gravitational lens.

\paragraph*{Shapiro-free lenses} We propose a `weightless' lens in \AE{}ST which --- though it is \emph{not} a wormhole --- is inspired by the line element of the massless Ellis--Bronnikov wormhole~\cite{Ellis1973Drainhole,Bronnikov1973ScalarTensor}. The Ellis--Bronnikov wormhole is given by the `$(+)$' configuration in
\begin{equation}\label{EllisWormhole}
	\mathrm{d}s^2 = -\mathrm{d}t^2 + \mathrm{d}r^2 + \left(r^2\mp \ell^2\right)\mathrm{d}\Omega^2 \,\, ,
\end{equation}
where~$t$ and~$r$ are now the proper time and radius. As popularised by Visser~\cite{Visser1989,Visser1995}, Morris and Thorne~\cite{MorrisThorne1988,MorrisThorneYurtsever1988}, the `$(+)$' configuration is a far less subtle wormhole than the BH mimicker, with correspondingly more applications in science fiction: it seemingly allows test geodesics to pass between two asymptotically flat regions via a `throat' of minimal area~$4\pi\ell^2$ at~$r=0$, which needs to be held open by phantom matter in the Einstein field equations. By contrast, our `Shapiro-free lens' is the `$(-)$' configuration of~\cref{EllisWormhole}, which requires no phantom matter, being an exact solution to the \AE{}ST field equations in which the wormhole throat closes up. This promotes the object from a formal curiosity to a subtle observational challenge. Its constant-time hypersurfaces are highly curved, as illustrated in~\cref{ShapiroFreeLenses} --- indeed this figure captures the \emph{full} Riemannian curvature, which has no `time part' whatever (so-called \emph{ultrastaticity}\footnote{Ultrastatic spacetime is not only static, in the sense that it has a global timelike Killing vector, but additionally this Killing vector has unit norm~\cite{Sonego2010}. The general ultrastatic line element takes the form~$\mathrm{d}s^2=-\mathrm{d}\mathrm{t}^2+\tensor{g}{_{ij}}\mathrm{d}\tensor{x}{^i}\mathrm{d}\tensor{x}{^j}$. On the slices defined by~$t=\mathrm{const.}$ additionally have no extrinsic curvature.}). The object thus has no mass~$M$ at infinity, no gravitational redshift, and indeed \emph{no gravity}. A naked singularity (or star-like matter configuration) sources the scalar hair and sets~$\ell$, but it does not accrete, and whilst its spatial curvature gives rise to lensing, the lensed rays experience no Shapiro delay. This is a result of the object being more compact than its Schwarzschild equivalent: its Weyl potential, relevant for lensing, scales as~$\sim 1/r^2$, rather than~$\sim 1/r$ as in GR. The Shapiro delay incurred by integrating the tails of this potential is thus not enhanced (as the case in GR) by a logarithmic divergence in the source-observer distance: it is this logarithm that usually dominates in cosmological lenses. Besides lacking the logarithmic contribution, Shapiro delay is suppressed in a more general sense, as follows. The delay in GR is proportional overall to~$2GM$, which is seen in~\cref{Schwarzschild} to be the gravitational scale of the lens. The new kind of lens in the `$(-)$' configuration of~\cref{EllisWormhole} has a gravitational scale of~$\ell$, but its Shapiro delay is proportional to~$\ell^2/b$, where~$b$ is the impact factor. Thus, for the new lens to induce comparable Shapiro delay at the same~$b$ to a GR lens, its scale must be~$\ell\sim \sqrt{2GMb}$. For all astronomical lenses~$b\gg 2GM$, implying a large geometric mean~$\ell\gg 2GM$. To give a concrete example, one would need~$\ell\sim\SI{e3}{\kilo\meter}$ to comparably perturb sun-grazing radar pulses (as in Shapiro's original experiment~\cite{Shapiro1964,Shapiro1968}), i.e., three orders of magnitude larger than a solar-mass black hole. This separation of scales is ten times more severe in the microlensing context, where we imagine actually searching for such objects. In this case, another concrete example is given by primordial BHs of~$M\sim\SI{10}{\solarmass}$, which are a popular focus in microlensing: their magnification of sources in the Milky Way region would be apparent within an Einstein radius of~$b\sim\SI{10}{\au}$. Such a microlensing event would need to be explained by a `large' object of~$\ell\sim\SI{e5}{\kilo\meter}$, i.e., four orders of magnitude larger than the BH. Thus, assuming that \AE{}ST gives rise to a population of more `reasonably sized' lenses, the microlensing implications are expected to be very subtle.\footnote{Ultimately, until the formation of Shapiro-free lenses is understood, the question of whether or not a given value of~$\ell$ is `reasonable' is still speculative, and the scales of primordial BHs simply offer a familiar benchmark.} Besides being weaker, the Weyl potential also has the `wrong' sign, akin to negative mass in GR. The Shapiro delays that do arise would thus be \emph{negative}, in that signals would be slightly advanced, not retarded. We can understand why this happens heuristically, from the `$(-)$' configuration in~\cref{EllisWormhole}. The lensed rays are traversing a spherical excision of radius~$\ell$, that has  then been `sewed-up'. In some sense, they are taking a short-cut through a piece of space that has been removed. The angular deflection also carries the `wrong' sign, so that the object is optically repulsive. To summarise, a population of such lenses (e.g. early-Universe relics) would thus seem hard --- but not impossible --- to observe, and their unusual characteristics would provide a clear signal for \AE{}ST.

\paragraph*{In this work} Our objective is to show that (i) the exact solutions proposed in~\cref{ElingJacobsonWormholeLineElement,eq:time_norm,RadiusInTermsOfSpaceFunc,EllisWormhole} do indeed satisfy the \AE{}ST field equations, and (ii) that they have the above phenomenological properties.
In particular whilst the mimicker solution does evidently have a Newtonian limit, we will not be overly concerned with matching either the mimicker or the lens to the MOND regime, because the question of strong- and weak-field matching in \AE{}ST is not yet well understood: there is a generic expectation that this matching process is intimately connected with the dynamical history of the system. Two aspects need to be investigated in further work. Firstly, dynamical production and stability against perturbations, in the context of cosmological/astrophysical backgrounds and timescales, leading to abundance predictions. Secondly, computation of precise observational signatures leading to multi-messenger forecasts. In~\cref{section: Equations of motion} we will obtain the field equations of \AE{}ST. In~\cref{section: wormhole solutions} we obtain the compact solutions, with a cosmological solution in~\cref{section: A Cosmological Solution}. We conclude in~\cref{section: Conclusion and discussion}. Technical appendices follow, including two of general interest. In~\cref{ConservedCurrent}, we show that the acceleration of static \ae{}ther is \emph{generally} a conserved current. In~\cref{appendix: Nature of the horizon} we show that the BH mimicker is a wormhole \emph{in name only}: instead of leading to another universe, it leads to a singular Killing horizon.

\paragraph*{Conventions} We try to match~\cite{Skordis:2020eui,Skordis:2021mry}, using the signature~$(-,+,+,+)$, and natural units~$c\equiv\hbar\equiv 1$. Further conventions are introduced as needed.

\section{Theoretical development}\label{TheoreticalDevelopment}

\subsection{Overview of theories}

\paragraph*{Scalar-tensor theory} The simplest extension of GR is the massless scalar-tensor theory
\begin{equation}\label{original_action_ST}
	\tensor*{S}{_{\text{ST}}}  \equiv\int \mathrm{d} ^4x \, \frac{\sqrt{-g}}{2\kappa}\bigg[R\mp\tensor{\nabla}{_\mu}\varphi\tensor{\nabla}{^\mu}\varphi\bigg],
\end{equation}
where~$R$ is the Ricci scalar,\footnote{Our conventions are~$R\equiv\tensor*{R}{^\mu_\mu}$ for the Ricci tensor~$\tensor{R}{_{\mu\nu}}\equiv\tensor{R}{^\sigma_{\mu\sigma\nu}}$, where the Riemann tensor is~$\tensor{R}{^\sigma_{\rho\mu\nu}}\equiv\tensor{\partial}{_\mu}\tensor*{\Gamma}{^\sigma_{\rho\nu}}-\tensor{\partial}{_\nu}\tensor*{\Gamma}{^\sigma_{\rho\mu}}+\dots$ and~$\tensor*{\Gamma}{^\mu_{\nu\rho}}\equiv\tensor{g}{^{\mu\sigma}}\big(\tensor{\partial}{_{(\nu}}\tensor{g}{_{\rho)\sigma}}-\frac{1}{2}\tensor{\partial}{_{\sigma}}\tensor{g}{_{\nu\rho}}\big)$ is the Christoffel symbol.} and~$\kappa \equiv 8 \pi G$ is the Einstein constant with~$G$ the Newton--Cavendish constant. The shift-symmetric scalar~$\varphi$ is made dimensionless in~\cref{original_action_ST}, and with our signature the `$(\mp)$' sign allows canonical `$(-)$' or phantom `$(+)$' character. One of our key findings will be that this primitive class of models goes a long way towards explaining the solutions of \AE{}ST, even though the latter is a far more intricate model. This is especially surprising, because whilst \AE{}ST also contains a shift-symmetric scalar~$\varphi$, the operators of this field are everywhere entangled with an additional dimensionless vector field~$\tensor{A}{^\mu}$, which has an \ae{}ther character.

\paragraph*{Einstein-\ae{}ther theory} By contrast with the scalar-tensor theory of~\cref{original_action_ST}, the simplest \ae{}ther model is Maxwell-type Einstein-\ae{}ther (E\AE{}) theory~\cite{Ae, Ae_review}. The action, using~$\tensor{F}{_\mu _\nu}\equiv \tensor{\nabla}{_\mu} \tensor{A}{_\nu} - \tensor{\nabla}{_\nu} \tensor{A}{_\mu}$, is given by
\begin{equation}\label{original_action_EAE}
\begin{aligned}
	\tensor*{S}{_{\text{E\AE{}}}}&\equiv\int \mathrm{d} ^4x \, \frac{\sqrt{-g}}{2\kappa}\Bigg[R -\frac{\KB{}}{2} \tensor{F}{^{\mu\nu}} \tensor{F}{_{\mu \nu}}
\\
&\hspace{80pt}- \lambda\left(\tensor{A}{^\mu} \tensor{A}{_\mu} +1\right) \Bigg], 
\end{aligned}
\end{equation}
where the dimensionless \ae{}ther coupling lies in the range
\begin{equation}\label{AetherCouplingRange}
    0 < \KB{} < 2 \,\, ,
\end{equation}
mandated by the stability of perturbations.\footnote{Note in particular that the solutions to the E\AE{} field equations are not always continuous as~$\KB{}\to 0$, though as a rule of thumb the E\AE{} phenomenology coincides with pure GR in this limit.} The general idea is to identify a preferred frame at each point in the spacetime as having a four-velocity given by the vector field~$\tensor{A}{^\mu}$. For this interpretation to hold, the vector must be everywhere timelike and with unit length: these conditions are ensured in~\cref{original_action_EAE} by the dynamics of~$\lambda$, which is a Lagrange multiplier. Ignoring the multiplier terms leads back to Einstein--Maxwell theory in which~$\sqrt{\KB{}/\kappa}\tensor{A}{^\mu}$ is the electromagnetic gauge potential.

\paragraph*{\AE{}ther-scalar-tensor theory} Going beyond scalar-tensor and E\AE{} theories, \AE{}ST blends the dynamics of~$\varphi$ and~$\tensor{A}{^\mu}$ through the action
\begin{equation}\label{original_action_AEST}
\begin{aligned}
	\tensor*{S}{_{\text{\AE{}ST}}} & \equiv\int \mathrm{d} ^4x \, \frac{\sqrt{-g}}{2\kappa} \Bigg[R -\frac{\KB{}}{2} \tensor{F}{^{\mu\nu}} \tensor{F}{_{\mu \nu}}
\\
&\hspace{30pt} +2\left(2-\KB{} \right) \tensor{J}{^\mu} \tensor{\nabla}{_\mu} \varphi - \left(2-\KB{}\right) \mathcal{Y}
\\
&\hspace{30pt} -\mathcal{F}\left(\mathcal{Y}, \mathcal{Q} \right) - \lambda\left(\tensor{A}{^\mu} \tensor{A}{_\mu} +1\right) \Bigg] \,\, , 
\end{aligned}
\end{equation}
where, as in E\AE{} theory, we assume~\cref{AetherCouplingRange} for reasons of stability~\cite{Lin_stab_in_mk_AeST}; the other quantities in~\cref{original_action_AEST} are defined
\begin{equation}\label{AuxiliaryQuantities}
\begin{gathered} 
\tensor{J}{_\mu} \equiv \tensor{A}{^\nu} \tensor{\nabla}{_\nu} \tensor{A}{_\mu},\quad 
\tensor{q}{_\mu_\nu} \equiv \tensor{g}{_\mu_\nu} + \tensor{A}{_\mu} \tensor{A}{_\nu},
    \\
\mathcal{Q} \equiv \tensor{A}{^\mu} \tensor{\nabla}{_\mu} \varphi, \quad
	\mathcal{Y} \equiv \tensor{q}{^\mu^\nu} \tensor{\nabla}{_\mu}\varphi   \tensor{\nabla}{_\nu} \varphi.
\end{gathered}
\end{equation} 
In particular, if~$\tensor{A}{^\mu}$ is interpreted as the four-velocity of the preferred frame, then it follows from~\cref{AuxiliaryQuantities} that~$\tensor{J}{^\mu}$ must be the four-acceleration of that frame. In general,~$\mathcal{F}(\mathcal{Y}, \mathcal{Q})$ is a free function of the scalars~$\mathcal{Y}$ and~$\mathcal{Q}$. One motivated form for this function was proposed in~\cite{Skordis:2020eui}, whereby
\begin{equation}\label{F_split_into_Q_and_Y}
	\mathcal{F}(\mathcal{Y}, \mathcal{Q}) = \left(2-\KB{}\right)\mathcal{J} \left(\mathcal{Y} \right)-\mathcal{F}_{20}\left(\mathcal{Q} - \mathcal{Q}_0 \right)^2  \,,
\end{equation}
where~$\mathcal{Q}_0$ has mass-dimension one and
\begin{equation}\label{FBound}
	\mathcal{F}_{20}>0,
\end{equation}
is dimensionless. The idea behind~\cref{F_split_into_Q_and_Y} is that the~$\mathcal{Q}$-dependent term is responsible for reproducing a cosmological CDM-like component, while the~$\mathcal{Y}$-dependent term is responsible for explaining the missing mass problem in galaxies~\cite{Skordis:2020eui}. Heuristically,~$\mathcal{Q}$ carries the time derivatives of the scalar field, while~$\mathcal{Y}$ carries the spatial derivatives, so that~$\mathcal{Q}$ is relevant in time-dependent situations (e.g. cosmology) while~$\mathcal{Y}$ is relevant to static or quasi-static situations (e.g. galaxies). To reproduce the successes of MOND in describing gravity on galactic scales, the function~$\mathcal{J}(\mathcal{Y})$ in~\cref{F_split_into_Q_and_Y} is (somewhat hazily) defined as 
\begin{equation}\label{J_function}
    \mathcal{J}\left( \mathcal{Y}\right) \equiv
	\begin{cases}
         \lambda_s   \mathcal{Y}, \quad  &\mathcal{J}(\mathcal{Y}) \gg a_0^2,\\
         \frac{2 \lambda_s\mathcal{Y}^{3/2}}{3 \left(1+\lambda_s \right)a_0}, \quad &\mathcal{J}(\mathcal{Y}) \ll a_0^2,
	\end{cases}
\end{equation}
where~$a_0 \sim \SI{1e-10}{\metre\per\second\squared}$ is the MOND acceleration scale, and the dimensionless parameter~$\lambda_s$ obeys
\begin{equation}\label{LambdaS}
    0 < \lambda_s,
\end{equation}
again to avoid an instability.\footnote{Note that~\cite{Lin_stab_in_mk_AeST} suggests the more conservative bound~$1 < \lambda_s$, though the bound given in~\cref{LambdaS} is the one most commonly assumed.}. In galaxies,~\cref{J_function} will reproduce Newtonian gravity at accelerations large compared to~$a_0$, and MOND-like gravity at accelerations small compared to~$a_0$. In the following we will mostly restrict ourselves to the Newtonian ansatz for~$\mathcal{J}\left( \mathcal{Y}\right)$, which is relevant to small radii, appropriate for BH and wormhole solutions. Without introducing new parameters to modify the Newtonian ansatz, we will find that the strong-field departures from Newtonian phenomena have remarkably promising observational implications. Having already restricted to the minimal quadratic~$\mathcal{Q}$ ansatz required for effective CDM, our decision to neglect the MOND ansatz allows us to avoid the further set of arbitrarily many parameters that would interpolate between the two ans\"atze for~$\mathcal{J}\left( \mathcal{Y}\right)$.\footnote{Note that, as initially formulated in terms of arbitrary functions, \AE{}ST is not a predictive model with finitely many parameters; the same problem is seen in~$f(R)$ theories.} Even with this maximally predictive approach, the prior constraints on the five remaining parameters in \AE{}ST, namely~$G$,~$\KB{}$,~$\lambda_s$,~$\mathcal{F}_{20}$ and~$\mathcal{Q}_0$, remain very weak. They include
\begin{equation}\label{GoldilocksEstimate}
	\frac{2\left(1+\lambda_s\right)}{\lambda_s\left(2-\KB{}\right)}\mathcal{F}_{20}\mathcal{Q}_0^2 \sim \SI{1}{\per\mega\parsec\squared} \,\, ,
\end{equation}
which was arrived at in~\cite{Mistele2023} as a `goldilocks' estimate for producing MOND in galaxies, and suppressing it on the scale of galaxy clusters (see also~\cite{Durakovic:2023out}).\footnote{In more detail, a value in~\cref{GoldilocksEstimate}~$\lesssim\SI{1}{\per\mega\parsec\squared}$ is preferred to avoid oscillatory accelerations in galaxies, which would be predicted by \AE{}ST but which are not observed. Meanwhile, galaxy clusters appear to exhibit an offset from the MOND acceleration relation: this may be explained by~$\gtrsim\SI{1}{\per\mega\parsec\squared}$. Only a value~$\sim\SI{1}{\per\mega\parsec\squared}$ is consistent with both regimes.} An extra subtlety also affects~$G$ (i.e.~$\kappa$) in~\cref{original_action_AEST}, which in \AE{}ST should be considered a \emph{bare} coupling.\footnote{The theories~\cref{original_action_ST,original_action_EAE,original_action_AEST} are all assumed to be low-energy effective theories. Our use of `bare' has nothing whatever to do with renormalisation, and means only that the classical dynamics of the \ae{}ther and scalar impose further modifications to the Newtonian limit, beyond GR.} The \emph{measured} Newtonian~$\GN{}\approx\SI{6.67e-11}{\metre\cubed\per\kilogram\per\second\squared}$ was derived in~\cite{Skordis:2020eui}, and found to be given by the formula
\begin{equation}\label{GN}
	\GN{} \equiv \frac{\left(1+\lambda_s\right)2G}{\lambda_s\left(2 - \KB{}\right)} \,\, ,
\end{equation}
where~$G$,~$\KB{}$ and~$\lambda_s$ are otherwise unconstrained.

\subsection{Relevant solutions}\label{subsection: Previous exact solutions}

\paragraph*{Ellis--Bronnikov drainhole} Solutions in scalar-tensor theory have been extensively studied~\cite{WH_ES,EB_WH1,EB_WH2}. The phantom `$(+)$' case of~\cref{original_action_ST} admits the celebrated two-parameter `drainhole' solution of Ellis~\cite{Ellis1973Drainhole} and Bronnikov~\cite{Bronnikov1973ScalarTensor} (see also~\cite{EB_WH1,EB_WH2,LvWormhole,Shadow_WH_ES}). Both sides of this asymmetric wormhole are asymptotically flat. The near side behaves like the Schwarzschild solution at spatial infinity, one of the parameters setting the attractive asymptotic mass; the far side is correspondingly repulsive. The throat of the wormhole is independently dilated by the second parameter. The massless limit is a symmetric, one-parameter wormhole. The absence of a mass means that, throughout the whole spacetime, static observers experience no acceleration. The resulting wormhole has no gravity, and geodesics `coast' freely from one side to the other only if they happen to intersect the throat.

\paragraph*{Fisher--JNW solution} Conversely, if the scalar is canonical, i.e. the `$(-)$' case of~\cref{original_action_ST}, there is a two-parameter analogue of the Ellis--Bronnikov solution attributed first to Fisher~\cite{Fisher1948} and rediscovered much later by Janis, Newman and Winicour (JNW)~\cite{JanisNewmanWinicour1968}. The analogy is exact, in the sense that analytic continuation of the Fisher--JNW solution to the phantom `$(+)$' case just yields the Ellis--Bronnikov drainhole in different coordinates. Non-phantom Fisher--JNW spacetime is also asymptotically flat, but it contains a naked singularity at its center, not a throat. As with the Ellis--Bronnikov solution, however, there exists a massless limit: the naked singularity persists, but it is weightless, so that static observers are not drawn towards it.

\paragraph*{Eling--Jacobson wormhole} Static, spherical solutions to E\AE{} theory in~\cref{original_action_EAE} were studied extensively by Eling and Jacobson in~\cite{BH_in_Ae}. The Eling--Jacobson solution is also massive, with an asymptotically flat near side (which we will think of as an `exterior') consistent with the Newtonian limit of the Schwarzschild solution. The Eling--Jacobson solution, however, does not quite describe a BH, but rather a highly asymmetric wormhole. On the far side (`interior') of its throat lies a very different region which is not asymptotically flat, but which ends at a singular Killing horizon. In the~$\KB{}\to 0$ limit, the throat radius approaches (from above) the value of the Schwarzschild radius associated with the mass, and in fact the whole `exterior'-side geometry approaches the Schwarzschild vacuum. Similar spacetime geometries have been found in
exact solutions to bumblebee gravity~\cite{BH_in_BB}, and more recently in exact solutions to \AE{}ST itself~\cite{costas_stealth_bh}.

\paragraph*{Stealth BH} Ignoring the multiplier term in~\cref{original_action_EAE} yields Einstein--Maxwell theory. It is sometimes forgotten that a unit-timelike condition on the electromagnetic gauge potential is actually a perfectly valid gauge choice in electromagnetism. Indeed, this \emph{Dirac gauge} was among the earliest motivations for \ae{}ther models such as the E\AE{} theory. It is unsurprising, therefore, that one branch of solutions to E\AE{} theory is actually Reissner--Nordström. Less obvious is the fact that this solution is also inherited by \AE{}ST, as shown recently in~\cite{costas_stealth_bh}. In the limit of vanishing charge, the Reissner--Nordström geometry cannot be observationally distinguished from that of Schwarzschild, and such field configurations in \AE{}ST have been dubbed `\emph{stealth}' BHs.

\subsection{Overview of equations}\label{section: Equations of motion}

\paragraph*{Simplified action} Our starting point for everything that follows will be to refine~\cref{original_action_AEST} by assuming the form proposed by Skordis and Złosnik in~\cref{F_split_into_Q_and_Y}. We will not, however, enforce the assumptions in~\cref{J_function}, which lead to MOND phenomenology. We rewrite the action as
\begin{equation}\label{Action_general}
\begin{aligned}
	\tensor*{S}{_{\text{\AE{}ST}}^{\text{SZ}}} & \equiv \int \mathrm{d} ^4x \, \frac{\sqrt{-g}}{2\kappa} \Bigg[{R} -\frac{\KB{}}{2} \tensor{F}{^{\mu\nu}}\tensor{F}{_{\mu \nu}}
\\
&\hspace{20pt}+2\left(2-\KB{} \right) \tensor{J}{^\mu} \tensor{\nabla}{_\mu} \varphi - \mathcal{V}\left( \mathcal{Y}\right)
\\
&\hspace{20pt}+\mathcal{F}_{20}\left(\mathcal{Q}-\mathcal{Q}_0 \right)^2 - \lambda\left(\tensor{A}{^\mu} \tensor{A}{_\mu} +1\right) \Bigg] \,\, , 
\end{aligned}
\end{equation}
so that~$\mathcal{J}(\mathcal{Y})$ from~\cref{F_split_into_Q_and_Y} and~$(2-\KB{}) \mathcal{Y}$ from~\cref{original_action_AEST} are combined into the `potential'~$\mathcal{V}(\mathcal{Y})$.

\paragraph*{Covariant equations} The field equations are obtained by taking variations of~\cref{Action_general} with respect to the dynamical degrees of freedom: the metric~$\tensor{g}{_{\mu\nu}}$, the \ae{}ther field~$\tensor{A}{^\mu}$, the scalar field~$\varphi$ and the Lagrange multiplier~$\lambda$. Variation with respect to~$\lambda$ yields the normalization constraint on the \ae{}ther field
\begin{equation}\label{UnitTimelike}
    \tensor{A}{^\mu} \tensor{A}{_\mu} = -1\,\, .
\end{equation}
Even after~\cref{UnitTimelike} has been imposed, the remaining field equations (which include among their components propagating equations as well as constraints) remain somewhat cumbersome, and we confine them to~\cref{appendix: Field equations}. Note that the~$A^\mu$-equation contains a term of the form~$\delta \tensor*{S}{_{\text{\AE{}ST}}^{\text{SZ}}}/\delta\tensor{A}{^\mu}\supset-\sqrt{-g}\lambda\tensor{A}{_\mu}/\kappa$, so that after contraction with an extra factor of~$A^\mu$ followed by an application of~\cref{UnitTimelike} it allows for the Lagrange multiplier~$\lambda$ to be determined algebraically (see~\cref{lag_exp}).

\paragraph*{Component equations} Initially, the line element in the spherically symmetric spacetime is not assumed to be Schwarzschild-like or isotropic; we write
\begin{equation}\label{sph_back}
	  \mathrm{d}  s^2 = -e^{2\mathcal{N}(r)}\,   \mathrm{d}  t^2 + e^{2\mathcal{M}(r)}\,   \mathrm{d}  r^2 + \mathcal{R}(r)^2\,   \mathrm{d}  \Omega^2\,\, , 
\end{equation}
and this provides our ansatz for the metric~$\tensor{g}{_{\mu\nu}}$. The scalar field~$\varphi$ and the \ae{}ther field~$\tensor{A}{^\mu}$ are assumed to share the same spherical symmetry with the spacetime. In the case of the \ae{}ther field, this means that only temporally and radially aligned components are allowed, and together with~\cref{UnitTimelike,sph_back} this yields 
\begin{equation}\label{ae_back}
\left[\tensor{A}{_\mu}\right] = \left[  \cosh \big(\alpha(r)\big)   e^{\mathcal{N}(r)}, \sinh \big(\alpha(r)\big)   e^{\mathcal{M}(r)}, 0, 0\right].
\end{equation}
Finally, we allow~$\varphi$ to additionally have some time dependence, which is related to the first term in~\cref{F_split_into_Q_and_Y}\footnote{It may seem unlikely that time dependence of~$\varphi$ will lead to exact solutions for which the spacetime geometry is static, but in fact this happens in \AE{}ST for the simplest case of Minkowski spacetime (see detailed discussion in~\cite{Lin_stab_in_mk_AeST}). Heuristically,~$\varphi = \mu t + \dots$ introduces a chemical potential~$\mu$ in a statistical mechanics description of the theory. From this perspective, it is natural to expect static spacetime at equilibrium.}
\begin{equation}\label{P_back}
\varphi = \mathcal{Q}_0   \chi(t) +  \psi(r), 
\end{equation}
accordingly, we refer to it as~$\varphi(t,r)$. The component-level equations following from~\cref{sph_back,ae_back,P_back} can be found in~\cref{appendix: Static spherical equations in vacuum}. From this point on, however, we will focus on the solutions with purely time-aligned \ae{}ther (the so-called `\emph{static \ae{}ther}' condition), by setting
\begin{equation}\label{EnforceTimelike}
	\alpha(r) = 0,
\end{equation}
in~\cref{sph_back}.\footnote{The consequences of relaxing this condition are discussed in~\cite{Hsu:2024ftc}.} Following from~\cref{EnforceTimelike} we then find (see discussion in~\cref{ConservedCurrent})
\begin{equation}\label{Q_identity}
\begin{gathered}
\tensor{F}{_{\mu\nu}} \tensor{F}{^{\mu\nu}} = -2\tensor{J}{_\mu}\tensor{J}{^\mu},  \quad \tensor{J}{_{\mu}} \tensor{\nabla}{^\mu}\varphi  = \tensor{J}{_{\mu}} \tensor{\nabla}{^\mu}\psi, \\
	\mathcal{Y} = \tensor{\nabla}{_\mu} \psi \, \tensor{\nabla}{^\mu} \psi, \quad \mathcal{Q} = -\mathcal{Q}_0\frac{\mathrm{d}\chi}{\mathrm{d}t}e^{-\mathcal{N}}.
 \end{gathered}
 \end{equation}
We will also impose the constraint on~\cref{sph_back}
\begin{equation}\label{RadialCoordinate}
	\mathcal{M}(r) = -\mathcal{N}(r) \,.
\end{equation}
The condition in~\cref{RadialCoordinate} can always be met by a suitable rescaling of~$r$. In general, this choice will not coincide with~$\mathcal{R}(r)=r$. Accordingly,~$r$ need not be tied to the area of enclosing two-spheres, so the resulting coordinates need not be Schwarzschild-like. Note, however, that in general, the~$r$ coordinate scaled according to~\cref{sph_back,RadialCoordinate} is proportional to the affine parameter of radial null geodesics. When~\cref{EnforceTimelike,P_back,RadialCoordinate} are imposed on the field equations, a constraint from~\cref{eqA1} yields (see~\cref{appendix: Wormhole solution: mathematical detail})
\begin{equation}\label{AdditionalCondition_0}
\begin{aligned}
	\mathcal{Q}_{0}\Bigg(
	& \left[ \left(2-\KB{}\right) \frac{{\mathrm{d}} \mathcal{N}}{{\mathrm{d}} r} + \left( \mathcal{F}_{20} - \frac{{\mathrm{d}} \mathcal{V}}{{\mathrm{d}} \mathcal{Y}} \right)\frac{{\mathrm{d}}\psi}{{\mathrm{d}} r} \right]\frac{{\mathrm{d}}\chi}{{\mathrm{d}} t}
 \\
	&\hspace{110pt}-\mathcal{F}_{20} e^{\mathcal{N}} \frac{{\mathrm{d}} \psi}{{\mathrm{d}}r} 
	\Bigg) = 0,
\end{aligned}
\end{equation}
and~\cref{AdditionalCondition_0} will be instrumental in guiding our search for solutions. We will show that the spherically symmetric, static solutions to the \AE{}ST field equations can have both scalar-tensor and E\AE{} character. In particular, we show that \AE{}ST in~\cref{original_action_AEST} inherits the Eling--Jacobson wormhole of E\AE{} theory in~\cref{original_action_EAE}, but with an `effective' \ae{}ther coupling~$\KBE{}$, which is shifted relative to the bare~$\KB{}$ by the scalar current. This solution forms the basis for our BH mimicker. We will uncover a second branch of solutions for which the \ae{}ther acceleration~$\tensor{J}{^\mu}$ vanishes: this describes an ultrastatic (massless) spacetime, and forms the basis for our Shapiro-free lens.

\section{Asymptotically flat cases}\label{section: wormhole solutions}
\paragraph*{Guiding equations} To obtain asymptotically flat solutions, we need to neglect the MOND regime. Without MOND, by referring to~\cref{J_function} we see that the function~$\mathcal{V}(\mathcal{Y})$ in~\cref{Action_general} is linear, i.e.
\begin{equation}\label{NoMOND}
	\mathcal{V}(\mathcal{Y}) = \left(2-\KB{}\right)\left(1+\lambda_s\right) \mathcal{Y}.
\end{equation}
Our discussion will be driven in particular by two component-level conditions that arise out of the field equations when the assumptions in~\cref{EnforceTimelike,P_back,RadialCoordinate,NoMOND} are applied. Firstly, is found that~\cref{eqG01} provides a necessary condition for the existence of non-trivial solutions
\begin{equation}
	\mathcal{F}_{20} \mathcal{Q}_{0} \mathcal{R} \frac{\mathrm{d}\psi}{\mathrm{d}r} \left( e^{ \mathcal{N}} - \frac{\mathrm{d}\chi}{\mathrm{d}t} \right) = 0 \,\, . \label{condition_for_timelike_A}
\end{equation}
We focus on the case where~\cref{condition_for_timelike_A} is satisfied because
\begin{equation}\label{NoDust}
\mathcal{F}_{20}\mathcal{Q}_0=0.
\end{equation}
Referring back to~\cref{F_split_into_Q_and_Y}, the condition in~\cref{NoDust} implies that we are neglecting the capacity of \AE{}ST to provide a CDM component in the cosmological background and perturbation theory --- again, this ought to be safe for strong fields. Secondly, under~\cref{NoDust,NoMOND} the complicated constraint in~\cref{AdditionalCondition_0} reduces to 
\begin{equation}\label{AdditionalCondition}
	\mathcal{Q}_{0}\frac{\mathrm{d}\chi}{\mathrm{d}t}\left(\frac{\mathrm{d}\mathcal{N}}{\mathrm{d}r} - \left(1+\lambda_s\right)\frac{\mathrm{d}\psi}{\mathrm{d}r} \right) = 0,
\end{equation}
which has a much simpler form.

\paragraph*{Heuristic action} When~\cref{sph_back,ae_back,P_back,EnforceTimelike,NoDust,NoMOND} hold,~\cref{Q_identity} can be used to re-write the action in~\cref{Action_general} as\footnote{Note that the existence of a covariant reduced action such as that given in~\cref{Effective_Action} is not guaranteed in general. For highly symmetric spacetimes, however, one is guaranteed to be able to write a non-covariant minisuperspace action which reproduces valid field equations~\cite{Palais1979ThePO,Fels_2002} --- we are not using this technique.}
\begin{align}
	&\tensor*{S}{_{\text{\AE{}ST}}^{\text{SZ}}} \cong\int \mathrm{d} ^4x \, \frac{\sqrt{-g}}{2\kappa} \Big[R +{\KB{}} \tensor{J}{^{\mu}} \tensor{J}{_{\mu}}- \lambda\left(\tensor{A}{^\mu} \tensor{A}{_\mu} +1\right)
	\nonumber \\
	&\hspace{35pt}+\left(2-\KB{}\right) \Big(2\tensor{J}{^\mu} - \left(1+\lambda_s\right) \tensor{\nabla}{^\mu} \psi\Big) \tensor{\nabla}{_\mu} \psi  \Big]. \label{Effective_Action}
\end{align}
We have verified explicitly that the equations of motion derived from~\cref{Effective_Action} reproduce the solutions of the original action in~\cref{Action_general}, expressed in terms of the metric~$g_{\mu \nu}$, the scalar~$\varphi$, and the vector~$A^\mu$, after~\cref{sph_back,ae_back,P_back,EnforceTimelike,NoDust,NoMOND} are imposed.\footnote{The Lagrange multiplier~$\lambda$ implied by~\cref{Effective_Action} differs by a constant factor from that implied by the original action. This affects the energy-momentum tensor, to which~$\lambda$ contributes. This change is, however, exactly compensated for by another change, due to the fact that the metric variation of the two (off-shell) expressions~$- 2\tensor{J}{^\mu}\tensor{J}{_\mu}$ and~$\tensor{F}{^\mu ^\nu}\tensor{F}{_\mu _\nu}$ are not identical, despite having the same on-shell values (see~\cref{Q_identity}). The full energy-momentum tensor implied by~\cref{Effective_Action} thus matches that implied by the original action given our assumptions.} We particularly notice in~\cref{Effective_Action} that all~$\chi(t)$ dependence has disappeared. This is because the `acceleration' of the \ae{}ther defined in~\cref{AuxiliaryQuantities} has only one non-vanishing component~$J_r=\mathrm{d}\mathcal{N}/\mathrm{d}r$, whilst~$q^{tt}$ vanishes due to~\cref{EnforceTimelike}. The absence of~$\chi(t)$ will be exploited in~\cref{subsection: Constantinos' solution extended}. From~\cref{Effective_Action}, the field equation associated with the one remaining scalar~$\psi(r)$ takes the form of a conservation law
\begin{equation}\label{ConservationLaw}
	\tensor{\nabla}{_\mu} \big(\tensor{J}{^\mu} -\left(1+\lambda_s\right)\tensor{\nabla}{^\mu} \psi \big) = 0 \,\,, 
\end{equation}
which reflects the surviving shift symmetry.

\paragraph*{Possible branches} There are evidently at least two ways to realize~\cref{NoDust},~$\mathcal{Q}_0 = 0$ and~$\mathcal{F}_{20} = 0$. Perhaps surprisingly, the~$\mathcal{F}_{20} (\mathcal{Q} - \mathcal{Q}_0)^2$ term in~\cref{Action_general} is absent from~\cref{Effective_Action} not just in the case~$\mathcal{F}_{20} = 0$ but also for~$\mathcal{Q}_0 = 0$. This is because the condition~$\mathcal{Q}_0 = 0$ is not only a statement about the constants parameterizing the action in~\cref{Action_general}, but also about the scalar field in~\cref{P_back}. In particular, it eliminates all time dependence in~$\varphi$, so that~\cref{Q_identity} leads to~$\mathcal{Q}$ vanishing identically, along with~$\mathcal{F}_{20} (\mathcal{Q} - \mathcal{Q}_0)^2$. Despite the action in~\cref{Effective_Action} having the same form for~$\mathcal{F}_{20} = 0$ and~$\mathcal{Q}_0 = 0$, the solutions in either case are not identical, as we will see below. 

\subsection{New branches of exact solutions}\label{subsection: New solution}

\paragraph*{General case} The first realisation of~\cref{NoDust} will be through the condition
\begin{equation}\label{NewCase}
 \mathcal{F}_{20}\ne0, \quad \mathcal{Q}_0 =0 \,.
\end{equation}
In this case, we satisfy~\cref{AdditionalCondition} by assuming no time dependence in~$\varphi(t,r)$, so that~\cref{P_back} implies simply
\begin{equation}\label{NoTimeDependence}
	\chi(t)=0 \,.
\end{equation}
The crucial point is that the reduced covariant action for \AE{}ST in~\cref{Effective_Action} reduces to the corresponding action for E\AE{} theory in~\cref{original_action_EAE} if~$\tensor{\nabla}{_\mu}\psi$ is proportional to~$\tensor{J}{_\mu}$. For example, if
\begin{equation}\label{scalar_current}
\tensor{\nabla}{_\mu} \psi = q \tensor{J}{_\mu},
\end{equation}
for some constant~$q$, then~\cref{Effective_Action} reduces to~\cref{original_action_EAE} in which~$\KB{}$ is replaced by an `effective' \ae{}ther coupling
\begin{equation}\label{modified_KB}
	\KBE{} \equiv \KB{}+\left(2-\KB{}\right)q\Big[2- \left(1+\lambda_s\right)q\Big] \,\, . 
\end{equation}
This raises the possibility of `lifting' solutions from the original E\AE{} theory to \AE{}ST. In component form, the condition in~\cref{scalar_current} implies
\begin{equation}\label{PsiBehaviour}
	\frac{\mathrm{d}\psi}{\mathrm{d}r}=q\frac{\mathrm{d}\mathcal{N}}{\mathrm{d}r},
\end{equation}
but~\cref{PsiBehaviour} is not the only condition on~$\psi(r)$. It is shown in~\cref{ConservedCurrent} that for static \ae{}ther in static geometries which solve the E\AE{} field equations, one always has the identity
\begin{equation}\label{ConservedCurrentEquation}
	\tensor{\nabla}{_\mu}\tensor{J}{^\mu} = 0.
\end{equation}
If~\cref{ConservedCurrentEquation} holds, then~\cref{ConservationLaw} requires that the scalar field must satisfy the massless Klein--Gordon equation
\begin{equation}\label{KleinGordon}
	\tensor{\nabla}{_\mu}\tensor{\nabla}{^\mu} \, \psi = 0 \,.
\end{equation}
We reiterate that the reason for suppressing time dependence in~\cref{NoTimeDependence} is that otherwise the heuristic action cannot reduce to E\AE{} theory. By combining~\cref{AetherCouplingRange,LambdaS} with~\cref{modified_KB} we obtain the physical range
\begin{equation}\label{NewAetherCouplingRange}
	2q\Big[2 - (1+\lambda_s)   q\Big] < \KBE{} < 2 \,\, .
\end{equation}
By comparing~\cref{AetherCouplingRange} with~\cref{NewAetherCouplingRange}, we see how the upper bounds of~$\KBE{}$ and~$\KB{}$ are the same, whilst the ratio of the spatial gradient of the scalar to the \ae{}ther acceleration in~\cref{scalar_current}, and the parameter~$\lambda_s$, alter the lower bound of~$\KBE{}$.

\paragraph*{Scalar hair} Note that any choice of~$q$ other than~$q=(1+\lambda_s)^{-1}$ implies that there is a non-vanishing scalar current at spatial infinity. For that reason, such solutions were dismissed in~\cite{Reyes:2024oha} in the context of neutron stars. Indeed, such a constraint is required for consistency in the neutron star case, because the current flowing out at spatial infinity would otherwise not be balanced by any incoming current elsewhere (as required by charge conservation); the same holds for discussions of galaxies and galaxy clusters in \AE{}ST~\cite{Mistele2023,Durakovic:2023out,Skordis:2020eui}. In our case, there \emph{is} always a balancing current and all equations of motion can consistently be satisfied with a non-zero scalar current. Thus, in the following, we allow general values of~$q$. That said, the balancing current often comes from regions where the spacetime is singular, so that the origin of these currents remains somewhat mysterious. For later reference, we record the value~$\KBEN{}$ following from~\cref{PsiBehaviour} of~$\KBE{}$ for the case of no current at infinity, as\footnote{Note that in~\cite{costas_stealth_bh} this corresponds precisely to~$\KBEN{}\equiv2\tilde{n}$.}
\begin{align}
    &{\KBEN{}} \equiv \frac{\lambda_s \KB{} + 2}{1+\lambda_s} \, . \label{KBE0}
\end{align}
Unlike for a more general choice of~$q$, the value in~\cref{KBE0} is strictly positive. Thus, the GR limit~$\KBE{}\to 0$ is only accessible when one allows a scalar current at infinity.

\paragraph*{Ellis--Bronnikov drainhole}
In the (presumably non-physical) case where~\cref{AetherCouplingRange} is violated such that
\begin{equation}\label{UnphysicalCase}
	\KB{} > 2,
\end{equation}
the logic used in~\cref{NewAetherCouplingRange}, shows that the `effective' \ae{}ther coupling also obeys
\begin{equation}\label{UnphysicalCase2}
	\KBE{} > 2.
\end{equation}
It is shown in~\cref{appendix: Wormhole solution: mathematical detail} that the solutions in this case are Ellis--Bronnikov drainholes, such that the line element in~\cref{sph_back} is constrained beyond~\cref{RadialCoordinate} to
\begin{subequations}\small
\begin{align}
	\mathcal{N}(r) &= -\frac{2}{\sqrt{2\KBE{}-4}} \cot^{-1} \left[ \frac{2r}{\GN{}M\sqrt{2\KBE{}-4}}\right], \label{EB_wormholeA}
    \\
	\mathcal{R}(r) &=\frac{1}{\sqrt{2}}e^{-\mathcal{N}(r)}\sqrt{2r^2 + \GN{}^2M^2\left(\KBE{}-2\right)}  \,\,.\label{EB_wormholeB}
\end{align} 
\end{subequations}
In~\cref{EB_wormholeA,EB_wormholeB} the constant~$M$ is the mass of the drainhole at spatial infinity.\footnote{Note that~$M$ appears everywhere in combination with~$\GN{}$. We have chosen it this way, in line with the expectation that the Newtonian limit of \AE{}ST be described by~\cref{GN}. Ultimately, however,~$M$ is an integration constant whose interpretation is flexible.} This spacetime has two asymmetric but asymptotically flat regions, and the embedding in Minkowski space is especially evident on the~$r\to\infty$ side, whereupon~$\mathcal{N}(r)\to 0$. From~\cref{EB_wormholeB} the radius of the throat, i.e. the minimum-area two-sphere, is found to be
\begin{equation}\label{EB_wormhole_throat_radius}
	\mathcal{R}_{\text{T}}=\sqrt{\frac{\KBE{}}{2}}\GN{}M\exp \left[\frac{\tan^{-1} \left( \sqrt{\frac{\KBE{}}{2}-1}\right)}{\sqrt{\frac{\KBE{}}{2}-1}} \right].
\end{equation}
As mentioned in~\cref{subsection: Previous exact solutions}, the geometry defined by~\cref{sph_back,RadialCoordinate,EB_wormholeA,EB_wormholeB} also solves scalar-tensor theory in~\cref{original_action_ST} in the phantom `$(+)$' case.

\paragraph*{First extended Eling--Jacobson solution} If~\cref{UnphysicalCase} is not met, but rather~$\KB{}$ lies within the physical range of~\cref{AetherCouplingRange}, then the solutions (see~\cref{appendix: Wormhole solution: mathematical detail}) correspond to the Eling--Jacobson wormhole~\cite{BH_in_Ae, Ae_BH_new_chart_analytical_Oost_2021, Ae_BH_new_chart_numerical_Gao_2013}. In this case~\cref{EB_wormholeA,EB_wormholeB} are replaced by
\begin{subequations} 
\begin{align}
	\mathcal{N}(r) &=\frac{\ln\left[ 1 - \sqrt{4-2\KBE{}}\frac{\GN{}M}{r} \right] }{\sqrt{4-2\KBE{}}}  \,\,,\label{JE_wormholeA}
    \\
	\mathcal{R}(r) &= r\left[ 1 - \sqrt{4-2\KBE{}}\frac{\GN{}M}{r} \right]^{\frac{1}{2}-\frac{1}{\sqrt{4-2\KBE{}}}},\label{JE_wormholeB}
\end{align} 
\end{subequations}
and it may be shown that the asymptotically flat portion of~\cref{sph_back,RadialCoordinate,JE_wormholeA,JE_wormholeB} corresponds to~\cref{ElingJacobsonWormholeLineElement,eq:time_norm,RadiusInTermsOfSpaceFunc} in different coordinates. Once again,~$M$ may be interpreted as the mass of the solution at spatial infinity~$r\to\infty$, where asymptotic flatness again follows from~$\mathcal{N}(r)\to 0$. From~\cref{JE_wormholeB} the formula for the throat radius given in~\cref{EB_wormhole_throat_radius} is now replaced by
\begin{equation} 
	\mathcal{R}_{\text{T}}=\sqrt{\frac{\KBE{}}{2}} \GN{}M \exp \left[\frac{\tanh^{-1} \left( \sqrt{1-\frac{\KBE{}}{2}}\right)}{\sqrt{1-\frac{\KBE{}}{2}}} \right]. \label{JE_wormhole_throat_radius}
\end{equation}
After passing through the throat, the volume element diverges as~$r\to\Rh$ from above, where
\begin{equation}\label{NullTime}
	\Rh{}\equiv 2\GN{}M\sqrt{1-\KBE{}/2}.
\end{equation}
It can be shown that this surface is a Killing horizon; in~\cite{BH_in_Ae}, it was also conjectured to be a null singularity. Moreover, this behavior is not uniquely seen in E\AE{} theory or in \AE{}ST. Instead, it seems to be a general phenomenon in other Lorentz-violating theories, such as bumblebee gravity~\cite{BH_in_BB}. Although it seems possible to extend the spacetime described by~\cref{sph_back,RadialCoordinate,JE_wormholeA,JE_wormholeB} to the region~$0< r < \Rh{}$ for certain values of~$\KBE{}$, we prove in~\cref{appendix: Nature of the horizon} that photon geodesics cannot be traced through the Killing horizon. As mentioned in~\cref{subsection: Previous exact solutions}, this geometry is best known as a solution to the E\AE{} theory in~\cref{original_action_EAE} though --- as we will see in~\cref{subsection: Constantinos' solution extended} --- it is already known to occur in \AE{}ST, albeit as part of a quite different solution branch to that considered here, in that the scalar is time-dependent~\cite{costas_stealth_bh}.\footnote{Note also that the~$\KBE=0$ (pure Schwarzschild) special case of~\cref{JE_wormholeA,JE_wormholeB}, with a truly static scalar as in our setup, was also idenfified in~\cite{costas_stealth_bh}.}

\paragraph*{Anti-Ellis--Bronnikov solution} We now relax the condition~\cref{scalar_current}, and so lose the connection to E\AE{} theory; as a consequence our results will no longer depend on~$\KBE{}$. We impose the new condition
\begin{equation}\label{NewCondition}
\tensor{J}{^\mu} = 0,
\end{equation}
and it can be shown (see~\cref{appendix: Wormhole solution: mathematical detail}) that~\cref{JE_wormholeA,JE_wormholeB} are now replaced by
\begin{equation}\label{Ellis_wormholeA}
\mathcal{N}(r)=0, \quad \mathcal{R}(r) = \sqrt{r^2-\ell^2} \,\, , 
\end{equation}
where~$\ell$ is an integration constant, whilst the condition in~\cref{PsiBehaviour,KleinGordon} is replaced by
\begin{equation}\label{Ellis_wormholeB}
	\psi(r) = \pm\frac{\sqrt{2}\tanh^{-1}\left(\ell/r\right)}{\sqrt{\left(2-\KB{}\right)\left(1+\lambda_s\right)}} \,\,.
\end{equation}
To give an interpretation for~$\ell$, it follows from~\cref{Ellis_wormholeA} that as~$r\to |\ell|$ from above the scalar~$\psi(r)$ will diverge. Thus, the geometry defined by~\cref{sph_back,RadialCoordinate,Ellis_wormholeA} --- which is evidently the `$(-)$' configuration of that described in~\cref{EllisWormhole} --- appears to describe a naked singularity. It is, moreover, \emph{ultrastatic}, in stark contrast to the geometries in~\cref{EB_wormholeA,JE_wormholeA}. Indeed, in the cases of the Ellis--Bronnikov drainhole and the Eling--Jacobson wormhole, the redshift function~$\mathcal{N}(r)$ has a non-trivial dependence on~$r$. Parametrically, by keeping~$\KBE{}$ constant, the ultrastatic limit~$\mathcal{N}(r)\to 0$ corresponds to the massless limit~$M\to 0$ in both solutions --- and in both solutions this limit is just Minkowski spacetime. As mentioned in~\cref{subsection: Previous exact solutions}, however, the Ellis--Bronnikov drainhole has another ultrastatic, massless limit in which the throat~$\mathcal{R}_{\text{T}}$ is held constant in~\cref{EB_wormhole_throat_radius} through a simultaneous tuning of~$\KBE{}$. The resulting symmetric wormhole is precisely described by the geometry in~\cref{Ellis_wormholeA} under the analytic continuation
\begin{equation}\label{AnalyticContinuation}
\ell\mapsto i\ell,
\end{equation}
--- i.e. the `$(+)$' configuration of~\cref{EllisWormhole} --- and for this reason we refer to~\cref{Ellis_wormholeA} as `anti-Ellis--Bronnikov' spacetime. Whilst the geometry remains real under this analytic continuation, by tracing~\cref{AnalyticContinuation} through~\cref{Ellis_wormholeB} we find that the scalar field~$\psi(r)$ becomes imaginary. It is natural, therefore, to conjecture that anti-Ellis--Bronnikov spacetime is an exact solution to the scalar-tensor theory in~\cref{original_action_ST} in the canonical `$(-)$' case. Referring back to our discussion in~\cref{subsection: Previous exact solutions}, this would suggest that it corresponds to an ultrastatic limit of Fisher--JNW spacetime.

\subsection{Connection to previous branches}\label{subsection: Constantinos' solution extended}

\paragraph*{Second extended Eling--Jacobson solution} We now enforce~\cref{NoDust} through the condition complementary to that given in~\cref{NewCase}, namely
\begin{equation}\label{SkordisCase}
 \mathcal{F}_{20} = 0\,, \quad \mathcal{Q}_0 \ne 0 \,.
\end{equation}
The scenario in~\cref{SkordisCase} was considered in~\cite{costas_stealth_bh}, where~\cref{NoTimeDependence} is replaced by assuming a linear time dependence of the scalar
\begin{equation}\label{LinearTime}
 \chi(t) = t \,.
\end{equation}
The motivation for~\cref{LinearTime} is that it matches the expected late-time cosmological behavior for the scenario envisioned originally in~\cite{Skordis:2020eui}. Here, we point out that a more general solution is possible, with an \emph{arbitrary} time dependence of the scalar replacing~\cref{LinearTime}. That this is possible may be expected from the fact that the existence of a time-dependent~$\chi(t)$ scalar field does not modify the heuristic action in~\cref{Effective_Action}. Due to the time dependence in the scalar, the constraint~\cref{AdditionalCondition} now becomes non-trivial, since the overall prefactor~$\mathrm{d}\chi/\mathrm{d}t$ no longer vanishes. In fact, the constraint now requires that the conserved current associated with the scalar's shift symmetry (see~\cref{ConservationLaw}) vanishes identically according to
\begin{align}\label{NewestCondition}
 \tensor{J}{^\mu}-\left(1+\lambda_s\right)\tensor{\nabla}{^\mu}\psi = 0 \,.
\end{align}
The on-shell condition in~\cref{NewestCondition} may be viewed as a special case of~\cref{scalar_current}, namely
\begin{align}\label{NewCondition2}
    q=(1+\lambda_s)^{-1} \,.
\end{align}
With~\cref{NewCondition2} satisfied, all the results of~\cref{subsection: New solution} now apply here. In particular, and assuming the physical range in~\cref{AetherCouplingRange}, the metric is that of~\cref{JE_wormholeA,JE_wormholeB}, but restricted to the case where~$\KBEN{}$ is the value of~$\KBE{}$ for our particular choice of~$q$ (see~\cref{KBE0})
\begin{subequations} 
\begin{align}
	\mathcal{N}(r) &=\frac{\ln\left[ 1 - \sqrt{4-2{\KBEN{}}}\frac{\GN{}M}{r} \right] }{\sqrt{4-2{\KBEN{}}}}  \,\,,\label{constantinos_solution_metricA}
    \\
	\mathcal{R}(r) &= r\left[ 1 - \sqrt{4-2{\KBEN{}}}\frac{\GN{}M}{r} \right]^{\frac{1}{2}-\frac{1}{\sqrt{4-2{\KBEN{}}}}} \,,\label{constantinos_solution_metricB}
\end{align} 
\end{subequations}
and the geometry defined by~\cref{sph_back,RadialCoordinate,constantinos_solution_metricA,constantinos_solution_metricB} matches the corresponding metric from~\cite{costas_stealth_bh}. According to~\cref{P_back}, the scalar is given by
\begin{equation}\label{constantinos_solution_scalar}
	\psi(r) = (1+\lambda_s)^{-1} \mathcal{N}(r) + \mathrm{const} \,.
\end{equation}
Since the spacetime geometry is expected to dictate most of the observables in the strong-field regime, it is especially interesting to notice how the BH mimicker geometry arisees across multiple field configurations for the scalar. Finally, we reiterate the differences between the first and second extended Eiling--Jacobson solutions, and the way in which they extend the solutions already presented in~\cite{costas_stealth_bh}. The first extended soluton follows from a truly static scalar field, and allows for an `effective' \ae{}ther coupling as defined in~\cref{modified_KB}, depending only on the static profile of the scalar. The second extended solution allows for a scalar with completely arbitrary time-dependence~$\chi(t)$: it also gives rise to an `effective' \ae{}ther coupling, but this coupling is \emph{not} derived from~\cref{modified_KB}, and is instead fixed to the special value in~\cref{KBE0} as a consequence of~$q=\left(1+\lambda_s\right)^{-1}$. The special case of the first extended solution in which~$\KBE{}=0$ was identified in~\cite{costas_stealth_bh}. The special case of the second extended solution, in which~$\chi(t)=t$, was also identified in~\cite{costas_stealth_bh}.

\section{Cosmological case}\label{section: A Cosmological Solution}

\paragraph*{Einstein static universe} Before concluding (and mostly for completeness), we discuss a cosmological solution that admits a purely timelike \ae{}ther field. While not asymptotically flat, this provides a contrast to the isolated solutions of~\cref{section: wormhole solutions} and connects to the large-scale phenomenology of the theory. We relax the conditions in~\cref{NoMOND,NoDust}, and so re-admit the CDM component to the \AE{}ST action. Introducing a new constant~$\Lambda$ of mass dimension four, the conditions in~\cref{sph_back,RadialCoordinate} are extended (see~\cref{appendix: Wormhole solution: mathematical detail}) by
\begin{align}\label{cosmological_solution_metric}
    \mathcal{N}(r)=0, \quad \mathcal{R}(r) = \frac{\sin\left( \sqrt{\kappa \Lambda}   r\right)}{\sqrt{\kappa \Lambda}},
\end{align}
and from~\cref{cosmological_solution_metric} we see that~\cref{AdditionalCondition_0} can be satisfied by precisely the linear running condition in~\cref{LinearTime}. Evidently, the spacetime geometry described by~\cref{sph_back,RadialCoordinate,cosmological_solution_metric} corresponds to the Einstein static universe, i.e., a solution of the Einstein equations with dust (or CDM) and a cosmological constant~$\Lambda$. With reference to~\cref{Action_general}, it is not therefore surprising that other field equations require~\cref{NoMOND} to be replaced by
\begin{equation}\label{CCOrigin}
\mathcal{V}(\mathcal{Y} ) = 2\kappa\Lambda.
\end{equation}
Accordingly, this solution is valid only when the parameters of \AE{}ST are carefully chosen to support~\cref{CCOrigin}. For other choices of the coupling parameters, the \ae{}ther must not be purely timelike. The field~$\psi(r)$ is meanwhile given by
\begin{equation}\label{cosmological_solution_scalar}
	\psi(r) = -\frac{\sqrt{\kappa \Lambda} r\cot \left( \sqrt{\kappa \Lambda}   r \right)}{\left(2-\KB{}\right)}  \,\,. 
\end{equation}
Although the spacetime in~\cref{sph_back,RadialCoordinate,cosmological_solution_metric} is homogeneous, this property is not shared by the scalar field in~\cref{cosmological_solution_scalar}.

\section{Conclusions}\label{section: Conclusion and discussion}

\paragraph*{Summary} Because \ae{}ther-scalar-tensor theory provides modified Newtonian dynamics, there is (somewhat unusually) no incentive to interpret the exotic compact objects found above as dark matter candidates. For the reasons given in~\cref{section: Introduction}, we identify the Eling--Jacobson-type solutions as \emph{black hole mimickers}, and the anti-Ellis--Bronnikov solutions as \emph{Shapiro-free lenses}. Respectively, we imagine these as secretly accounting for some fraction of the apparently observed black hole population, and as a quiet source of anomalous lensing events. The question of production, stability and detection of these objects is left for future work.

\paragraph*{Formal \ae{}ther-scalar-tensor results} We have obtained several spherically symmetric exact solutions to the \ae{}ther-scalar-tensor theory of gravity with static spacetime geometry, and an \ae{}ther field which is aligned with the timelike Killing vector. When the modified Newtonian dynamics and effective cold dark matter sectors are neglected (as may be expected in the strong-field regime) we find three branches of asymptotically flat solutions:
\begin{itemize}
\item If the gradient of the scalar is static, and is aligned with the \ae{}ther acceleration, there is an exact correspondence with Einstein-\ae{}ther theory in which the \ae{}ther coupling is renormalised by the scalar flux at infinity. For stable values of the bare \ae{}ther coupling, this correspondence leads to the famous Eling--Jacobson wormhole; an unstable \ae{}ther coupling leads to an Ellis--Bronnikov drainhole.

\item Alternatively, if the scalar is static but the \ae{}ther has no acceleration, then the Einstein-\ae{}ther correspondence is broken. This leads to the line element of anti-Ellis--Bronnikov spacetime. In real terms, static observers in this geometry feel no acceleration, but there is a naked scalar singularity.

\item The Eling--Jacobson solution also reappears when the scalar rolls linearly in time (as motivated by the requirements of cosmology), and this scenario may be extended to arbitrary time dependence of the scalar.

\item When the cold dark matter sector is included, along with a cosmological constant, a non-asymptotically-flat solution is found, which has the same spacetime geometry as the Einstein static universe. This geometry is homogeneous, but the scalar field exhibits a profile, introducing a preferred centre.
\end{itemize}

\paragraph*{Formal Einstein-\ae{}ther results} The Einstein-\ae{}ther correspondence also leads us to obtain several formal results in that theory. Firstly, we confirm previous conjectures that Eling--Jacobson spacetime is inextensible beyond the Killing horizon on the `interior' of the throat, which is shown to be a null singularity. This result may be relevant to bumblebee gravity~\cite{BH_in_BB}, which features solutions very similar to the Eling--Jacobson wormhole. Secondly, we show that when the \ae{}ther field is aligned with a timelike Killing vector field, the \ae{}ther acceleration is always a conserved current.

\paragraph*{Stability of the solutions} Solution of the classical field equations is a necessary but wholly insufficient criterion for the spacetimes considered here to be physically relevant. In particular, the parameter constraints quoted in~\cref{AetherCouplingRange,LambdaS} are known only to be relevant for the stability of Minkowski spacetime. Efficient means have been developed for determining the stability of compact exact solutions in modified gravity theories, most notably those of Siefert and Wald~\cite{Seifert:2006kv,Seifert:2007fr}, and also Luz and Carloni~\cite{Luz:2024yjm}. The general approach takes advantage of the spherical (or axisymmetric) symmetry to cast the perturbation equations in a general operator form. The square eigenfrequencies of the operator then determine the stability of each mode. It should be noted that even when the linear perturbations are found to be stable, the possibility remains for non-linear instabilities to arise. This is especially true when the exact solutions are found to be singular surfaces in the canonical phase space of the theory: in such cases perturbation theory is not a priori appropriate, and non-perturbative analysis should be used. This is an important question for models such as \AE{}ST, for which the field content is quite rich despite the derivative order of the Lagrangian being low: certain undesirable modes may be found to propagate generically in the bulk of the phase space. The analysis of linear stability, and of the non-linear canonical structure, is left for future work.

\paragraph*{Potential for misuse} Before discussing observational prospects, we briefly comment on the potential for misuse of \AE{}ST in the context of compact objects. The \AE{}ST model was originally put forward as a consistent classical field theory for explaining phenomena on the largest of scales, without recourse to particle dark matter. From a conservative standpoint, it is reasonable to assume that phenomenological models which serve this purpose should be ultra-low-energy effective field theories, and that their regime of validity should in fact be restricted to the scales on which their phenomenology is required. On intermediate scales, there is no need for \AE{}ST, since GR not only explains all the observations at the classical level but (together with its unique derivative expansion) also constitutes a self-consistent and universal infrared foundation for the true quantum theory of gravity (the details of which remain unknown). To make sense of this picture, one should consider mechanistically how the action in~\cref{original_action_AEST} can emerge at large scales from what is already known about fundamental theory, and why it interpolates at intermediate scales to GR. This is not a classical exercise, and requires techniques which are completely beyond the scope of this work.\footnote{This is similar in spirit to what is done in~\cite{Deser:2007jk}.} In this work, we have followed~\cite{costas_stealth_bh} in brazenly applying the classical \AE{}ST field equations to intermediate scales, regardless of the above concerns, which is tantamount to putting \AE{}ST forward as a replacement for (or extension of) GR.\footnote{We stress that when \AE{}ST was originally put forwards, the choice of interpretation was left open~\cite{Skordis:2020eui}. Moreover, \AE{}ST claims to describe at least part of the interpolation process by providing for the Newtonian regime.} It is interesting, therefore, that together with~\cite{costas_stealth_bh} the results of this exercise do not openly contradict existing observations, i.e., that the strong-field solutions do not by themselves force us towards the ultra-low-energy interpretation of \AE{}ST --- even if the requirements of an eventual quantum framing seem to do so.

\paragraph*{Why scalar hair is allowed} One possible objection to the new solutions is that they are hairy: they emit a flux of the conserved current associated with the shift symmetry of the scalar: the model --- being completely shift-symmetric --- seemingly provides no charge to source this flux. This objection can be disregarded, because even if \AE{}ST is valid on the intermediate scales of compact objects it must still be an effective theory. Once the theory is quantised there is a generic expectation that symmetry-breaking operators will be introduced at the one-loop level. This runs contrary to notable counterexamples of relevance to no-hair theorems: in Galileon/Horndeski theory, the derivative interactions are such that loops from Galileon vertices never generate lower-derivative symmetry-breaking operators. More generally, however, coupling of shift-symmetric scalars to matter or gravity \emph{is} expected to result in symmetry breaking. This effect would need to be taken into account when matching our solutions to extreme matter configurations. Indeed, despite our observations that both the black hole mimicker and the Shapiro-free lens solutions terminate at singular regions, there is no obvious reason why exotic matter sources cannot be introduced to avoid this. Such an attempt seems particularly desirable in the case of the lens, which would otherwise have a naked singularity, though the non-self-gravitating nature of the solution may give rise to particular challenges. In either case, hair may be sourced by quantum corrections that arise in the extreme environments on the approach to singular regions, where all current models eventually break down.

\paragraph*{Observational outlook: mimickers} From an observational perspective, such \ae{}ther-inspired compact objects could produce distinctive signatures in upcoming surveys and multi-messenger data. For example, mergers of black hole mimickers might display non-standard gravitational wave ringdown: several studies predict late-time echoes or `anti-chirps' rather than the standard Kerr quasinormal modes~\cite{Bao2023,Cardoso2016}. Detecting these subtle signals will require the high sensitivity of LIGO/Virgo/KAGRA’s O4 run and next-generation observatories (Einstein Telescope, Cosmic Explorer), and space-based missions such as LISA could also probe extreme-mass-ratio inspirals around supermassive mimickers. Moreover, if horizons are absent, infalling matter could emit prompt electromagnetic or neutrino bursts, making coordinated gravitational/electromagnetic/neutrino searches (e.g. IceCube, Fermi/Swift) a promising avenue. At the most basic level, we provide the rudimentary formulae corresponding to the radii (in the Schwarzschild-like coordinates of~\cref{ElingJacobsonWormholeLineElement}) corresponding to the photon sphere~$r_\gamma$ and the innermost stable circular orbit~$r_{\text{ISCO}}$, for the mimicker proposed in~\cref{ElingJacobsonWormholeLineElement,eq:time_norm,RadiusInTermsOfSpaceFunc}. These formulae, which are used to obtain the slightly dilated radii illustrated in~\cref{BlackHoleMimickers}, are respectively
\begin{widetext}
\begin{subequations}
\begin{align}
	r_{\gamma}&\equiv\GN{}M\sqrt{3+\frac{\KBE{}}{2}}\left[\left(\frac{\sqrt{2}+\sqrt{2-\KBE{}}}{\sqrt{2}-\sqrt{2-\KBE{}}}\right)\left(\frac{2+\KBE{}-\sqrt{2}\sqrt{2-\KBE{}}}{2+\KBE{}+\sqrt{2}\sqrt{2-\KBE{}}}\right)\right]^{\frac{1}{\sqrt{4-2\KBE{}}}},\label{ValuePhotonRing}
	\\
	r_{\text{ISCO}}&\equiv\frac{\GN{}M\sqrt{\KBE{}}\sqrt{4+\KBE{}-\sqrt{2}\sqrt{8+\KBE{}}}}{\sqrt{10+\KBE{}-2\sqrt{2}\sqrt{8+\KBE{}}}-\sqrt{2}}\left[\left(\frac{\sqrt{2}+\sqrt{2-\KBE{}}}{\sqrt{2}-\sqrt{2-\KBE{}}}\right)\right.
	\nonumber\\
	&\hspace{150pt}\times\left.\left(\frac{\sqrt{10+\KBE{}-2\sqrt{2}\sqrt{8+\KBE{}}}-\sqrt{2-\KBE{}}}{\sqrt{10+\KBE{}-2\sqrt{2}\sqrt{8+\KBE{}}}+\sqrt{2-\KBE{}}}\right)\right]^{\frac{1}{\sqrt{4-2\KBE{}}}}.\label{ValueISCO}
\end{align}
\end{subequations}
\end{widetext}
To reiterate the physical parameters, besides the Newtonian mass~$M$, the formulae in~\cref{ValuePhotonRing,ValueISCO} refer to the measured Newton--Cavendish constant~$\GN{}$ as defined in~\cref{GN}, and the dimensionless number~$\KBE{}$ as defined in~\cref{modified_KB}, with particular physical significance granted the `special' value~$\KBEN{}$ as defined in~\cref{KBE0}. A possible objection to the possibility that some observed black holes may be mimickers is that the accreting matter is not hidden behind an infinite-redshift horizon. Indeed, if matter were to `loiter' at the throat, it may give rise to observable X-ray emissions. To address this, we note that the throat is guaranteed to be an unstable environment for orbiting matter, since it always lies well within the ISCO. To address this question more broadly, we provide the formula for the proper acceleration experienced by massive, static observers at the wormhole throat:\footnote{Note that for the non-physical case~$\KBE{}>2$ we would have~${|\bf{a}|}=2\exp\left[-\tan^{-1}\left(2/\sqrt{-4+2\KBE{}}\right)/\sqrt{-4+2\KBE{}}\right]/\left(\GN{}M\KBE{}\right)$ in place of~\cref{AccelerationThroat}.}
\begin{align}
	{|\bf{a}|} = \frac{2(2-\KBE{})}{\GN{} M \KBE{}} 
	\left(\frac{\sqrt{2}-\sqrt{2-\KBE{}}}{\sqrt{2}+\sqrt{2-\KBE{}}}\right)^{\frac{1}{\sqrt{4-2\KBE{}}}}.\label{AccelerationThroat}
\end{align}
The important point is that~\cref{AccelerationThroat} approaches zero in the extremal case~$\KBE{}\to 2$, and diverges in the `stealth' limit of GR as~$\KBE{}\to 0$. For all finite values of~$\KBE{}$, however, matter loitering at the throat would need to be sustained by an outward acceleration of~${|\bf{a}|}\sim1/\GN{}M$. In the case of a solar-mass mimicker, this is comparable to the surface gravity of a neutron star at~${|\bf{a}|}\sim\SI{6e13}{\meter\per\second\squared}$. For a supermassive mimicker, the acceleration may be suppressed by a factor of~$10^{6-9}$. For much of the parameter space, therefore, it is reasonable to expect that mimicker throats would not be visible, though for large masses and extremal values of~$\KBE{}$, observable signatures may be expected and should be investigated. 

\paragraph*{Observational outlook: lenses} Meanwhile, Shapiro-free (massless) lenses may behave like cosmic strings: multiple images with identical magnification and essentially zero Shapiro time delay. Indeed, to leading order, the Shapiro time delay in the weak-field limit is
\begin{equation} \label{shapiro_weakfield}
	\Delta t \approx -b \frac{\ell^2}{2 b^2}\left[ \arccos\left(\frac{b}{r_\text{s}}\right) + \arccos\left(\frac{b}{r_\text{o}}\right) \right] \,,
\end{equation}
where~$b$ is the impact parameter,~$r_\text{s}$ is the coordinate distance of the source to the singularity and~$r_\text{o}$ the distance of the observer. By contrast the GR expression, computed from the Schwarzschild line element in~\cref{Schwarzschild}, is
\begin{equation}\label{shapiro_GR}
	\Delta t \approx 2 b \frac{G M}{b} \left[1 + \ln\left(\frac{4 r_\text{o} r_\text{s}}{b^2}\right)\right].
\end{equation}
When we compare~\cref{shapiro_weakfield} with~\cref{shapiro_GR}, we notice that the former has the opposite sign and, more importantly, does not diverge for far away observers or far away sources. Indeed, in the limit~$r_\text{o} \to \infty$ and~$r_\text{s} \to \infty$, the exact expression for the time delay, not assuming the weak-field limit, converges to
\begin{equation}
	\Delta t = 2 b \left[ K\left(-\frac{\ell^2}{b^2} \right) - E\left(-\frac{\ell^2}{b^2}\right) \right]\,,
\end{equation}
where~$K$ and~$E$ are the complete elliptic integrals of the first and second kind, respectively. Thus, for typical astrophysical and cosmological scales, we expect these lenses to be practically Shapiro-less. Regarding the sign of~\cref{shapiro_weakfield}, implying a Shapiro advance rather than a delay, we note that this phenomenology is likely not unique to the Shapiro-less lenses discussed here. Indeed, the gravitational acceleration around galaxies and galaxy clusters in \ae{}ther-scalar-tensor theory is expected to oscillate at very large radii~\cite{Skordis:2020eui,Mistele2023,Durakovic:2023out}. This corresponds to an oscillating effective mass profile that would, in some regions, likely induce a Shapiro advance. The implications and phenomenology of these negative Shapiro delays in \ae{}ther-scalar-tensor theory will be investigated in future work. Note that, whilst we focus on the Shapiro time delay in this work, other lensing characteristics are comparably reduced and inverted. Working again at the most basic level, the `$(-)$' configuration of~\cref{EllisWormhole}, which describes the lens, can be used to compute the effective Weyl potential relevant to gravitational lensing~$\Phi_{\text{lens}}\equiv (\Phi_{\text{N}}+\Psi_{\text{N}})/2$ in the Newtonian limit~$\mathrm{d}s^2\approx-\left(1+2\Phi_{\text{N}}\right)\mathrm{d}t^2+\left(1-2\Psi_{\text{N}}\right)(\mathrm{d}r^2+r^2\mathrm{d}\Omega^2)$.\footnote{Note that the conventional line element for weak-field lensing calculations corresponds to a choice of \emph{isotropic} coordinates, which differ from the Schwarzschild coordinates in~\cref{Schwarzschild} and proper-radial coordinates in~\cref{EllisWormhole} by some rescaling of the radii.} We find that ultrastaticity of the lens leads to a vanishing redshift potential~$\Phi_{\text{N}}=0$, such that to leading order in~$\ell/r$
\begin{equation}\label{LensingPotential}
	\Phi_{\text{lens}} = \frac{1}{2}\Psi_{\text{N}} \approx \frac{\ell^2}{8r^2} \,.
\end{equation}
Note once again that~\cref{LensingPotential} contrasts strongly in terms of sign and scaling with the GR case, computed from the Schwarzschild line element in~\cref{Schwarzschild}, for which~$\Phi_{\text{lens}}=\Phi_{\text{N}}=\Psi_{\text{N}}=-G_N M/r$. Indeed, if one were to interpret the lensing signal implied by~\cref{LensingPotential} in terms of a non-relativistic matter source in GR, one would infer a mass profile
\begin{equation}\label{MassProfile}
	M(r) \approx - \frac{\ell^2}{4 \GN{} r}.
\end{equation}
Evidently, the profile in~\cref{MassProfile} is grossly anomalous. It corresponds to a compact negative mass surrounded (and exactly cancelled) by a diffuse positive~$\sim 1/r^4$ density profile, such as those used to model the stellar mass density in elliptical galaxies and bulges~\cite{Jaffe1983,Hernquist1990,Dehnen1993} (see also~\cite{Kassiola1993}). Such an object would not be physical in the GR context, but it could be explained by \ae{}ther-scalar-tensor theory without recourse to exotic matter of any kind. Extending the approximations in~\cref{LensingPotential,MassProfile}, the exact lensing deflection angle~$\Delta \phi$, not assuming the weak-field limit, is given by
\begin{equation}\label{LensingDeflection}
	\Delta \phi = 2 K\left(-\frac{\ell^2}{b^2}\right) - \pi,
\end{equation}
where~$b$ is the impact parameter and~$K$ is once again the complete elliptical integral of the first kind. Time-domain instruments such as the Rubin Observatory (LSST) and Roman Space Telescope, together with radio facilities (SKA, CHIME), will monitor large numbers of transients and strongly lensed systems. These campaigns can test for anomalous lensing. Strongly lensed gravitational wave signals could exhibit interference fringes or multiple chirps with unexpected timing. In summary, a coordinated multi-messenger strategy is key: gravitational wave detectors (O4 and beyond) searching for non-GR ringdowns, alongside precise timing and imaging surveys for synchronized lensing transients, may constrain or reveal these phenomena. The advent of SKA, LSST/Rubin, Roman, LISA and IceCube thus provides concrete opportunities towards a smoking gun for \ae{}ther-scalar-tensor black hole mimickers and Shapiro-free lenses.

\begin{acknowledgments}
This work was improved by many useful discussions with Justin Feng, Mike Hobson, Yi-Hsiung Hsu, Ted Jacobson, Anthony Lasenby, Constantinos Skordis, David Vokrouhlický, Richard Woodard, Jingbo Yang and Tom Złosnik.

This work used the DiRAC Data Intensive service~(CSD3 \href{www.csd3.cam.ac.uk}{www.csd3.cam.ac.uk}) at the University of Cambridge, managed by the University of Cambridge University Information Services on behalf of the STFC DiRAC HPC Facility~(\href{www.dirac.ac.uk}{www.dirac.ac.uk}). The DiRAC component of CSD3 at Cambridge was funded by BEIS, UKRI and STFC capital funding and STFC operations grants. DiRAC is part of the UKRI Digital Research Infrastructure.

This work also used the Newton server, access to which was provisioned by Will Handley using an ERC grant.

WB was supported by the research environment and infrastructure of the Handley Lab at the University of Cambridge.

WB is grateful for the support of Girton College, Cambridge, Marie Skłodowska-Curie Actions and the Institute of Physics of the Czech Academy of Sciences. AD was supported by the European Regional Development Fund and the Czech Ministry of Education, Youth and Sports: project MSCA Fellowship CZ FZU I — CZ.02.01.01/00/22{\_}010/0002906. TM was supported by the DFG (German Research Foundation) – 514562826.

Co-funded by the European Union (Physics for Future -- Grant Agreement No.~101081515). Views and opinions expressed are however those of the author(s) only and do not necessarily reflect those of the European Union or European Research Executive Agency. Neither the European Union nor the granting authority can be held responsible for them.

\end{acknowledgments}

\bibliographystyle{apsrev4-2}
\bibliography{Manuscript.bib}

\begin{thebibliography}{82}%
\makeatletter
\providecommand \@ifxundefined [1]{%
 \@ifx{#1\undefined}
}%
\providecommand \@ifnum [1]{%
 \ifnum #1\expandafter \@firstoftwo
 \else \expandafter \@secondoftwo
 \fi
}%
\providecommand \@ifx [1]{%
 \ifx #1\expandafter \@firstoftwo
 \else \expandafter \@secondoftwo
 \fi
}%
\providecommand \natexlab [1]{#1}%
\providecommand \enquote  [1]{``#1''}%
\providecommand \bibnamefont  [1]{#1}%
\providecommand \bibfnamefont [1]{#1}%
\providecommand \citenamefont [1]{#1}%
\providecommand \href@noop [0]{\@secondoftwo}%
\providecommand \href [0]{\begingroup \@sanitize@url \@href}%
\providecommand \@href[1]{\@@startlink{#1}\@@href}%
\providecommand \@@href[1]{\endgroup#1\@@endlink}%
\providecommand \@sanitize@url [0]{\catcode `\\12\catcode `\$12\catcode
  `\&12\catcode `\#12\catcode `\^12\catcode `\_12\catcode `\%12\relax}%
\providecommand \@@startlink[1]{}%
\providecommand \@@endlink[0]{}%
\providecommand \url  [0]{\begingroup\@sanitize@url \@url }%
\providecommand \@url [1]{\endgroup\@href {#1}{\urlprefix }}%
\providecommand \urlprefix  [0]{URL }%
\providecommand \Eprint [0]{\href }%
\providecommand \doibase [0]{https://doi.org/}%
\providecommand \selectlanguage [0]{\@gobble}%
\providecommand \bibinfo  [0]{\@secondoftwo}%
\providecommand \bibfield  [0]{\@secondoftwo}%
\providecommand \translation [1]{[#1]}%
\providecommand \BibitemOpen [0]{}%
\providecommand \bibitemStop [0]{}%
\providecommand \bibitemNoStop [0]{.\EOS\space}%
\providecommand \EOS [0]{\spacefactor3000\relax}%
\providecommand \BibitemShut  [1]{\csname bibitem#1\endcsname}%
\let\auto@bib@innerbib\@empty
\bibitem [{\citenamefont {{McGaugh}}\ \emph {et~al.}(2000)\citenamefont
  {{McGaugh}}, \citenamefont {{Schombert}}, \citenamefont {{Bothun}},\ and\
  \citenamefont {{ de Blok}}}]{McGaugh2000}%
  \BibitemOpen
  \bibfield  {author} {\bibinfo {author} {\bibfnamefont {S.~S.}\ \bibnamefont
  {{McGaugh}}}, \bibinfo {author} {\bibfnamefont {J.~M.}\ \bibnamefont
  {{Schombert}}}, \bibinfo {author} {\bibfnamefont {G.~D.}\ \bibnamefont
  {{Bothun}}},\ and\ \bibinfo {author} {\bibfnamefont {W.~J.~G.}\ \bibnamefont
  {{ de Blok}}},\ }\href {https://doi.org/10.1086/312628} {\bibfield  {journal}
  {\bibinfo  {journal} {Astrophys. J. Lett.}\ }\textbf {\bibinfo {volume}
  {533}},\ \bibinfo {pages} {L99} (\bibinfo {year} {2000})},\ \Eprint
  {https://arxiv.org/abs/astro-ph/0003001} {arXiv:astro-ph/0003001 [astro-ph]}
  \BibitemShut {NoStop}%
\bibitem [{\citenamefont {{Mistele}}\ \emph {et~al.}(2024)\citenamefont
  {{Mistele}}, \citenamefont {{McGaugh}}, \citenamefont {{Lelli}},
  \citenamefont {{ Schombert}},\ and\ \citenamefont {{Li}}}]{Mistele2024}%
  \BibitemOpen
  \bibfield  {author} {\bibinfo {author} {\bibfnamefont {T.}~\bibnamefont
  {{Mistele}}}, \bibinfo {author} {\bibfnamefont {S.}~\bibnamefont
  {{McGaugh}}}, \bibinfo {author} {\bibfnamefont {F.}~\bibnamefont {{Lelli}}},
  \bibinfo {author} {\bibfnamefont {J.}~\bibnamefont {{ Schombert}}},\ and\
  \bibinfo {author} {\bibfnamefont {P.}~\bibnamefont {{Li}}},\ }\href
  {https://doi.org/10.3847/2041-8213/ad54b0} {\bibfield  {journal} {\bibinfo
  {journal} {Astrophys. J. Lett.}\ }\textbf {\bibinfo {volume} {969}},\
  \bibinfo {eid} {L3} (\bibinfo {year} {2024})},\ \Eprint
  {https://arxiv.org/abs/2406.09685} {arXiv:2406.09685 [astro-ph.GA]}
  \BibitemShut {NoStop}%
\bibitem [{\citenamefont {{Mistele}}(2024)}]{Mistele2023b}%
  \BibitemOpen
  \bibfield  {author} {\bibinfo {author} {\bibfnamefont {T.}~\bibnamefont
  {{Mistele}}},\ }\href {https://doi.org/10.1103/PhysRevD.110.024062}
  {\bibfield  {journal} {\bibinfo  {journal} {\prd}\ }\textbf {\bibinfo
  {volume} {110}},\ \bibinfo {eid} {024062} (\bibinfo {year} {2024})},\ \Eprint
  {https://arxiv.org/abs/2305.07742} {arXiv:2305.07742 [gr-qc]} \BibitemShut
  {NoStop}%
\bibitem [{\citenamefont {{Brouwer}}\ \emph {et~al.}(2021)\citenamefont
  {{Brouwer}}, \citenamefont {{Oman}}, \citenamefont {{Valentijn}},
  \citenamefont {{Bilicki}}, \citenamefont {{Heymans}}, \citenamefont
  {{Hoekstra}}, \citenamefont {{Napolitano}}, \citenamefont {{Roy}},
  \citenamefont {{Tortora}}, \citenamefont {{Wright}}, \citenamefont
  {{Asgari}}, \citenamefont {{van den Busch}}, \citenamefont {{Dvornik}},
  \citenamefont {{Erben}}, \citenamefont {{ Giblin}}, \citenamefont {{Graham}},
  \citenamefont {{Hildebrandt}}, \citenamefont {{Hopkins}}, \citenamefont
  {{Kannawadi}}, \citenamefont {{Kuijken }}, \citenamefont {{Liske}},
  \citenamefont {{Shan}}, \citenamefont {{Tr{\"o}ster }}, \citenamefont
  {{Verlinde}},\ and\ \citenamefont {{Visser}}}]{Brouwer2021}%
  \BibitemOpen
  \bibfield  {author} {\bibinfo {author} {\bibfnamefont {M.~M.}\ \bibnamefont
  {{Brouwer}}}, \bibinfo {author} {\bibfnamefont {K.~A.}\ \bibnamefont
  {{Oman}}}, \bibinfo {author} {\bibfnamefont {E.~A.}\ \bibnamefont
  {{Valentijn}}}, \bibinfo {author} {\bibfnamefont {M.}~\bibnamefont
  {{Bilicki}}}, \bibinfo {author} {\bibfnamefont {C.}~\bibnamefont
  {{Heymans}}}, \bibinfo {author} {\bibfnamefont {H.}~\bibnamefont
  {{Hoekstra}}}, \bibinfo {author} {\bibfnamefont {N.~R.}\ \bibnamefont
  {{Napolitano}}}, \bibinfo {author} {\bibfnamefont {N.}~\bibnamefont {{Roy}}},
  \bibinfo {author} {\bibfnamefont {C.}~\bibnamefont {{Tortora}}}, \bibinfo
  {author} {\bibfnamefont {A.~H.}\ \bibnamefont {{Wright}}}, \bibinfo {author}
  {\bibfnamefont {M.}~\bibnamefont {{Asgari}}}, \bibinfo {author}
  {\bibfnamefont {J.~L.}\ \bibnamefont {{van den Busch}}}, \bibinfo {author}
  {\bibfnamefont {A.}~\bibnamefont {{Dvornik}}}, \bibinfo {author}
  {\bibfnamefont {T.}~\bibnamefont {{Erben}}}, \bibinfo {author} {\bibfnamefont
  {B.}~\bibnamefont {{ Giblin}}}, \bibinfo {author} {\bibfnamefont {A.~W.}\
  \bibnamefont {{Graham}}}, \bibinfo {author} {\bibfnamefont {H.}~\bibnamefont
  {{Hildebrandt}}}, \bibinfo {author} {\bibfnamefont {A.~M.}\ \bibnamefont
  {{Hopkins}}}, \bibinfo {author} {\bibfnamefont {A.}~\bibnamefont
  {{Kannawadi}}}, \bibinfo {author} {\bibfnamefont {K.}~\bibnamefont {{Kuijken
  }}}, \bibinfo {author} {\bibfnamefont {J.}~\bibnamefont {{Liske}}}, \bibinfo
  {author} {\bibfnamefont {H.}~\bibnamefont {{Shan}}}, \bibinfo {author}
  {\bibfnamefont {T.}~\bibnamefont {{Tr{\"o}ster }}}, \bibinfo {author}
  {\bibfnamefont {E.}~\bibnamefont {{Verlinde}}},\ and\ \bibinfo {author}
  {\bibfnamefont {M.}~\bibnamefont {{Visser}}},\ }\href
  {https://doi.org/10.1051/0004-6361/202040108} {\bibfield  {journal} {\bibinfo
   {journal} {Astronomy \& Astrophysics}\ }\textbf {\bibinfo {volume} {650}},\
  \bibinfo {eid} {A113} (\bibinfo {year} {2021})},\ \Eprint
  {https://arxiv.org/abs/2106.11677} {arXiv:2106.11677 [astro-ph.GA]}
  \BibitemShut {NoStop}%
\bibitem [{\citenamefont {{Lelli}}\ \emph {et~al.}(2017)\citenamefont
  {{Lelli}}, \citenamefont {{McGaugh}}, \citenamefont {{Schombert}},\ and\
  \citenamefont {{Pawlowski}}}]{Lelli2017b}%
  \BibitemOpen
  \bibfield  {author} {\bibinfo {author} {\bibfnamefont {F.}~\bibnamefont
  {{Lelli}}}, \bibinfo {author} {\bibfnamefont {S.~S.}\ \bibnamefont
  {{McGaugh}}}, \bibinfo {author} {\bibfnamefont {J.~M.}\ \bibnamefont
  {{Schombert}}},\ and\ \bibinfo {author} {\bibfnamefont {M.~S.}\ \bibnamefont
  {{Pawlowski}}},\ }\href {https://doi.org/10.3847/1538-4357/836/2/152}
  {\bibfield  {journal} {\bibinfo  {journal} {Astrophys. J.}\ }\textbf
  {\bibinfo {volume} {836}},\ \bibinfo {eid} {152} (\bibinfo {year} {2017})},\
  \Eprint {https://arxiv.org/abs/1610.08981} {arXiv:1610.08981 [astro-ph.GA]}
  \BibitemShut {NoStop}%
\bibitem [{\citenamefont {{Milgrom}}(1983{\natexlab{a}})}]{MOND3}%
  \BibitemOpen
  \bibfield  {author} {\bibinfo {author} {\bibfnamefont {M.}~\bibnamefont
  {{Milgrom}}},\ }\href {https://doi.org/10.1086/161130} {\bibfield  {journal}
  {\bibinfo  {journal} {\apj}\ }\textbf {\bibinfo {volume} {270}},\ \bibinfo
  {pages} {365} (\bibinfo {year} {1983}{\natexlab{a}})}\BibitemShut {NoStop}%
\bibitem [{\citenamefont {{Milgrom}}(1983{\natexlab{b}})}]{MOND4}%
  \BibitemOpen
  \bibfield  {author} {\bibinfo {author} {\bibfnamefont {M.}~\bibnamefont
  {{Milgrom}}},\ }\href {https://doi.org/10.1086/161131} {\bibfield  {journal}
  {\bibinfo  {journal} {\apj}\ }\textbf {\bibinfo {volume} {270}},\ \bibinfo
  {pages} {371} (\bibinfo {year} {1983}{\natexlab{b}})}\BibitemShut {NoStop}%
\bibitem [{\citenamefont {{Milgrom}}(1983{\natexlab{c}})}]{MOND5}%
  \BibitemOpen
  \bibfield  {author} {\bibinfo {author} {\bibfnamefont {M.}~\bibnamefont
  {{Milgrom}}},\ }\href {https://doi.org/10.1086/161132} {\bibfield  {journal}
  {\bibinfo  {journal} {\apj}\ }\textbf {\bibinfo {volume} {270}},\ \bibinfo
  {pages} {384} (\bibinfo {year} {1983}{\natexlab{c}})}\BibitemShut {NoStop}%
\bibitem [{\citenamefont {Famaey}\ and\ \citenamefont
  {McGaugh}(2012)}]{Famaey:2011kh}%
  \BibitemOpen
  \bibfield  {author} {\bibinfo {author} {\bibfnamefont {B.}~\bibnamefont
  {Famaey}}\ and\ \bibinfo {author} {\bibfnamefont {S.}~\bibnamefont
  {McGaugh}},\ }\href {https://doi.org/10.12942/lrr-2012-10} {\bibfield
  {journal} {\bibinfo  {journal} {Living Rev. Rel.}\ }\textbf {\bibinfo
  {volume} {15}},\ \bibinfo {pages} {10} (\bibinfo {year} {2012})},\ \Eprint
  {https://arxiv.org/abs/1112.3960} {arXiv:1112.3960 [astro-ph.CO]}
  \BibitemShut {NoStop}%
\bibitem [{\citenamefont {Skordis}\ and\ \citenamefont
  {Zlosnik}(2021)}]{Skordis:2020eui}%
  \BibitemOpen
  \bibfield  {author} {\bibinfo {author} {\bibfnamefont {C.}~\bibnamefont
  {Skordis}}\ and\ \bibinfo {author} {\bibfnamefont {T.}~\bibnamefont
  {Zlosnik}},\ }\href {https://doi.org/10.1103/PhysRevLett.127.161302}
  {\bibfield  {journal} {\bibinfo  {journal} {Phys. Rev. Lett.}\ }\textbf
  {\bibinfo {volume} {127}},\ \bibinfo {pages} {161302} (\bibinfo {year}
  {2021})},\ \Eprint {https://arxiv.org/abs/2007.00082} {arXiv:2007.00082
  [astro-ph.CO]} \BibitemShut {NoStop}%
\bibitem [{\citenamefont {{Mistele}}\ \emph {et~al.}(2023)\citenamefont
  {{Mistele}}, \citenamefont {{McGaugh}},\ and\ \citenamefont
  {{Hossenfelder}}}]{Mistele2023}%
  \BibitemOpen
  \bibfield  {author} {\bibinfo {author} {\bibfnamefont {T.}~\bibnamefont
  {{Mistele}}}, \bibinfo {author} {\bibfnamefont {S.}~\bibnamefont
  {{McGaugh}}},\ and\ \bibinfo {author} {\bibfnamefont {S.}~\bibnamefont
  {{Hossenfelder}}},\ }\href {https://doi.org/10.1051/0004-6361/202346025}
  {\bibfield  {journal} {\bibinfo  {journal} {Astronomy \& Astrophysics}\
  }\textbf {\bibinfo {volume} {676}},\ \bibinfo {eid} {A100} (\bibinfo {year}
  {2023})},\ \Eprint {https://arxiv.org/abs/2301.03499} {arXiv:2301.03499
  [astro-ph.GA]} \BibitemShut {NoStop}%
\bibitem [{\citenamefont {Reyes}\ and\ \citenamefont
  {Sakstein}(2024)}]{Reyes:2024oha}%
  \BibitemOpen
  \bibfield  {author} {\bibinfo {author} {\bibfnamefont {C.}~\bibnamefont
  {Reyes}}\ and\ \bibinfo {author} {\bibfnamefont {J.}~\bibnamefont
  {Sakstein}},\ }\href {https://doi.org/10.1103/PhysRevD.110.084019} {\bibfield
   {journal} {\bibinfo  {journal} {Phys. Rev. D}\ }\textbf {\bibinfo {volume}
  {110}},\ \bibinfo {pages} {084019} (\bibinfo {year} {2024})},\ \Eprint
  {https://arxiv.org/abs/2406.18225} {arXiv:2406.18225 [gr-qc]} \BibitemShut
  {NoStop}%
\bibitem [{\citenamefont {Reyes}\ and\ \citenamefont
  {Sakstein}(2025)}]{Reyes:2025oet}%
  \BibitemOpen
  \bibfield  {author} {\bibinfo {author} {\bibfnamefont {C.}~\bibnamefont
  {Reyes}}\ and\ \bibinfo {author} {\bibfnamefont {J.}~\bibnamefont
  {Sakstein}},\ }\href@noop {} {\bibfield  {journal} {\bibinfo  {journal}
  {arXiv preprints}\ } (\bibinfo {year} {2025})},\ \Eprint
  {https://arxiv.org/abs/2505.03527} {arXiv:2505.03527 [gr-qc]} \BibitemShut
  {NoStop}%
\bibitem [{\citenamefont {Bernardo}\ and\ \citenamefont
  {Chen}(2023)}]{AeST_BH1}%
  \BibitemOpen
  \bibfield  {author} {\bibinfo {author} {\bibfnamefont {R.~C.}\ \bibnamefont
  {Bernardo}}\ and\ \bibinfo {author} {\bibfnamefont {C.-Y.}\ \bibnamefont
  {Chen}},\ }\bibfield  {journal} {\bibinfo  {journal} {General Relativity and
  Gravitation}\ }\textbf {\bibinfo {volume} {55}},\ \href
  {https://doi.org/10.1007/s10714-023-03075-x} {10.1007/s10714-023-03075-x}
  (\bibinfo {year} {2023})\BibitemShut {NoStop}%
\bibitem [{\citenamefont {Skordis}\ and\ \citenamefont
  {Vokrouhlicky}(2025)}]{costas_stealth_bh}%
  \BibitemOpen
  \bibfield  {author} {\bibinfo {author} {\bibfnamefont {C.}~\bibnamefont
  {Skordis}}\ and\ \bibinfo {author} {\bibfnamefont {D.~M.~J.}\ \bibnamefont
  {Vokrouhlicky}},\ }\href {https://doi.org/10.1088/1475-7516/2025/03/035}
  {\bibfield  {journal} {\bibinfo  {journal} {JCAP}\ }\textbf {\bibinfo
  {volume} {03}},\ \bibinfo {pages} {035}},\ \Eprint
  {https://arxiv.org/abs/2412.15395} {arXiv:2412.15395 [gr-qc]} \BibitemShut
  {NoStop}%
\bibitem [{\citenamefont {Bolton}(1972)}]{Bolton1972}%
  \BibitemOpen
  \bibfield  {author} {\bibinfo {author} {\bibfnamefont {C.~T.}\ \bibnamefont
  {Bolton}},\ }\href {https://doi.org/10.1038/235271b0} {\bibfield  {journal}
  {\bibinfo  {journal} {Nature}\ }\textbf {\bibinfo {volume} {235}},\ \bibinfo
  {pages} {271} (\bibinfo {year} {1972})}\BibitemShut {NoStop}%
\bibitem [{\citenamefont {Remillard}\ and\ \citenamefont
  {McClintock}(2006)}]{Remillard2006}%
  \BibitemOpen
  \bibfield  {author} {\bibinfo {author} {\bibfnamefont {R.~A.}\ \bibnamefont
  {Remillard}}\ and\ \bibinfo {author} {\bibfnamefont {J.~E.}\ \bibnamefont
  {McClintock}},\ }\href
  {https://doi.org/10.1146/annurev.astro.44.051905.092532} {\bibfield
  {journal} {\bibinfo  {journal} {Annu. Rev. Astron. Astrophys.}\ }\textbf
  {\bibinfo {volume} {44}},\ \bibinfo {pages} {49} (\bibinfo {year}
  {2006})}\BibitemShut {NoStop}%
\bibitem [{\citenamefont {Sch{\"o}del}\ \emph {et~al.}(2002)\citenamefont
  {Sch{\"o}del}, \citenamefont {Ott}, \citenamefont {Genzel}, \citenamefont
  {Eckart}, \citenamefont {Mouawad},\ and\ \citenamefont
  {Alexander}}]{Schodel2002}%
  \BibitemOpen
  \bibfield  {author} {\bibinfo {author} {\bibfnamefont {R.}~\bibnamefont
  {Sch{\"o}del}}, \bibinfo {author} {\bibfnamefont {T.}~\bibnamefont {Ott}},
  \bibinfo {author} {\bibfnamefont {R.}~\bibnamefont {Genzel}}, \bibinfo
  {author} {\bibfnamefont {A.}~\bibnamefont {Eckart}}, \bibinfo {author}
  {\bibfnamefont {N.}~\bibnamefont {Mouawad}},\ and\ \bibinfo {author}
  {\bibfnamefont {T.}~\bibnamefont {Alexander}},\ }\href
  {https://doi.org/10.1038/nature01121} {\bibfield  {journal} {\bibinfo
  {journal} {Nature}\ }\textbf {\bibinfo {volume} {419}},\ \bibinfo {pages}
  {694} (\bibinfo {year} {2002})}\BibitemShut {NoStop}%
\bibitem [{\citenamefont {Ghez}\ \emph {et~al.}(2008)\citenamefont {Ghez},
  \citenamefont {Salim}, \citenamefont {Weinberg},\ and\ \citenamefont {{et
  al.}}}]{Ghez2008}%
  \BibitemOpen
  \bibfield  {author} {\bibinfo {author} {\bibfnamefont {A.~M.}\ \bibnamefont
  {Ghez}}, \bibinfo {author} {\bibfnamefont {S.}~\bibnamefont {Salim}},
  \bibinfo {author} {\bibfnamefont {N.~N.}\ \bibnamefont {Weinberg}},\ and\
  \bibinfo {author} {\bibnamefont {{et al.}}},\ }\href
  {https://doi.org/10.1086/592738} {\bibfield  {journal} {\bibinfo  {journal}
  {Astrophys. J.}\ }\textbf {\bibinfo {volume} {689}},\ \bibinfo {pages} {1044}
  (\bibinfo {year} {2008})}\BibitemShut {NoStop}%
\bibitem [{\citenamefont {Lynden-Bell}\ and\ \citenamefont
  {Rees}(1971)}]{LyndenBell1971}%
  \BibitemOpen
  \bibfield  {author} {\bibinfo {author} {\bibfnamefont {D.}~\bibnamefont
  {Lynden-Bell}}\ and\ \bibinfo {author} {\bibfnamefont {M.~J.}\ \bibnamefont
  {Rees}},\ }\href@noop {} {\bibfield  {journal} {\bibinfo  {journal} {Mon.
  Not. Roy. Astron. Soc.}\ }\textbf {\bibinfo {volume} {152}},\ \bibinfo
  {pages} {461} (\bibinfo {year} {1971})}\BibitemShut {NoStop}%
\bibitem [{\citenamefont {Kormendy}\ and\ \citenamefont
  {Ho}(2013)}]{Kormendy2013}%
  \BibitemOpen
  \bibfield  {author} {\bibinfo {author} {\bibfnamefont {J.}~\bibnamefont
  {Kormendy}}\ and\ \bibinfo {author} {\bibfnamefont {L.~C.}\ \bibnamefont
  {Ho}},\ }\href {https://doi.org/10.1146/annurev-astro-082708-101811}
  {\bibfield  {journal} {\bibinfo  {journal} {Annu. Rev. Astron. Astrophys.}\
  }\textbf {\bibinfo {volume} {51}},\ \bibinfo {pages} {511} (\bibinfo {year}
  {2013})}\BibitemShut {NoStop}%
\bibitem [{\citenamefont {Narayan}\ and\ \citenamefont
  {Heyl}(2002)}]{Narayan2002}%
  \BibitemOpen
  \bibfield  {author} {\bibinfo {author} {\bibfnamefont {R.}~\bibnamefont
  {Narayan}}\ and\ \bibinfo {author} {\bibfnamefont {J.~S.}\ \bibnamefont
  {Heyl}},\ }\href {https://doi.org/10.1086/342502} {\bibfield  {journal}
  {\bibinfo  {journal} {Astrophys. J.}\ }\textbf {\bibinfo {volume} {574}},\
  \bibinfo {pages} {L139} (\bibinfo {year} {2002})}\BibitemShut {NoStop}%
\bibitem [{\citenamefont {Narayan}\ and\ \citenamefont
  {McClintock}(2008)}]{Narayan2008}%
  \BibitemOpen
  \bibfield  {author} {\bibinfo {author} {\bibfnamefont {R.}~\bibnamefont
  {Narayan}}\ and\ \bibinfo {author} {\bibfnamefont {J.~E.}\ \bibnamefont
  {McClintock}},\ }\href {https://doi.org/10.1016/j.newar.2008.03.002}
  {\bibfield  {journal} {\bibinfo  {journal} {New Astron. Rev.}\ }\textbf
  {\bibinfo {volume} {51}},\ \bibinfo {pages} {733} (\bibinfo {year}
  {2008})}\BibitemShut {NoStop}%
\bibitem [{\citenamefont {Abbott}\ \emph {et~al.}(2016)\citenamefont {Abbott},
  \citenamefont {Abbott}, \citenamefont {Abbott}, \citenamefont
  {Collaboration},\ and\ \citenamefont {Collaboration)}}]{Abbott2016}%
  \BibitemOpen
  \bibfield  {author} {\bibinfo {author} {\bibfnamefont {B.}~\bibnamefont
  {Abbott}}, \bibinfo {author} {\bibfnamefont {R.}~\bibnamefont {Abbott}},
  \bibinfo {author} {\bibfnamefont {T.}~\bibnamefont {Abbott}}, \bibinfo
  {author} {\bibfnamefont {e.~L.~S.}\ \bibnamefont {Collaboration}},\ and\
  \bibinfo {author} {\bibfnamefont {V.}~\bibnamefont {Collaboration)}},\ }\href
  {https://doi.org/10.1103/PhysRevLett.116.221101} {\bibfield  {journal}
  {\bibinfo  {journal} {Phys. Rev. Lett.}\ }\textbf {\bibinfo {volume} {116}},\
  \bibinfo {pages} {221101} (\bibinfo {year} {2016})}\BibitemShut {NoStop}%
\bibitem [{\citenamefont {Cardoso}\ \emph {et~al.}(2016)\citenamefont
  {Cardoso}, \citenamefont {Franzin},\ and\ \citenamefont
  {Pani}}]{Cardoso2016}%
  \BibitemOpen
  \bibfield  {author} {\bibinfo {author} {\bibfnamefont {V.}~\bibnamefont
  {Cardoso}}, \bibinfo {author} {\bibfnamefont {E.}~\bibnamefont {Franzin}},\
  and\ \bibinfo {author} {\bibfnamefont {P.}~\bibnamefont {Pani}},\ }\href
  {https://doi.org/10.1103/PhysRevLett.116.171101} {\bibfield  {journal}
  {\bibinfo  {journal} {Phys. Rev. Lett.}\ }\textbf {\bibinfo {volume} {116}},\
  \bibinfo {pages} {171101} (\bibinfo {year} {2016})}\BibitemShut {NoStop}%
\bibitem [{\citenamefont {Falcke}\ \emph {et~al.}(2000)\citenamefont {Falcke},
  \citenamefont {Melia},\ and\ \citenamefont {Agol}}]{Falcke2000}%
  \BibitemOpen
  \bibfield  {author} {\bibinfo {author} {\bibfnamefont {H.}~\bibnamefont
  {Falcke}}, \bibinfo {author} {\bibfnamefont {F.}~\bibnamefont {Melia}},\ and\
  \bibinfo {author} {\bibfnamefont {E.}~\bibnamefont {Agol}},\ }\href
  {https://doi.org/10.1086/312423} {\bibfield  {journal} {\bibinfo  {journal}
  {Astrophys. J. Lett.}\ }\textbf {\bibinfo {volume} {528}},\ \bibinfo {pages}
  {L13} (\bibinfo {year} {2000})}\BibitemShut {NoStop}%
\bibitem [{\citenamefont {Event Horizon Telescope
  Collaboration;~Akiyama}(2019)}]{EHT2019}%
  \BibitemOpen
  \bibfield  {author} {\bibinfo {author} {\bibfnamefont {K.~e.}\ \bibnamefont
  {Event Horizon Telescope Collaboration;~Akiyama}},\ }\href
  {https://doi.org/10.3847/2041-8213/ab0ec7} {\bibfield  {journal} {\bibinfo
  {journal} {Astrophys. J. Lett.}\ }\textbf {\bibinfo {volume} {875}},\
  \bibinfo {pages} {L1} (\bibinfo {year} {2019})}\BibitemShut {NoStop}%
\bibitem [{\citenamefont {Event Horizon Telescope
  Collaboration;~Akiyama}(2022)}]{EHT2022}%
  \BibitemOpen
  \bibfield  {author} {\bibinfo {author} {\bibfnamefont {K.~e.}\ \bibnamefont
  {Event Horizon Telescope Collaboration;~Akiyama}},\ }\href
  {https://doi.org/10.3847/2041-8213/ac6674} {\bibfield  {journal} {\bibinfo
  {journal} {Astrophys. J. Lett.}\ }\textbf {\bibinfo {volume} {930}},\
  \bibinfo {pages} {L12} (\bibinfo {year} {2022})}\BibitemShut {NoStop}%
\bibitem [{\citenamefont {Tanaka}\ \emph {et~al.}(1995)\citenamefont {Tanaka},
  \citenamefont {Nandra}, \citenamefont {Fabian},\ and\ \citenamefont {{et
  al.}}}]{Tanaka1995}%
  \BibitemOpen
  \bibfield  {author} {\bibinfo {author} {\bibfnamefont {Y.}~\bibnamefont
  {Tanaka}}, \bibinfo {author} {\bibfnamefont {K.}~\bibnamefont {Nandra}},
  \bibinfo {author} {\bibfnamefont {A.~C.}\ \bibnamefont {Fabian}},\ and\
  \bibinfo {author} {\bibnamefont {{et al.}}},\ }\href
  {https://doi.org/10.1038/375659a0} {\bibfield  {journal} {\bibinfo  {journal}
  {Nature}\ }\textbf {\bibinfo {volume} {375}},\ \bibinfo {pages} {659}
  (\bibinfo {year} {1995})}\BibitemShut {NoStop}%
\bibitem [{\citenamefont {Fabian}\ \emph {et~al.}(2000)\citenamefont {Fabian},
  \citenamefont {Iwasawa}, \citenamefont {Reynolds},\ and\ \citenamefont
  {Young}}]{Fabian2000}%
  \BibitemOpen
  \bibfield  {author} {\bibinfo {author} {\bibfnamefont {A.~C.}\ \bibnamefont
  {Fabian}}, \bibinfo {author} {\bibfnamefont {K.}~\bibnamefont {Iwasawa}},
  \bibinfo {author} {\bibfnamefont {C.~S.}\ \bibnamefont {Reynolds}},\ and\
  \bibinfo {author} {\bibfnamefont {A.~J.}\ \bibnamefont {Young}},\ }\href
  {https://doi.org/10.1086/316610} {\bibfield  {journal} {\bibinfo  {journal}
  {Publ. Astron. Soc. Pac.}\ }\textbf {\bibinfo {volume} {112}},\ \bibinfo
  {pages} {1145} (\bibinfo {year} {2000})}\BibitemShut {NoStop}%
\bibitem [{\citenamefont {{GRAVITY
  Collaboration}~(Abuter}(2018)}]{GRAVITY2018}%
  \BibitemOpen
  \bibfield  {author} {\bibinfo {author} {\bibfnamefont {R.~e.}\ \bibnamefont
  {{GRAVITY Collaboration}~(Abuter}},\ }\href
  {https://doi.org/10.1051/0004-6361/201833718} {\bibfield  {journal} {\bibinfo
   {journal} {Astron. Astrophys.}\ }\textbf {\bibinfo {volume} {615}},\
  \bibinfo {pages} {L15} (\bibinfo {year} {2018})}\BibitemShut {NoStop}%
\bibitem [{\citenamefont {{GRAVITY
  Collaboration}~(Abuter}(2020)}]{GRAVITY2020}%
  \BibitemOpen
  \bibfield  {author} {\bibinfo {author} {\bibfnamefont {R.~e.}\ \bibnamefont
  {{GRAVITY Collaboration}~(Abuter}},\ }\href
  {https://doi.org/10.1051/0004-6361/202037813} {\bibfield  {journal} {\bibinfo
   {journal} {Astron. Astrophys.}\ }\textbf {\bibinfo {volume} {636}},\
  \bibinfo {pages} {L5} (\bibinfo {year} {2020})}\BibitemShut {NoStop}%
\bibitem [{\citenamefont {Gott}(1985)}]{Gott1985}%
  \BibitemOpen
  \bibfield  {author} {\bibinfo {author} {\bibfnamefont {I.}~\bibnamefont
  {Gott}, \bibfnamefont {J.~Richard}},\ }\href@noop {} {\bibfield  {journal}
  {\bibinfo  {journal} {Astrophysical Journal}\ }\textbf {\bibinfo {volume}
  {288}},\ \bibinfo {pages} {422} (\bibinfo {year} {1985})}\BibitemShut
  {NoStop}%
\bibitem [{\citenamefont {Huterer}\ and\ \citenamefont
  {Vachaspati}(2003)}]{Huterer2003}%
  \BibitemOpen
  \bibfield  {author} {\bibinfo {author} {\bibfnamefont {D.}~\bibnamefont
  {Huterer}}\ and\ \bibinfo {author} {\bibfnamefont {T.}~\bibnamefont
  {Vachaspati}},\ }\href {https://doi.org/10.1103/PhysRevD.68.041301}
  {\bibfield  {journal} {\bibinfo  {journal} {Physical Review D}\ }\textbf
  {\bibinfo {volume} {68}},\ \bibinfo {pages} {041301} (\bibinfo {year}
  {2003})}\BibitemShut {NoStop}%
\bibitem [{\citenamefont {Tisserand}\ \emph {et~al.}(2007)\citenamefont
  {Tisserand}, \citenamefont {Le~Guillou}, \citenamefont {Afonso},\ and\
  \citenamefont {et~al.}}]{Tisserand2007}%
  \BibitemOpen
  \bibfield  {author} {\bibinfo {author} {\bibfnamefont {P.}~\bibnamefont
  {Tisserand}}, \bibinfo {author} {\bibfnamefont {L.}~\bibnamefont
  {Le~Guillou}}, \bibinfo {author} {\bibfnamefont {C.}~\bibnamefont {Afonso}},\
  and\ \bibinfo {author} {\bibnamefont {et~al.}},\ }\href
  {https://doi.org/10.1051/0004-6361:20066017} {\bibfield  {journal} {\bibinfo
  {journal} {Astronomy \& Astrophysics}\ }\textbf {\bibinfo {volume} {469}},\
  \bibinfo {pages} {387} (\bibinfo {year} {2007})}\BibitemShut {NoStop}%
\bibitem [{\citenamefont {Mr{\'o}z}\ \emph {et~al.}(2024)\citenamefont
  {Mr{\'o}z}, \citenamefont {Udalski}, \citenamefont {Szyma{\'n}ski},
  \citenamefont {Wyrzykowski},\ and\ \citenamefont {et~al.}}]{Mroz2024}%
  \BibitemOpen
  \bibfield  {author} {\bibinfo {author} {\bibfnamefont {P.}~\bibnamefont
  {Mr{\'o}z}}, \bibinfo {author} {\bibfnamefont {A.}~\bibnamefont {Udalski}},
  \bibinfo {author} {\bibfnamefont {M.~K.}\ \bibnamefont {Szyma{\'n}ski}},
  \bibinfo {author} {\bibfnamefont {{\L}.}~\bibnamefont {Wyrzykowski}},\ and\
  \bibinfo {author} {\bibnamefont {et~al.}},\ }\href
  {https://doi.org/10.3847/1538-4365/ad452e} {\bibfield  {journal} {\bibinfo
  {journal} {Astrophysical Journal Supplement}\ }\textbf {\bibinfo {volume}
  {273}},\ \bibinfo {pages} {4} (\bibinfo {year} {2024})}\BibitemShut {NoStop}%
\bibitem [{\citenamefont {Inoue}(2018)}]{Inoue2017}%
  \BibitemOpen
  \bibfield  {author} {\bibinfo {author} {\bibfnamefont {K.~T.}\ \bibnamefont
  {Inoue}},\ }\href {https://doi.org/10.1016/j.newast.2017.07.006} {\bibfield
  {journal} {\bibinfo  {journal} {New Astronomy}\ }\textbf {\bibinfo {volume}
  {58}},\ \bibinfo {pages} {47} (\bibinfo {year} {2018})}\BibitemShut {NoStop}%
\bibitem [{\citenamefont {Niikura}\ \emph {et~al.}(2019)\citenamefont
  {Niikura}, \citenamefont {Takada}, \citenamefont {Yasuda}, \citenamefont
  {Lupton}, \citenamefont {Sumi}, \citenamefont {More}, \citenamefont {Kurita},
  \citenamefont {Sugiyama}, \citenamefont {More}, \citenamefont {Oguri},\ and\
  \citenamefont {Chiba}}]{Niikura2019}%
  \BibitemOpen
  \bibfield  {author} {\bibinfo {author} {\bibfnamefont {H.}~\bibnamefont
  {Niikura}}, \bibinfo {author} {\bibfnamefont {M.}~\bibnamefont {Takada}},
  \bibinfo {author} {\bibfnamefont {N.}~\bibnamefont {Yasuda}}, \bibinfo
  {author} {\bibfnamefont {R.~H.}\ \bibnamefont {Lupton}}, \bibinfo {author}
  {\bibfnamefont {T.}~\bibnamefont {Sumi}}, \bibinfo {author} {\bibfnamefont
  {S.}~\bibnamefont {More}}, \bibinfo {author} {\bibfnamefont {T.}~\bibnamefont
  {Kurita}}, \bibinfo {author} {\bibfnamefont {S.}~\bibnamefont {Sugiyama}},
  \bibinfo {author} {\bibfnamefont {A.}~\bibnamefont {More}}, \bibinfo {author}
  {\bibfnamefont {M.}~\bibnamefont {Oguri}},\ and\ \bibinfo {author}
  {\bibfnamefont {M.}~\bibnamefont {Chiba}},\ }\href
  {https://doi.org/10.1038/s41550-019-0723-1} {\bibfield  {journal} {\bibinfo
  {journal} {Nature Astronomy}\ }\textbf {\bibinfo {volume} {3}},\ \bibinfo
  {pages} {524} (\bibinfo {year} {2019})}\BibitemShut {NoStop}%
\bibitem [{\citenamefont {Taylor}\ and\ \citenamefont
  {Weisberg}(1989)}]{Taylor1989}%
  \BibitemOpen
  \bibfield  {author} {\bibinfo {author} {\bibfnamefont {J.~H.}\ \bibnamefont
  {Taylor}}\ and\ \bibinfo {author} {\bibfnamefont {J.~M.}\ \bibnamefont
  {Weisberg}},\ }\href {https://doi.org/10.1086/167917} {\bibfield  {journal}
  {\bibinfo  {journal} {Astrophysical Journal}\ }\textbf {\bibinfo {volume}
  {345}},\ \bibinfo {pages} {434} (\bibinfo {year} {1989})}\BibitemShut
  {NoStop}%
\bibitem [{\citenamefont {Kramer}\ \emph {et~al.}(2021)\citenamefont {Kramer},
  \citenamefont {Stairs}, \citenamefont {Manchester}, \citenamefont {Wex},\
  and\ \citenamefont {et~al.}}]{Kramer2021}%
  \BibitemOpen
  \bibfield  {author} {\bibinfo {author} {\bibfnamefont {M.}~\bibnamefont
  {Kramer}}, \bibinfo {author} {\bibfnamefont {I.~H.}\ \bibnamefont {Stairs}},
  \bibinfo {author} {\bibfnamefont {R.~N.}\ \bibnamefont {Manchester}},
  \bibinfo {author} {\bibfnamefont {N.}~\bibnamefont {Wex}},\ and\ \bibinfo
  {author} {\bibnamefont {et~al.}},\ }\href
  {https://doi.org/10.1103/PhysRevX.11.041050} {\bibfield  {journal} {\bibinfo
  {journal} {Physical Review X}\ }\textbf {\bibinfo {volume} {11}},\ \bibinfo
  {pages} {041050} (\bibinfo {year} {2021})}\BibitemShut {NoStop}%
\bibitem [{\citenamefont {Lazaridis}\ \emph {et~al.}(2009)\citenamefont
  {Lazaridis}, \citenamefont {Wex}, \citenamefont {Jessner}, \citenamefont
  {Kramer}, \citenamefont {Stappers}, \citenamefont {Verbiest}, \citenamefont
  {Tauris},\ and\ \citenamefont {et~al.}}]{Lazaridis2009}%
  \BibitemOpen
  \bibfield  {author} {\bibinfo {author} {\bibfnamefont {K.}~\bibnamefont
  {Lazaridis}}, \bibinfo {author} {\bibfnamefont {N.}~\bibnamefont {Wex}},
  \bibinfo {author} {\bibfnamefont {A.}~\bibnamefont {Jessner}}, \bibinfo
  {author} {\bibfnamefont {M.}~\bibnamefont {Kramer}}, \bibinfo {author}
  {\bibfnamefont {B.~W.}\ \bibnamefont {Stappers}}, \bibinfo {author}
  {\bibfnamefont {J.~P.~W.}\ \bibnamefont {Verbiest}}, \bibinfo {author}
  {\bibfnamefont {T.~M.}\ \bibnamefont {Tauris}},\ and\ \bibinfo {author}
  {\bibnamefont {et~al.}},\ }\href
  {https://doi.org/10.1111/j.1365-2966.2009.15481.x} {\bibfield  {journal}
  {\bibinfo  {journal} {Monthly Notices of the Royal Astronomical Society}\
  }\textbf {\bibinfo {volume} {400}},\ \bibinfo {pages} {805} (\bibinfo {year}
  {2009})}\BibitemShut {NoStop}%
\bibitem [{\citenamefont {Freire}\ \emph {et~al.}(2012)\citenamefont {Freire},
  \citenamefont {Wex}, \citenamefont {Esposito-Far{\`e}se}, \citenamefont
  {Verbiest}, \citenamefont {Bailes}, \citenamefont {Kramer}, \citenamefont
  {Stairs}, \citenamefont {Antoniadis},\ and\ \citenamefont
  {Janssen}}]{Freire2012}%
  \BibitemOpen
  \bibfield  {author} {\bibinfo {author} {\bibfnamefont {P.~C.~C.}\
  \bibnamefont {Freire}}, \bibinfo {author} {\bibfnamefont {N.}~\bibnamefont
  {Wex}}, \bibinfo {author} {\bibfnamefont {G.}~\bibnamefont
  {Esposito-Far{\`e}se}}, \bibinfo {author} {\bibfnamefont {J.~P.~W.}\
  \bibnamefont {Verbiest}}, \bibinfo {author} {\bibfnamefont {M.}~\bibnamefont
  {Bailes}}, \bibinfo {author} {\bibfnamefont {M.}~\bibnamefont {Kramer}},
  \bibinfo {author} {\bibfnamefont {I.~H.}\ \bibnamefont {Stairs}}, \bibinfo
  {author} {\bibfnamefont {J.}~\bibnamefont {Antoniadis}},\ and\ \bibinfo
  {author} {\bibfnamefont {G.~H.}\ \bibnamefont {Janssen}},\ }\href
  {https://doi.org/10.1111/j.1365-2966.2012.21253.x} {\bibfield  {journal}
  {\bibinfo  {journal} {Monthly Notices of the Royal Astronomical Society}\
  }\textbf {\bibinfo {volume} {423}},\ \bibinfo {pages} {3328} (\bibinfo {year}
  {2012})}\BibitemShut {NoStop}%
\bibitem [{\citenamefont {Agazie}\ \emph {et~al.}(2023)\citenamefont {Agazie},
  \citenamefont {Arzoumanian}, \citenamefont {Baker}, \citenamefont {Bécsy},\
  and\ \citenamefont {et~al.}}]{NANOGrav2023}%
  \BibitemOpen
  \bibfield  {author} {\bibinfo {author} {\bibfnamefont {G.}~\bibnamefont
  {Agazie}}, \bibinfo {author} {\bibfnamefont {Z.}~\bibnamefont {Arzoumanian}},
  \bibinfo {author} {\bibfnamefont {P.~T.}\ \bibnamefont {Baker}}, \bibinfo
  {author} {\bibfnamefont {B.}~\bibnamefont {Bécsy}},\ and\ \bibinfo {author}
  {\bibnamefont {et~al.}},\ }\href {https://doi.org/10.3847/2041-8213/acdac6}
  {\bibfield  {journal} {\bibinfo  {journal} {Astrophysical Journal Letters}\
  }\textbf {\bibinfo {volume} {951}},\ \bibinfo {pages} {L8} (\bibinfo {year}
  {2023})}\BibitemShut {NoStop}%
\bibitem [{\citenamefont {Antoniadis}\ \emph {et~al.}(2023)\citenamefont
  {Antoniadis}, \citenamefont {Bailes}, \citenamefont {Brem},\ and\
  \citenamefont {et~al.}}]{EPTA2023}%
  \BibitemOpen
  \bibfield  {author} {\bibinfo {author} {\bibfnamefont {J.}~\bibnamefont
  {Antoniadis}}, \bibinfo {author} {\bibfnamefont {M.}~\bibnamefont {Bailes}},
  \bibinfo {author} {\bibfnamefont {P.}~\bibnamefont {Brem}},\ and\ \bibinfo
  {author} {\bibnamefont {et~al.}},\ }\href
  {https://doi.org/10.1051/0004-6361/202346844} {\bibfield  {journal} {\bibinfo
   {journal} {Astronomy \& Astrophysics}\ }\textbf {\bibinfo {volume} {678}},\
  \bibinfo {pages} {A50} (\bibinfo {year} {2023})}\BibitemShut {NoStop}%
\bibitem [{\citenamefont {Ellis}(1973)}]{Ellis1973Drainhole}%
  \BibitemOpen
  \bibfield  {author} {\bibinfo {author} {\bibfnamefont {H.~G.}\ \bibnamefont
  {Ellis}},\ }\href {https://doi.org/10.1063/1.1666161} {\bibfield  {journal}
  {\bibinfo  {journal} {Journal of Mathematical Physics}\ }\textbf {\bibinfo
  {volume} {14}},\ \bibinfo {pages} {104} (\bibinfo {year} {1973})}\BibitemShut
  {NoStop}%
\bibitem [{\citenamefont {Bronnikov}(1973)}]{Bronnikov1973ScalarTensor}%
  \BibitemOpen
  \bibfield  {author} {\bibinfo {author} {\bibfnamefont {K.~A.}\ \bibnamefont
  {Bronnikov}},\ }\href@noop {} {\bibfield  {journal} {\bibinfo  {journal}
  {Acta Physica Polonica B}\ }\textbf {\bibinfo {volume} {4}},\ \bibinfo
  {pages} {251} (\bibinfo {year} {1973})}\BibitemShut {NoStop}%
\bibitem [{\citenamefont {Visser}(1989)}]{Visser1989}%
  \BibitemOpen
  \bibfield  {author} {\bibinfo {author} {\bibfnamefont {M.}~\bibnamefont
  {Visser}},\ }\href {https://doi.org/10.1016/0550-3213(89)90337-9} {\bibfield
  {journal} {\bibinfo  {journal} {Nuclear Physics B}\ }\textbf {\bibinfo
  {volume} {328}},\ \bibinfo {pages} {203} (\bibinfo {year}
  {1989})}\BibitemShut {NoStop}%
\bibitem [{\citenamefont {Visser}(1995)}]{Visser1995}%
  \BibitemOpen
  \bibfield  {author} {\bibinfo {author} {\bibfnamefont {M.}~\bibnamefont
  {Visser}},\ }\href@noop {} {\emph {\bibinfo {title} {Lorentzian Wormholes:
  From Einstein to Hawking}}}\ (\bibinfo  {publisher} {AIP Press},\ \bibinfo
  {address} {New York},\ \bibinfo {year} {1995})\BibitemShut {NoStop}%
\bibitem [{\citenamefont {Morris}\ and\ \citenamefont
  {Thorne}(1988)}]{MorrisThorne1988}%
  \BibitemOpen
  \bibfield  {author} {\bibinfo {author} {\bibfnamefont {M.~S.}\ \bibnamefont
  {Morris}}\ and\ \bibinfo {author} {\bibfnamefont {K.~S.}\ \bibnamefont
  {Thorne}},\ }\href {https://doi.org/10.1119/1.15620} {\bibfield  {journal}
  {\bibinfo  {journal} {American Journal of Physics}\ }\textbf {\bibinfo
  {volume} {56}},\ \bibinfo {pages} {395} (\bibinfo {year} {1988})}\BibitemShut
  {NoStop}%
\bibitem [{\citenamefont {Morris}\ \emph {et~al.}(1988)\citenamefont {Morris},
  \citenamefont {Thorne},\ and\ \citenamefont
  {Yurtsever}}]{MorrisThorneYurtsever1988}%
  \BibitemOpen
  \bibfield  {author} {\bibinfo {author} {\bibfnamefont {M.~S.}\ \bibnamefont
  {Morris}}, \bibinfo {author} {\bibfnamefont {K.~S.}\ \bibnamefont {Thorne}},\
  and\ \bibinfo {author} {\bibfnamefont {U.}~\bibnamefont {Yurtsever}},\ }\href
  {https://doi.org/10.1103/PhysRevLett.61.1446} {\bibfield  {journal} {\bibinfo
   {journal} {Physical Review Letters}\ }\textbf {\bibinfo {volume} {61}},\
  \bibinfo {pages} {1446} (\bibinfo {year} {1988})}\BibitemShut {NoStop}%
\bibitem [{\citenamefont {Sonego}(2010)}]{Sonego2010}%
  \BibitemOpen
  \bibfield  {author} {\bibinfo {author} {\bibfnamefont {S.}~\bibnamefont
  {Sonego}},\ }\href {https://doi.org/10.1063/1.3485599} {\bibfield  {journal}
  {\bibinfo  {journal} {Journal of Mathematical Physics}\ }\textbf {\bibinfo
  {volume} {51}},\ \bibinfo {pages} {092502} (\bibinfo {year}
  {2010})}\BibitemShut {NoStop}%
\bibitem [{\citenamefont {Shapiro}(1964)}]{Shapiro1964}%
  \BibitemOpen
  \bibfield  {author} {\bibinfo {author} {\bibfnamefont {I.~I.}\ \bibnamefont
  {Shapiro}},\ }\href {https://doi.org/10.1103/PhysRevLett.13.789} {\bibfield
  {journal} {\bibinfo  {journal} {Physical Review Letters}\ }\textbf {\bibinfo
  {volume} {13}},\ \bibinfo {pages} {789} (\bibinfo {year} {1964})}\BibitemShut
  {NoStop}%
\bibitem [{\citenamefont {Shapiro}\ \emph {et~al.}(1968)\citenamefont
  {Shapiro}, \citenamefont {Pettengill}, \citenamefont {Ash}, \citenamefont
  {Stone}, \citenamefont {Smith}, \citenamefont {Ingalls},\ and\ \citenamefont
  {Brockelman}}]{Shapiro1968}%
  \BibitemOpen
  \bibfield  {author} {\bibinfo {author} {\bibfnamefont {I.~I.}\ \bibnamefont
  {Shapiro}}, \bibinfo {author} {\bibfnamefont {G.~H.}\ \bibnamefont
  {Pettengill}}, \bibinfo {author} {\bibfnamefont {M.~E.}\ \bibnamefont {Ash}},
  \bibinfo {author} {\bibfnamefont {M.~L.}\ \bibnamefont {Stone}}, \bibinfo
  {author} {\bibfnamefont {W.~B.}\ \bibnamefont {Smith}}, \bibinfo {author}
  {\bibfnamefont {R.~P.}\ \bibnamefont {Ingalls}},\ and\ \bibinfo {author}
  {\bibfnamefont {R.~A.}\ \bibnamefont {Brockelman}},\ }\href
  {https://doi.org/10.1103/PhysRevLett.20.1265} {\bibfield  {journal} {\bibinfo
   {journal} {Physical Review Letters}\ }\textbf {\bibinfo {volume} {20}},\
  \bibinfo {pages} {1265} (\bibinfo {year} {1968})}\BibitemShut {NoStop}%
\bibitem [{\citenamefont {Skordis}\ and\ \citenamefont
  {Zlosnik}(2022{\natexlab{a}})}]{Skordis:2021mry}%
  \BibitemOpen
  \bibfield  {author} {\bibinfo {author} {\bibfnamefont {C.}~\bibnamefont
  {Skordis}}\ and\ \bibinfo {author} {\bibfnamefont {T.}~\bibnamefont
  {Zlosnik}},\ }\href {https://doi.org/10.1103/PhysRevD.106.104041} {\bibfield
  {journal} {\bibinfo  {journal} {Phys. Rev. D}\ }\textbf {\bibinfo {volume}
  {106}},\ \bibinfo {pages} {104041} (\bibinfo {year} {2022}{\natexlab{a}})},\
  \Eprint {https://arxiv.org/abs/2109.13287} {arXiv:2109.13287 [gr-qc]}
  \BibitemShut {NoStop}%
\bibitem [{\citenamefont {Jacobson}\ and\ \citenamefont
  {Mattingly}(2001)}]{Ae}%
  \BibitemOpen
  \bibfield  {author} {\bibinfo {author} {\bibfnamefont {T.}~\bibnamefont
  {Jacobson}}\ and\ \bibinfo {author} {\bibfnamefont {D.}~\bibnamefont
  {Mattingly}},\ }\bibfield  {journal} {\bibinfo  {journal} {Physical Review
  D}\ }\textbf {\bibinfo {volume} {64}},\ \href
  {https://doi.org/10.1103/physrevd.64.024028} {10.1103/physrevd.64.024028}
  (\bibinfo {year} {2001})\BibitemShut {NoStop}%
\bibitem [{\citenamefont {Eling}\ \emph {et~al.}(2005)\citenamefont {Eling},
  \citenamefont {Jacobson},\ and\ \citenamefont {Mattingly}}]{Ae_review}%
  \BibitemOpen
  \bibfield  {author} {\bibinfo {author} {\bibfnamefont {C.}~\bibnamefont
  {Eling}}, \bibinfo {author} {\bibfnamefont {T.}~\bibnamefont {Jacobson}},\
  and\ \bibinfo {author} {\bibfnamefont {D.}~\bibnamefont {Mattingly}},\
  }\href@noop {} {\bibinfo {title} {Einstein-aether theory}} (\bibinfo {year}
  {2005}),\ \Eprint {https://arxiv.org/abs/gr-qc/0410001} {arXiv:gr-qc/0410001
  [gr-qc]} \BibitemShut {NoStop}%
\bibitem [{\citenamefont {Skordis}\ and\ \citenamefont
  {Zlosnik}(2022{\natexlab{b}})}]{Lin_stab_in_mk_AeST}%
  \BibitemOpen
  \bibfield  {author} {\bibinfo {author} {\bibfnamefont {C.}~\bibnamefont
  {Skordis}}\ and\ \bibinfo {author} {\bibfnamefont {T.}~\bibnamefont
  {Zlosnik}},\ }\href {https://doi.org/10.1103/PhysRevD.106.104041} {\bibfield
  {journal} {\bibinfo  {journal} {Phys. Rev. D}\ }\textbf {\bibinfo {volume}
  {106}},\ \bibinfo {pages} {104041} (\bibinfo {year}
  {2022}{\natexlab{b}})}\BibitemShut {NoStop}%
\bibitem [{\citenamefont {Durakovic}\ and\ \citenamefont
  {Skordis}(2024)}]{Durakovic:2023out}%
  \BibitemOpen
  \bibfield  {author} {\bibinfo {author} {\bibfnamefont {A.}~\bibnamefont
  {Durakovic}}\ and\ \bibinfo {author} {\bibfnamefont {C.}~\bibnamefont
  {Skordis}},\ }\href {https://doi.org/10.1088/1475-7516/2024/04/040}
  {\bibfield  {journal} {\bibinfo  {journal} {JCAP}\ }\textbf {\bibinfo
  {volume} {04}},\ \bibinfo {pages} {040}},\ \Eprint
  {https://arxiv.org/abs/2312.00889} {arXiv:2312.00889 [astro-ph.CO]}
  \BibitemShut {NoStop}%
\bibitem [{\citenamefont {Huang}\ \emph {et~al.}(2022)\citenamefont {Huang},
  \citenamefont {Lü},\ and\ \citenamefont {Yang}}]{WH_ES}%
  \BibitemOpen
  \bibfield  {author} {\bibinfo {author} {\bibfnamefont {H.}~\bibnamefont
  {Huang}}, \bibinfo {author} {\bibfnamefont {H.}~\bibnamefont {Lü}},\ and\
  \bibinfo {author} {\bibfnamefont {J.}~\bibnamefont {Yang}},\ }\href
  {https://doi.org/10.1088/1361-6382/ac8266} {\bibfield  {journal} {\bibinfo
  {journal} {Classical and Quantum Gravity}\ }\textbf {\bibinfo {volume}
  {39}},\ \bibinfo {pages} {185009} (\bibinfo {year} {2022})}\BibitemShut
  {NoStop}%
\bibitem [{\citenamefont {Yazadjiev}(2017)}]{EB_WH1}%
  \BibitemOpen
  \bibfield  {author} {\bibinfo {author} {\bibfnamefont {S.}~\bibnamefont
  {Yazadjiev}},\ }\href {https://doi.org/10.1103/PhysRevD.96.044045} {\bibfield
   {journal} {\bibinfo  {journal} {Phys. Rev. D}\ }\textbf {\bibinfo {volume}
  {96}},\ \bibinfo {pages} {044045} (\bibinfo {year} {2017})}\BibitemShut
  {NoStop}%
\bibitem [{\citenamefont {Ellis}(1979)}]{EB_WH2}%
  \BibitemOpen
  \bibfield  {author} {\bibinfo {author} {\bibfnamefont {H.~G.}\ \bibnamefont
  {Ellis}},\ }\href {https://doi.org/10.1007/BF00756794} {\bibfield  {journal}
  {\bibinfo  {journal} {Gen. Rel. Grav.}\ }\textbf {\bibinfo {volume} {10}},\
  \bibinfo {pages} {105} (\bibinfo {year} {1979})}\BibitemShut {NoStop}%
\bibitem [{\citenamefont {Geng}\ and\ \citenamefont {L\"u}(2016)}]{LvWormhole}%
  \BibitemOpen
  \bibfield  {author} {\bibinfo {author} {\bibfnamefont {W.-J.}\ \bibnamefont
  {Geng}}\ and\ \bibinfo {author} {\bibfnamefont {H.}~\bibnamefont {L\"u}},\
  }\href {https://doi.org/10.1103/PhysRevD.93.044035} {\bibfield  {journal}
  {\bibinfo  {journal} {Phys. Rev. D}\ }\textbf {\bibinfo {volume} {93}},\
  \bibinfo {pages} {044035} (\bibinfo {year} {2016})}\BibitemShut {NoStop}%
\bibitem [{\citenamefont {Huang}\ \emph {et~al.}(2023)\citenamefont {Huang},
  \citenamefont {Kunz}, \citenamefont {Yang},\ and\ \citenamefont
  {Zhang}}]{Shadow_WH_ES}%
  \BibitemOpen
  \bibfield  {author} {\bibinfo {author} {\bibfnamefont {H.}~\bibnamefont
  {Huang}}, \bibinfo {author} {\bibfnamefont {J.}~\bibnamefont {Kunz}},
  \bibinfo {author} {\bibfnamefont {J.}~\bibnamefont {Yang}},\ and\ \bibinfo
  {author} {\bibfnamefont {C.}~\bibnamefont {Zhang}},\ }\href
  {https://doi.org/10.1103/PhysRevD.107.104060} {\bibfield  {journal} {\bibinfo
   {journal} {Phys. Rev. D}\ }\textbf {\bibinfo {volume} {107}},\ \bibinfo
  {pages} {104060} (\bibinfo {year} {2023})}\BibitemShut {NoStop}%
\bibitem [{\citenamefont {Fisher}(1948)}]{Fisher1948}%
  \BibitemOpen
  \bibfield  {author} {\bibinfo {author} {\bibfnamefont {I.~Z.}\ \bibnamefont
  {Fisher}},\ }\href@noop {} {\bibfield  {journal} {\bibinfo  {journal}
  {Zh. Eksp. Teor. Fiz.}\ }\textbf {\bibinfo {volume} {18}},\ \bibinfo
  {pages} {636} (\bibinfo {year} {1948})}\BibitemShut {NoStop}%
\bibitem [{\citenamefont {Janis}\ \emph {et~al.}(1968)\citenamefont {Janis},
  \citenamefont {Newman},\ and\ \citenamefont
  {Winicour}}]{JanisNewmanWinicour1968}%
  \BibitemOpen
  \bibfield  {author} {\bibinfo {author} {\bibfnamefont {A.~I.}\ \bibnamefont
  {Janis}}, \bibinfo {author} {\bibfnamefont {E.~T.}\ \bibnamefont {Newman}},\
  and\ \bibinfo {author} {\bibfnamefont {J.}~\bibnamefont {Winicour}},\ }\href
  {https://doi.org/10.1103/PhysRevLett.20.878} {\bibfield  {journal} {\bibinfo
  {journal} {Physical Review Letters}\ }\textbf {\bibinfo {volume} {20}},\
  \bibinfo {pages} {878} (\bibinfo {year} {1968})}\BibitemShut {NoStop}%
\bibitem [{\citenamefont {Eling}\ and\ \citenamefont
  {Jacobson}(2006)}]{BH_in_Ae}%
  \BibitemOpen
  \bibfield  {author} {\bibinfo {author} {\bibfnamefont {C.}~\bibnamefont
  {Eling}}\ and\ \bibinfo {author} {\bibfnamefont {T.}~\bibnamefont
  {Jacobson}},\ }\href@noop {} {\bibfield  {journal} {\bibinfo  {journal}
  {Classical and Quantum Gravity}\ }\textbf {\bibinfo {volume} {23}},\ \bibinfo
  {pages} {5643} (\bibinfo {year} {2006})}\BibitemShut {NoStop}%
\bibitem [{\citenamefont {Xu}\ \emph {et~al.}(2023)\citenamefont {Xu},
  \citenamefont {Liang},\ and\ \citenamefont {Shao}}]{BH_in_BB}%
  \BibitemOpen
  \bibfield  {author} {\bibinfo {author} {\bibfnamefont {R.}~\bibnamefont
  {Xu}}, \bibinfo {author} {\bibfnamefont {D.}~\bibnamefont {Liang}},\ and\
  \bibinfo {author} {\bibfnamefont {L.}~\bibnamefont {Shao}},\ }\href
  {https://doi.org/10.1103/PhysRevD.107.024011} {\bibfield  {journal} {\bibinfo
   {journal} {Phys. Rev. D}\ }\textbf {\bibinfo {volume} {107}},\ \bibinfo
  {pages} {024011} (\bibinfo {year} {2023})}\BibitemShut {NoStop}%
\bibitem [{\citenamefont {Hsu}\ \emph {et~al.}(2024)\citenamefont {Hsu},
  \citenamefont {Lasenby}, \citenamefont {Barker}, \citenamefont {Durakovic},\
  and\ \citenamefont {Hobson}}]{Hsu:2024ftc}%
  \BibitemOpen
  \bibfield  {author} {\bibinfo {author} {\bibfnamefont {Y.-H.}\ \bibnamefont
  {Hsu}}, \bibinfo {author} {\bibfnamefont {A.}~\bibnamefont {Lasenby}},
  \bibinfo {author} {\bibfnamefont {W.}~\bibnamefont {Barker}}, \bibinfo
  {author} {\bibfnamefont {A.}~\bibnamefont {Durakovic}},\ and\ \bibinfo
  {author} {\bibfnamefont {M.}~\bibnamefont {Hobson}},\ }\href@noop {}
  {\bibfield  {journal} {\bibinfo  {journal} {arXiv e-Prints}\ } (\bibinfo
  {year} {2024})},\ \Eprint {https://arxiv.org/abs/2411.02550}
  {arXiv:2411.02550 [gr-qc]} \BibitemShut {NoStop}%
\bibitem [{\citenamefont {Palais}(1979)}]{Palais1979ThePO}%
  \BibitemOpen
  \bibfield  {author} {\bibinfo {author} {\bibfnamefont {R.}~\bibnamefont
  {Palais}},\ }\href {https://api.semanticscholar.org/CorpusID:59127987}
  {\bibfield  {journal} {\bibinfo  {journal} {Communications in Mathematical
  Physics}\ }\textbf {\bibinfo {volume} {69}},\ \bibinfo {pages} {19} (\bibinfo
  {year} {1979})}\BibitemShut {NoStop}%
\bibitem [{\citenamefont {Fels}\ and\ \citenamefont {Torre}(2002)}]{Fels_2002}%
  \BibitemOpen
  \bibfield  {author} {\bibinfo {author} {\bibfnamefont {M.~E.}\ \bibnamefont
  {Fels}}\ and\ \bibinfo {author} {\bibfnamefont {C.~G.}\ \bibnamefont
  {Torre}},\ }\href {https://doi.org/10.1088/0264-9381/19/4/303} {\bibfield
  {journal} {\bibinfo  {journal} {Classical and Quantum Gravity}\ }\textbf
  {\bibinfo {volume} {19}},\ \bibinfo {pages} {641–675} (\bibinfo {year}
  {2002})}\BibitemShut {NoStop}%
\bibitem [{\citenamefont {Oost}\ \emph {et~al.}(2021)\citenamefont {Oost},
  \citenamefont {Mukohyama},\ and\ \citenamefont
  {Wang}}]{Ae_BH_new_chart_analytical_Oost_2021}%
  \BibitemOpen
  \bibfield  {author} {\bibinfo {author} {\bibfnamefont {J.}~\bibnamefont
  {Oost}}, \bibinfo {author} {\bibfnamefont {S.}~\bibnamefont {Mukohyama}},\
  and\ \bibinfo {author} {\bibfnamefont {A.}~\bibnamefont {Wang}},\ }\href
  {https://doi.org/10.3390/universe7080272} {\bibfield  {journal} {\bibinfo
  {journal} {Universe}\ }\textbf {\bibinfo {volume} {7}},\ \bibinfo {pages}
  {272} (\bibinfo {year} {2021})}\BibitemShut {NoStop}%
\bibitem [{\citenamefont {Gao}\ and\ \citenamefont
  {Shen}(2013)}]{Ae_BH_new_chart_numerical_Gao_2013}%
  \BibitemOpen
  \bibfield  {author} {\bibinfo {author} {\bibfnamefont {C.}~\bibnamefont
  {Gao}}\ and\ \bibinfo {author} {\bibfnamefont {Y.-G.}\ \bibnamefont {Shen}},\
  }\bibfield  {journal} {\bibinfo  {journal} {Physical Review D}\ }\textbf
  {\bibinfo {volume} {88}},\ \href {https://doi.org/10.1103/physrevd.88.103508}
  {10.1103/physrevd.88.103508} (\bibinfo {year} {2013})\BibitemShut {NoStop}%
\bibitem [{\citenamefont {Seifert}\ and\ \citenamefont
  {Wald}(2007)}]{Seifert:2006kv}%
  \BibitemOpen
  \bibfield  {author} {\bibinfo {author} {\bibfnamefont {M.~D.}\ \bibnamefont
  {Seifert}}\ and\ \bibinfo {author} {\bibfnamefont {R.~M.}\ \bibnamefont
  {Wald}},\ }\href {https://doi.org/10.1103/PhysRevD.75.084029} {\bibfield
  {journal} {\bibinfo  {journal} {Phys. Rev. D}\ }\textbf {\bibinfo {volume}
  {75}},\ \bibinfo {pages} {084029} (\bibinfo {year} {2007})},\ \Eprint
  {https://arxiv.org/abs/gr-qc/0612121} {arXiv:gr-qc/0612121} \BibitemShut
  {NoStop}%
\bibitem [{\citenamefont {Seifert}(2007)}]{Seifert:2007fr}%
  \BibitemOpen
  \bibfield  {author} {\bibinfo {author} {\bibfnamefont {M.~D.}\ \bibnamefont
  {Seifert}},\ }\href {https://doi.org/10.1103/PhysRevD.76.064002} {\bibfield
  {journal} {\bibinfo  {journal} {Phys. Rev. D}\ }\textbf {\bibinfo {volume}
  {76}},\ \bibinfo {pages} {064002} (\bibinfo {year} {2007})},\ \Eprint
  {https://arxiv.org/abs/gr-qc/0703060} {arXiv:gr-qc/0703060} \BibitemShut
  {NoStop}%
\bibitem [{\citenamefont {Luz}\ and\ \citenamefont
  {Carloni}(2024)}]{Luz:2024yjm}%
  \BibitemOpen
  \bibfield  {author} {\bibinfo {author} {\bibfnamefont {P.}~\bibnamefont
  {Luz}}\ and\ \bibinfo {author} {\bibfnamefont {S.}~\bibnamefont {Carloni}},\
  }\href {https://doi.org/10.1088/1361-6382/ad8a14} {\bibfield  {journal}
  {\bibinfo  {journal} {Class. Quant. Grav.}\ }\textbf {\bibinfo {volume}
  {41}},\ \bibinfo {pages} {235012} (\bibinfo {year} {2024})},\ \Eprint
  {https://arxiv.org/abs/2405.05321} {arXiv:2405.05321 [gr-qc]} \BibitemShut
  {NoStop}%
\bibitem [{\citenamefont {Deser}\ and\ \citenamefont
  {Woodard}(2007)}]{Deser:2007jk}%
  \BibitemOpen
  \bibfield  {author} {\bibinfo {author} {\bibfnamefont {S.}~\bibnamefont
  {Deser}}\ and\ \bibinfo {author} {\bibfnamefont {R.~P.}\ \bibnamefont
  {Woodard}},\ }\href {https://doi.org/10.1103/PhysRevLett.99.111301}
  {\bibfield  {journal} {\bibinfo  {journal} {Phys. Rev. Lett.}\ }\textbf
  {\bibinfo {volume} {99}},\ \bibinfo {pages} {111301} (\bibinfo {year}
  {2007})},\ \Eprint {https://arxiv.org/abs/0706.2151} {arXiv:0706.2151
  [astro-ph]} \BibitemShut {NoStop}%
\bibitem [{\citenamefont {Bao}\ \emph {et~al.}(2023)\citenamefont {Bao},
  \citenamefont {Hou},\ and\ \citenamefont {Zhang}}]{Bao2023}%
  \BibitemOpen
  \bibfield  {author} {\bibinfo {author} {\bibfnamefont {S.-s.}\ \bibnamefont
  {Bao}}, \bibinfo {author} {\bibfnamefont {S.}~\bibnamefont {Hou}},\ and\
  \bibinfo {author} {\bibfnamefont {H.}~\bibnamefont {Zhang}},\ }\href
  {https://doi.org/10.1140/epjc/s10052-023-11281-9} {\bibfield  {journal}
  {\bibinfo  {journal} {Eur. Phys. J. C}\ }\textbf {\bibinfo {volume} {83}},\
  \bibinfo {pages} {127} (\bibinfo {year} {2023})}\BibitemShut {NoStop}%
\bibitem [{\citenamefont {Jaffe}(1983)}]{Jaffe1983}%
  \BibitemOpen
  \bibfield  {author} {\bibinfo {author} {\bibfnamefont {W.}~\bibnamefont
  {Jaffe}},\ }\href {https://doi.org/10.1093/mnras/202.4.995} {\bibfield
  {journal} {\bibinfo  {journal} {Monthly Notices of the Royal Astronomical
  Society}\ }\textbf {\bibinfo {volume} {202}},\ \bibinfo {pages} {995}
  (\bibinfo {year} {1983})}\BibitemShut {NoStop}%
\bibitem [{\citenamefont {Hernquist}(1990)}]{Hernquist1990}%
  \BibitemOpen
  \bibfield  {author} {\bibinfo {author} {\bibfnamefont {L.}~\bibnamefont
  {Hernquist}},\ }\href {https://doi.org/10.1086/168845} {\bibfield  {journal}
  {\bibinfo  {journal} {The Astrophysical Journal}\ }\textbf {\bibinfo {volume}
  {356}},\ \bibinfo {pages} {359} (\bibinfo {year} {1990})}\BibitemShut
  {NoStop}%
\bibitem [{\citenamefont {Dehnen}(1993)}]{Dehnen1993}%
  \BibitemOpen
  \bibfield  {author} {\bibinfo {author} {\bibfnamefont {W.}~\bibnamefont
  {Dehnen}},\ }\href {https://doi.org/10.1093/mnras/265.1.250} {\bibfield
  {journal} {\bibinfo  {journal} {Monthly Notices of the Royal Astronomical
  Society}\ }\textbf {\bibinfo {volume} {265}},\ \bibinfo {pages} {250}
  (\bibinfo {year} {1993})}\BibitemShut {NoStop}%
\bibitem [{\citenamefont {Kassiola}\ and\ \citenamefont
  {Kovner}(1993)}]{Kassiola1993}%
  \BibitemOpen
  \bibfield  {author} {\bibinfo {author} {\bibfnamefont {A.}~\bibnamefont
  {Kassiola}}\ and\ \bibinfo {author} {\bibfnamefont {I.}~\bibnamefont
  {Kovner}},\ }\href {https://doi.org/10.1086/173325} {\bibfield  {journal}
  {\bibinfo  {journal} {The Astrophysical Journal}\ }\textbf {\bibinfo {volume}
  {417}},\ \bibinfo {pages} {450} (\bibinfo {year} {1993})}\BibitemShut
  {NoStop}%
\bibitem [{\citenamefont {Eling}\ \emph {et~al.}(2007)\citenamefont {Eling},
  \citenamefont {Jacobson},\ and\ \citenamefont {Miller}}]{NS_Ae}%
  \BibitemOpen
  \bibfield  {author} {\bibinfo {author} {\bibfnamefont {C.}~\bibnamefont
  {Eling}}, \bibinfo {author} {\bibfnamefont {T.}~\bibnamefont {Jacobson}},\
  and\ \bibinfo {author} {\bibfnamefont {M.~C.}\ \bibnamefont {Miller}},\
  }\href {https://api.semanticscholar.org/CorpusID:29588513} {\bibfield
  {journal} {\bibinfo  {journal} {Physical Review D}\ }\textbf {\bibinfo
  {volume} {76}},\ \bibinfo {pages} {042003} (\bibinfo {year}
  {2007})}\BibitemShut {NoStop}%
\end{thebibliography}%

\onecolumngrid

\appendix

\section{Throat conditions}\label{ThroatConds}

\paragraph*{Stationarity condition} Working from~\cref{JE_wormholeB,JE_wormhole_throat_radius}, the position of the throat is found to be
\begin{equation}\label{ThroatRadiusCoord}
        r_{\text{T}} \equiv \GN{} M \left( 1+ \sqrt{1-\KBE{}/2}\right).
\end{equation}
Taking again~\cref{JE_wormholeB} and differentiating, and then plugging in~\cref{ThroatRadiusCoord}, the stationarity~$\mathcal{R}'\left( r_{\text{T}}\right) = 0$ of the two-sphere area at the throat is readily confirmed. In particular,~\cref{ThroatRadiusCoord} is valid for (physical) bounds leading to~$\KBE{}<2$, as given in~\cref{NewAetherCouplingRange}. For~$\KBE{}>2$, the solution changes character to the Ellis--Bronnikov drainhole, as explained in~\cref{appendix: Wormhole solution: mathematical detail}.

\paragraph*{Flare-out condition} The stationarity of the two-sphere area can be promoted to minimality~$\mathcal{R}''\left( r_{\text{T}}\right) > 0$, where the formula --- again from~\cref{JE_wormholeB,ThroatRadiusCoord} --- is
\begin{equation}
        \mathcal{R}''\left( r_{\text{T}}\right) = \frac{\KBE{}\mathcal{R}_{\text{T}}}{2} \cdot \left(
        \frac{
        \GN{} M
        }{
        r_{\text{T}} \left(
            r_{\text{T}}-r_{\text{S}}
        \right)
        } 
        \right) ^2.
\end{equation}
Together, these conditions confirm that the object is indeed a wormhole for physical values of the parameters.

\section{Details of covariant equations}\label{appendix: Field equations}
\paragraph*{Definitions} In this appendix, we present the covariant field equations extracted from the action~\cref{original_action_AEST}, which we here write as
\begin{equation}\label{ActionAESTSecondAbbreviation}
	\tensor*{S}{_{\text{\AE{}ST}}} \equiv \int \mathrm{d} ^4x \, \frac{\sqrt{-g}}{2\kappa} \Bigg[R -\frac{\KB{}}{2} \tensor{F}{^{\mu\nu}} \tensor{F}{_{\mu \nu}}
+b  \tensor{J}{^\mu} \tensor{\nabla}{_\mu} \varphi
-\mathcal{G}\left(\mathcal{Y}, \mathcal{Q} \right) - \lambda\left(\tensor{A}{^\mu} \tensor{A}{_\mu} +1\right) \Bigg] \,\, ,
\end{equation}
where we define the quantities
\begin{equation}\label{GABDefinitions}
	\mathcal{G}\left(\mathcal{Y}, \mathcal{Q} \right) \equiv (2-\KB{}) \mathcal{Y}+{\mathcal{F}}\left(\mathcal{Y}, \mathcal{Q} \right), \quad a \equiv \frac{b}{2}(1+\lambda_s), \quad b \equiv 2(2-\KB{}) \,. 
\end{equation}
The shorthand~$a$ does not appear explicitly in this form of the action but will be useful below.
Unlike in~\cref{Action_general}, which assumes the specific form~\cref{F_split_into_Q_and_Y} of the function~$\mathcal{F}(\mathcal{Y}, \mathcal{Q})$, here we do not make any such assumptions.

\paragraph*{Equations} The equations for the vector~$A_\mu$, the scalar~$\varphi$, and the metric~$g_{\mu \nu}$  obtained from~\cref{ActionAESTSecondAbbreviation} are:
\begin{subequations}\small
    \begin{align}
	    \frac{\delta \tensor*{S}{_{\text{\AE{}ST}}}}{\delta\tensor{A}{_\mu}}\propto
&\ -2\lambda  \tensor{A}{^{\mu}}+b\left(\tensor{\nabla}{^{\mu}} \tensor{A}{^{\nu}}\right)\left(\tensor{\nabla}{_{\nu}} {\varphi}\right)-2 \KB{}\left(\tensor{\nabla}{_{\nu}} \tensor{\nabla}{^{\mu}} \tensor{A}{^{\nu}}\right)-b \tensor{A}{^{\nu}}\left(\tensor{\nabla}{_{\nu}} \tensor{\nabla}{^{\mu}} {\varphi}\right)+2 \KB{}\left(\tensor{\nabla}{_{\nu}} \tensor{\nabla}{^{\nu}} \tensor{A}{^{\mu}}\right)  \phantom{\Big)}\nonumber \\
&\ -\left(\tensor{\nabla}{^{\mu}} {\varphi}\right)\left(b\left(\tensor{\nabla}{_{\nu}} \tensor{A}{^{\nu}}\right)+{\mathcal{G}}^{(0, 1)}+2 \tensor{A}{^{\nu}}\left(\tensor{\nabla}{_{\nu}} {\varphi}\right)\left(2-\KB{}+{\mathcal{G}}^{(1, 0)}\right)\right)=0
	    ,\label{CovariantAEquation}\\
	    \frac{\delta \tensor*{S}{_{\text{\AE{}ST}}}}{\delta\varphi}\propto
&\ -b \tensor{A}{^{\mu}}\left(\tensor{\nabla}{_{\nu}} \tensor{\nabla}{_{\mu}} \tensor{A}{^{\nu}}\right)-b\left(\tensor{\nabla}{_{\mu}} \tensor{A}{_{\nu}}\right)\left(\tensor{\nabla}{^{\nu}} \tensor{A}{^{\mu}}\right)+\left(\tensor{\nabla}{_{\mu}} \tensor{A}{^{\mu}}\right) {\mathcal{G}}^{(0,1)}+\tensor{A}{^{\mu}}\left(\tensor{\nabla}{_{\mu}} \tensor{A}{^{\nu}}\right)\left(\tensor{\nabla}{_{\nu}} {\varphi}\right) {\mathcal{G}}^{(0,2)}+ \tensor{A}{^{\mu}} \tensor{A}{^{\nu}}\left(\tensor{\nabla}{_{\nu}} \tensor{\nabla}{_{\mu}} {\varphi}\right) {\mathcal{G}}^{(0, 2)}  \phantom{\Big)}\nonumber \\
&\ +2\left(\tensor{\nabla}{_{\mu}} \tensor{\nabla}{^{\mu}} {\varphi}\right) {\mathcal{G}}^{(1,0)}+2 \tensor{A}{^{\mu}}\left(\tensor{\nabla}{_{\mu}} {\varphi}\right)\left(\tensor{\nabla}{_{\nu}} \tensor{A}{^{\nu}}\right) {\mathcal{G}}^{(1, 0)}+ 2 \tensor{A}{^{\mu}}\left(\tensor{\nabla}{_{\mu}} \tensor{A}{^{\nu}}\right)\left(\tensor{\nabla}{_{\nu}} {\varphi}\right) {\mathcal{G}}^{(1, 0)}+2 \tensor{A}{^{\mu}} \tensor{A}{^{\nu}}\left(\tensor{\nabla}{_{\nu}} \tensor{\nabla}{_{\mu}} {\varphi}\right) {\mathcal{G}}^{(1, 0)} \phantom{\Big)}\nonumber \\
&\ +2\left(\tensor{\nabla}{_{\mu}} {\varphi}\right)\left(\tensor{\nabla}{_{\nu}} {\varphi}\right)\left(\tensor{\nabla}{^{\nu}} \tensor{A}{^{\mu}}\right) {\mathcal{G}}^{(1,1)}+ 2 \tensor{A}{^{\mu}}\left(\tensor{\nabla}{_{\mu}} \tensor{\nabla}{_{\nu}} {\varphi}\right)\left(\tensor{\nabla}{^{\nu}} {\varphi}\right) {\mathcal{G}}^{(1,1)}+2 \tensor{A}{^{\mu}}\left(\tensor{\nabla}{_{\nu}} \tensor{\nabla}{_{\mu}} {\varphi}\right)\left(\tensor{\nabla}{^{\nu}} {\varphi}\right) {\mathcal{G}}^{(1,1)} \phantom{\Big)}\nonumber \\
&\ +4 \tensor{A}{^{\mu}} \tensor{A}{^{\nu}}\left(\tensor{\nabla}{_{\mu}} \tensor{A}{^{\rho}}\right)\left(\tensor{\nabla}{_{\nu}} {\varphi}\right)\left(\tensor{\nabla}{_{\rho}} {\varphi}\right) {\mathcal{G}}^{(1,1)}+ 4 \tensor{A}{^{\mu}} \tensor{A}{^{\nu}} \tensor{A}{^{\rho}}\left(\tensor{\nabla}{_{\mu}} {\varphi}\right)\left(\tensor{\nabla}{_{\rho}} \tensor{\nabla}{_{\nu}} {\varphi}\right) {\mathcal{G}}^{(1,1)}+4\left(\tensor{\nabla}{^{\mu}} {\varphi}\right)\left(\tensor{\nabla}{_{\nu}} \tensor{\nabla}{_{\mu}} {\varphi}\right)\left(\tensor{\nabla}{^{\nu}} {\varphi}\right){\mathcal{G}}^{(2,0)}  \phantom{\Big)}\nonumber \\
&+4\tensor{A}{^{\mu}}\tensor{A}{^{\nu}} \tensor{A}{^{\rho}}\left(\tensor{\nabla}{_{\mu}} \tensor{A}{^{\sigma}}\right)\left(\tensor{\nabla}{_{\nu}} {\varphi}\right)\left(\tensor{\nabla}{_{\rho}} {\varphi}\right)\left(\tensor{\nabla}{_{\sigma}} {\varphi}\right){\mathcal{G}}^{(2,0)}+4\tensor{A}{^{\mu}} \tensor{A}{^{\nu}}\left(\tensor{\nabla}{_{\mu}} {\varphi}\right)\left(\tensor{\nabla}{^{\rho}} {\varphi}\right)\left(\tensor{\nabla}{_{\nu}} \tensor{\nabla}{_{\rho}} {\varphi}+\tensor{\nabla}{_{\rho}} \tensor{\nabla}{_{\nu}} {\varphi}\right){\mathcal{G}}^{(2,0)} \phantom{\Big)}\nonumber \\
&\ +4\tensor{A}{^{\mu}}\left(\tensor{\nabla}{_{\mu}} {\varphi}\right)\left(\tensor{\nabla}{_{\nu}} {\varphi}\right)\left(\left(\tensor{\nabla}{_{\rho}} {\varphi}\right)\left(\tensor{\nabla}{^{\rho}} \tensor{A}{^{\nu}}\right)+\tensor{A}{^{\nu}} \tensor{A}{^{\rho}} \tensor{A}{^{\sigma}}\left(\tensor{\nabla}{_{\sigma}} \tensor{\nabla}{_{\rho}} {\varphi}\right)\right)  {\mathcal{G}}^{(2,0)}=0
,\label{CovariantPhiEquation}\\
	    \frac{\delta \tensor*{S}{_{\text{\AE{}ST}}}}{\delta\tensor{g}{^{\mu\nu}}}\propto
&\ 2 \tensor{R}{_{\mu\nu}}+\tensor{g}{_{\mu\nu}}\left(-R+2 {\kappa}\left({\mathcal{G}}-b \tensor{A}{^{\rho}}\left(\tensor{\nabla}{_{\rho}} \tensor{A}{^{\sigma}}\right)\left(\tensor{\nabla}{_{\sigma}} {\varphi}\right)+\KB{}\left(-\left(\tensor{\nabla}{_{\rho}} \tensor{A}{_{\sigma}}\right)+\tensor{\nabla}{_{\sigma}} \tensor{A}{_{\rho}}\right)\left(\tensor{\nabla}{^{\sigma}} \tensor{A}{^{\rho}}\right)\right)\right) \phantom{\Big)}  \nonumber \\
&\ + 2 {\kappa} \Big(-2 \tensor{A}{_{\mu}} \tensor{A}{_{\nu}} \lambda+b \tensor{A}{^{\rho}}\left(\left(\tensor{\nabla}{_{\nu}} {\varphi}\right)\left(\tensor{\nabla}{_{\nu}} \tensor{A}{_{\mu}}\right)+\left(\tensor{\nabla}{_{\mu}} {\varphi}\right)\left(\tensor{\nabla}{_{\rho}} \tensor{A}{_{\nu}}\right)\right)+ b \tensor{A}{_{\nu}}\left(\tensor{\nabla}{_{\mu}} \tensor{A}{^{\rho}}\right)\left(\tensor{\nabla}{_{\rho}} {\varphi}\right)  \nonumber \\
&\ \phantom{+ 2 {\kappa}} +b \tensor{A}{_{\mu}}\left(\tensor{\nabla}{_{\nu}} \tensor{A}{^{\rho}}\right)\left(\tensor{\nabla}{_{\rho}} {\varphi}\right)-b \tensor{A}{_{\mu}} \tensor{A}{_{\nu}}\left(\tensor{\nabla}{_{\rho}} \tensor{\nabla}{^{\rho}} {\varphi}\right)-b \tensor{A}{_{\nu}}\left(\tensor{\nabla}{_{\rho}} {\varphi}\right)\left(\tensor{\nabla}{^{\rho}} \tensor{A}{_{\mu}}\right)- b \tensor{A}{_{\mu}}\left(\tensor{\nabla}{_{\rho}} {\varphi}\right)\left(\tensor{\nabla}{^{\rho}} \tensor{A}{_{\nu}}\right)  \nonumber \\
&\ \phantom{+ 2 {\kappa}} +2 \KB{}\left(-\left(\tensor{\nabla}{_{\mu}} \tensor{A}{^{\rho}}\right)\left(\tensor{\nabla}{_{\nu}} \tensor{A}{_{\rho}}\right)+\left(\tensor{\nabla}{_{\nu}} \tensor{A}{_{\rho}}-\tensor{\nabla}{_{\rho}} \tensor{A}{_{\nu}}\right)\left(\tensor{\nabla}{^{\rho}} \tensor{A}{_{\mu}}\right)+\left(\tensor{\nabla}{_{\mu}} \tensor{A}{_{\rho}}\right)\left(\tensor{\nabla}{^{\rho}} \tensor{A}{_{\nu}}\right)\right)  \nonumber \\
&\ \phantom{+ 2 {\kappa}} - \left(\tensor{A}{_{\nu}}\left(\tensor{\nabla}{_{\mu}} {\varphi}\right)+\tensor{A}{_{\mu}}\left(\tensor{\nabla}{_{\nu}} {\varphi}\right)\right) {\mathcal{G}}^{(0,1)}-2\left(\left(\tensor{\nabla}{_{\mu}} {\varphi}\right)\left(\tensor{\nabla}{_{\nu}} {\varphi}\right)+\tensor{A}{^{\rho}}\left(\tensor{A}{_{\nu}}\left(\tensor{\nabla}{_{\mu}} {\varphi}\right)+\tensor{A}{_{\mu}}\left(\tensor{\nabla}{_{\nu}} {\varphi}\right)\right)\left(\tensor{\nabla}{_{\rho}} {\varphi}\right)\right) {\mathcal{G}}^{(1,0)}\Big)=0
.\label{CovariantGEquation}
\end{align}
\end{subequations}
In~\cref{CovariantAEquation,CovariantPhiEquation,CovariantGEquation} we use the notation~${\mathcal{G}}^{(m,n)}$ to represent the~$m$th partial derivative with respect to the first argument and the~$n$th partial derivative with respect to the second argument of~${\mathcal{G}}\left( \mathcal{Y}, \mathcal{Q} \right)$. By contracting~\cref{CovariantAEquation} with an extra factor of~$\tensor{A}{_\mu}$ and using~\cref{UnitTimelike}, we obtain an algebraic equation for the Lagrange multiplier~$\lambda$,
\begin{align}
    2\lambda   = &- b   J^\nu \left(\nabla_\nu \varphi\right)
    + 2\KB{}  A^\mu \left(\nabla_\nu \nabla_\mu A^\nu\right) 
    + b  A^\mu A^\nu \left(\nabla_\nu \nabla_\mu \varphi\right)
    - 2\KB{}  A^\mu \left(\nabla_\nu \nabla^\nu A_\mu\right) 
    \nonumber\\
& - \left(A^\mu \nabla_\mu \varphi\right)  \bigg[ -b\left(\nabla_\nu A^\nu\right)  - {\mathcal{G}}^{(0,1)} +2 A^\nu \left(\nabla_\nu\varphi\right)   \left( \KB{}-2 - {\mathcal{G}}^{(0,1)} \right) \bigg]\,\, . \label{lag_exp}
\end{align}

\section{Details of component equations}\label{appendix: Static spherical equations in vacuum}

\paragraph*{Definitions} In this appendix, we present all the non-trivial components of the field equations~\cref{CovariantAEquation,CovariantPhiEquation,CovariantGEquation}, and at the same time we restrict the action~\cref{ActionAESTSecondAbbreviation} to the specific form~\cref{F_split_into_Q_and_Y} for the function~$\mathcal{F}(\mathcal{Y}, \mathcal{Q})$. The assumptions made about the behaviour of the fields are the minimal ones in~\cref{sph_back,ae_back} required for spherical symmetry, along with the Ansatz~\cref{P_back} for the scalar. The arguments of the fields~$\psi(r)$,~$\chi(t)$,~$\mathcal{M}(r)$,~$\mathcal{N}(r)$,~$\mathcal{R}(r)$,~$\alpha(r)$ and~$\mathcal{V}\left( \mathcal{Y}\right)$ are suppressed, and a prime denotes differentiation with respect to each argument. We also define~$\mathrm{S}_k \equiv \sinh(k\alpha(r))$ and~$\mathrm{C}_k \equiv \cosh(k\alpha(r))$.

\paragraph*{Equations} There are seven equations in total:
\begin{subequations}\small
\begin{align}
	\frac{\delta \tensor*{S}{_{\text{\AE{}ST}}^{\text{SZ}}}}{\delta\tensor{A}{_t}}\propto
&\ \ {\mathrm{S}}_{1}  \Big(4 \KB{} \mathcal{R} {\mathrm{C}}_{1} {\mathrm{S}}_{1} {\mathcal{M}}^{\prime} {\mathcal{N}}^{\prime}-8 \KB{} {\mathrm{C}}_{1} {\mathrm{S}}_{1} {\mathcal{N}}^{\prime} \mathcal{R}^{\prime}+4 \KB{} \mathcal{R} {\mathrm{S}}_{1}^2 {\mathcal{M}}^{\prime} {\alpha}^{\prime}-4 \KB{} \mathcal{R} {\mathrm{S}}_{1}^2 {\mathcal{N}}^{\prime} {\alpha}^{\prime}-8 \KB{} {\mathrm{S}}_{1}^2 \mathcal{R}^{\prime} {\alpha}^{\prime}-4 \KB{} \mathcal{R} {\mathrm{C}}_{1} {\mathrm{S}}_{1} {\alpha}^{\prime 2} \phantom{\Big)} \nonumber \\
&\ \phantom{{\mathrm{S}}_{1}   \Big(}- 4 e^{2 {\mathcal{M}}-{\mathcal{N}}} \mathcal{R} {\mathrm{S}}_{1} \mathcal{F}_{20} Q_0^2 {\chi}^{\prime}+2 e^{{\mathcal{M}}-{\mathcal{N}}} b \mathcal{R} {\mathrm{C}}_{2} Q_0 {\mathcal{N}}^{\prime} {\chi}^{\prime}-4 e^{{\mathcal{M}}-{\mathcal{N}}} b {\mathrm{S}}_{1}^2 Q_0 \mathcal{R}^{\prime} {\chi}^{\prime}-4 e^{2 {\mathcal{M}}-2 {\mathcal{N}}} \mathcal{R} {\mathrm{S}}_{2} Q_0^2 {\chi}^{\prime 2} \phantom{\Big)} \nonumber \\
&\ \phantom{{\mathrm{S}}_{1}   \Big(}
+2 e^{2 {\mathcal{M}}-2 {\mathcal{N}}} \KB{} \mathcal{R} {\mathrm{S}}_{2} Q_0^2 {\chi}^{\prime 2}
+ 2 e^{2 {\mathcal{M}}-2 {\mathcal{N}}} \mathcal{R} {\mathrm{S}}_{2} \mathcal{F}_{20} Q_0^2 {\chi}^{\prime 2}
+ 4 e^{{\mathcal{M}}-{\mathcal{N}}} \KB{} \mathcal{R} {\mathrm{C}}_{2} Q_0 {\chi}^{\prime} \psi^{\prime}
- 4 \KB{} \mathcal{R} {\mathrm{S}}_{1}^2 {\alpha}^{\prime \prime}
\phantom{\Big)} \nonumber \\
&\ \phantom{{\mathrm{S}}_{1}   \Big(}-4 e^{\mathcal{M}} \mathcal{R} {\mathrm{C}}_{1} \mathcal{F}_{20} Q_0 \psi^{\prime}+2 b \mathcal{R} {\mathrm{C}}_{1} {\mathrm{S}}_{1} {\mathcal{M}}^{\prime} \psi^{\prime}+2 b \mathcal{R} {\mathrm{C}}_{1} {\mathrm{S}}_{1} {\mathcal{N}}^{\prime} \psi^{\prime}-4 b {\mathrm{C}}_{1} {\mathrm{S}}_{1} \mathcal{R}^{\prime} \psi^{\prime}-8 e^{{\mathcal{M}}-{\mathcal{N}}} \mathcal{R} {\mathrm{C}}_{2} Q_0 {\chi}^{\prime} \psi^{\prime} \phantom{\Big)} \nonumber \\
&\ \phantom{{\mathrm{S}}_{1}   \Big(}+4 e^{{\mathcal{M}}-{\mathcal{N}}} \mathcal{R} {\mathrm{C}}_{2} \mathcal{F}_{20} Q_0 {\chi}^{\prime} \psi^{\prime}-8 \mathcal{R} {\mathrm{C}}_{1} {\mathrm{S}}_{1} \psi^{\prime 2}+4 \KB{} \mathcal{R} {\mathrm{C}}_{1} {\mathrm{S}}_{1} \psi^{\prime 2}+4 \mathcal{R} {\mathrm{C}}_{1} {\mathrm{S}}_{1} \mathcal{F}_{20} \psi^{\prime 2}-4 \KB{} \mathcal{R} {\mathrm{C}}_{1} {\mathrm{S}}_{1} {\mathcal{N}}^{\prime \prime} \phantom{\Big)} \nonumber \\
&\ \phantom{{\mathrm{S}}_{1}   \Big(}-e^{2 {\mathcal{M}}-2 {\mathcal{N}}} b \mathcal{R} {\mathrm{S}}_{2} Q_0 {\chi}^{\prime \prime}-2 b \mathcal{R} {\mathrm{C}}_{1} {\mathrm{S}}_{1} \psi^{\prime \prime}-2 e^{2 {\mathcal{M}}-2 {\mathcal{N}}} \mathcal{R} {\mathrm{S}}_{2} Q_0^2 {\chi}^{\prime 2} \mathcal{V}^{\prime}\nonumber \\
&\ \phantom{{\mathrm{S}}_{1}   \Big(}-4 e^{{\mathcal{M}}-{\mathcal{N}}} \mathcal{R} {\mathrm{C}}_{2} Q_0 {\chi}^{\prime} \psi^{\prime} \mathcal{V}^{\prime}-4 \mathcal{R} {\mathrm{C}}_{1} {\mathrm{S}}_{1} \psi^{\prime 2} \mathcal{V}^{\prime}\Big)=0, \label{eqA0}
\\
	\frac{\delta \tensor*{S}{_{\text{\AE{}ST}}^{\text{SZ}}}}{\delta\tensor{A}{_r}}\propto
	&\  {\mathrm{C}}_{1}  \Big(4 \KB{} \mathcal{R} {\mathrm{C}}_{1} {\mathrm{S}}_{1} {\mathcal{M}}^{\prime} {\mathcal{N}}^{\prime}-8 \KB{} {\mathrm{C}}_{1} {\mathrm{S}}_{1} {\mathcal{N}}^{\prime} \mathcal{R}^{\prime}+4 \KB{} \mathcal{R} {\mathrm{S}}_{1}^2 {\mathcal{M}}^{\prime} {\alpha}^{\prime}-4 \KB{} \mathcal{R} {\mathrm{S}}_{1}^2 {\mathcal{N}}^{\prime} {\alpha}^{\prime}-8 \KB{} {\mathrm{S}}_{1}^2 \mathcal{R}^{\prime} {\alpha}^{\prime}-4 \KB{} \mathcal{R} {\mathrm{C}}_{1} {\mathrm{S}}_{1} {\alpha}^{\prime 2} \phantom{\Big)} \nonumber \\
&\ \phantom{{\mathrm{C}}_{1}   \Big(}- 4 e^{2 {\mathcal{M}}-{\mathcal{N}}} \mathcal{R} {\mathrm{S}}_{1} \mathcal{F}_{20} Q_0^2 {\chi}^{\prime}+2 e^{{\mathcal{M}}-{\mathcal{N}}} b \mathcal{R} {\mathrm{C}}_{2} Q_0 {\mathcal{N}}^{\prime} {\chi}^{\prime}-4 e^{{\mathcal{M}}-{\mathcal{N}}} b {\mathrm{S}}_{1}^2 Q_0 \mathcal{R}^{\prime} {\chi}^{\prime}-4 e^{2 {\mathcal{M}}-2 {\mathcal{N}}} \mathcal{R} {\mathrm{S}}_{2} Q_0^2 {\chi}^{\prime 2} \phantom{\Big)} \nonumber \\
&\ \phantom{{\mathrm{C}}_{1}   \Big(}
+2 e^{2 {\mathcal{M}}-2 {\mathcal{N}}} \KB{} \mathcal{R} {\mathrm{S}}_{2} Q_0^2 {\chi}^{\prime 2}
+ 2 e^{2 {\mathcal{M}}-2 {\mathcal{N}}} \mathcal{R} {\mathrm{S}}_{2} \mathcal{F}_{20} Q_0^2 {\chi}^{\prime 2}
+ 4 e^{{\mathcal{M}}-{\mathcal{N}}} \KB{} \mathcal{R} {\mathrm{C}}_{2} Q_0 {\chi}^{\prime} \psi^{\prime}
- 4 \KB{} \mathcal{R} {\mathrm{S}}_{1}^2 {\alpha}^{\prime \prime}
\phantom{\Big)} \nonumber \\
&\ \phantom{{\mathrm{C}}_{1}   \Big(}-4 e^{\mathcal{M}} \mathcal{R} {\mathrm{C}}_{1} \mathcal{F}_{20} Q_0 \psi^{\prime}+2 b \mathcal{R} {\mathrm{C}}_{1} {\mathrm{S}}_{1} {\mathcal{M}}^{\prime} \psi^{\prime}+2 b \mathcal{R} {\mathrm{C}}_{1} {\mathrm{S}}_{1} {\mathcal{N}}^{\prime} \psi^{\prime}-4 b {\mathrm{C}}_{1} {\mathrm{S}}_{1} \mathcal{R}^{\prime} \psi^{\prime}-8 e^{{\mathcal{M}}-{\mathcal{N}}} \mathcal{R} {\mathrm{C}}_{2} Q_0 {\chi}^{\prime} \psi^{\prime} \phantom{\Big)} \nonumber \\
&\ \phantom{{\mathrm{C}}_{1}   \Big(}+4 e^{{\mathcal{M}}-{\mathcal{N}}} \mathcal{R} {\mathrm{C}}_{2} \mathcal{F}_{20} Q_0 {\chi}^{\prime} \psi^{\prime}-8 \mathcal{R} {\mathrm{C}}_{1} {\mathrm{S}}_{1} \psi^{\prime 2}+4 \KB{} \mathcal{R} {\mathrm{C}}_{1} {\mathrm{S}}_{1} \psi^{\prime 2}+4 \mathcal{R} {\mathrm{C}}_{1} {\mathrm{S}}_{1} \mathcal{F}_{20} \psi^{\prime 2}-4 \KB{} \mathcal{R} {\mathrm{C}}_{1} {\mathrm{S}}_{1} {\mathcal{N}}^{\prime \prime} \phantom{\Big)} \nonumber \\
&\ \phantom{{\mathrm{C}}_{1}   \Big(}-e^{2 {\mathcal{M}}-2 {\mathcal{N}}} b \mathcal{R} {\mathrm{S}}_{2} Q_0 {\chi}^{\prime \prime}-2 b \mathcal{R} {\mathrm{C}}_{1} {\mathrm{S}}_{1} \psi^{\prime \prime}-2 e^{2 {\mathcal{M}}-2 {\mathcal{N}}} \mathcal{R} {\mathrm{S}}_{2} Q_0^2 {\chi}^{\prime 2} \mathcal{V}^{\prime}\nonumber\\
&\ \phantom{{\mathrm{C}}_{1}   \Big(}-4 e^{{\mathcal{M}}-{\mathcal{N}}} \mathcal{R} {\mathrm{C}}_{2} Q_0 {\chi}^{\prime} \psi^{\prime} \mathcal{V}^{\prime}-4 \mathcal{R} {\mathrm{C}}_{1} {\mathrm{S}}_{1} \psi^{\prime 2} \mathcal{V}^{\prime}\Big)=0, \label{eqA1}
\\
\frac{\delta \tensor*{S}{_{\text{\AE{}ST}}^{\text{SZ}}}}{\delta\varphi}\propto
	&\  4 e^{3 {\mathcal{M}}} \mathcal{R} {\mathrm{S}}_{1} \mathcal{F}_{20} Q_0 {\mathcal{N}}^{\prime}+2 e^{2 {\mathcal{M}}} b \mathcal{R} {\mathrm{C}}_{1}^2 {\mathcal{M}}^{\prime} {\mathcal{N}}^{\prime}-2 e^{2 {\mathcal{M}}} b \mathcal{R} {\mathrm{C}}_{1}^2 {\mathcal{N}}^{\prime 2}+8 e^{3 {\mathcal{M}}} {\mathrm{S}}_{1} \mathcal{F}_{20} Q_0 \mathcal{R}^{\prime}-4 e^{2 {\mathcal{M}}} b {\mathrm{C}}_{1}^2 {\mathcal{N}}^{\prime} \mathcal{R}^{\prime}+4 e^{3 {\mathcal{M}}} \mathcal{R} {\mathrm{C}}_{1} \mathcal{F}_{20} Q_0 {\alpha}^{\prime}  \phantom{\Big)} \nonumber \\
&\ +2 e^{2 {\mathcal{M}}} b \mathcal{R} {\mathrm{C}}_{1} {\mathrm{S}}_{1} {\mathcal{M}}^{\prime} {\alpha}^{\prime}-3 e^{2 {\mathcal{M}}} b \mathcal{R} {\mathrm{S}}_{2} {\mathcal{N}}^{\prime} {\alpha}^{\prime}-2 e^{2 {\mathcal{M}}} b {\mathrm{S}}_{2} \mathcal{R}^{\prime} {\alpha}^{\prime}-2 e^{2 {\mathcal{M}}} b \mathcal{R} {\mathrm{C}}_{2} {\alpha}^{\prime 2}-4 e^{3 {\mathcal{M}}-{\mathcal{N}}} {\mathrm{S}}_{2} \mathcal{F}_{20} Q_0 \mathcal{R}^{\prime} {\chi}^{\prime}+ 8 e^{2 {\mathcal{M}}} {\mathrm{C}}_{1}^2 \mathcal{R}^{\prime} \psi^{\prime} \mathcal{V}^{\prime}  \phantom{\Big)} \nonumber \\
&\  -4 e^{3 {\mathcal{M}}-{\mathcal{N}}} \mathcal{R} {\mathrm{C}}_{2} \mathcal{F}_{20} Q_0 {\alpha}^{\prime} {\chi}^{\prime}+ 4 e^{2 {\mathcal{M}}} \mathcal{R} {\mathrm{S}}_{1}^2 \mathcal{F}_{20} {\mathcal{M}}^{\prime} \psi^{\prime}-4 e^{2 {\mathcal{M}}} \mathcal{R} {\mathrm{S}}_{1}^2 \mathcal{F}_{20} {\mathcal{N}}^{\prime} \psi^{\prime}-8 e^{2 {\mathcal{M}}} {\mathrm{S}}_{1}^2 \mathcal{F}_{20} \mathcal{R}^{\prime} \psi^{\prime}-4 e^{2 {\mathcal{M}}} \mathcal{R} {\mathrm{S}}_{2} \mathcal{F}_{20} \alpha^{\prime} \psi^{\prime}  \phantom{\Big)} \nonumber \\
&\ - e^{2 {\mathcal{M}}} b \mathcal{R} {\mathcal{N}}^{\prime \prime}-e^{2 {\mathcal{M}}} b \mathcal{R} {\mathrm{C}}_{2} {\mathcal{N}}^{\prime \prime}-e^{2 {\mathcal{M}}} b \mathcal{R} {\mathrm{S}}_{2} {\alpha}^{\prime \prime}-4 e^{4 {\mathcal{M}}-2 {\mathcal{N}}} \mathcal{R} {\mathrm{C}}_{1}^2 \mathcal{F}_{20} Q_0 {\chi}^{\prime \prime}-4 e^{2 {\mathcal{M}}} \mathcal{R} {\mathrm{S}}_{1}^2 \mathcal{F}_{20} \psi^{\prime \prime}+4 e^{2 {\mathcal{M}}} \mathcal{R} {\mathrm{S}}_{2} {\alpha}^{\prime} \psi^{\prime} \mathcal{V}^{\prime}  \phantom{\Big)} \nonumber \\
&\ + 4 e^{3 {\mathcal{M}}-{\mathcal{N}}} {\mathrm{S}}_{2} Q_0 \mathcal{R}^{\prime} \lambda^{\prime} \mathcal{V}^{\prime}+4 e^{3 {\mathcal{M}}-{\mathcal{N}}} \mathcal{R} {\mathrm{C}}_{2} Q_0 \alpha^{\prime} \lambda^{\prime} \mathcal{V}^{\prime}-4 e^{2 {\mathcal{M}}} \mathcal{R} {\mathrm{C}}_{1}^2 {\mathcal{M}}^{\prime} \psi^{\prime} \mathcal{V}^{\prime}+4 e^{2 {\mathcal{M}}} \mathcal{R} {\mathrm{C}}_{1}^2 {\mathcal{N}}^{\prime} \psi^{\prime} \mathcal{V}^{\prime}+4 e^{2 {\mathcal{M}}} \mathcal{R} {\mathrm{C}}_{1}^2 \psi^{\prime \prime} \mathcal{V}^{\prime}  \phantom{\Big)} \nonumber \\
&\ +4 e^{4 {\mathcal{M}}-2 {\mathcal{N}}} \mathcal{R} {\mathrm{S}}_{1}^2 Q_0 {\chi}^{\prime \prime} \mathcal{V}^{\prime}- 8 e^{3 {\mathcal{M}}-3 {\mathcal{N}}} \mathcal{R} {\mathrm{C}}_{1} {\mathrm{S}}_{1}^3 Q_0^3 {\mathcal{N}}^{\prime} {\chi}^{\prime 3} \mathcal{V}^{\prime \prime}+8 e^{3 {\mathcal{M}}-3 {\mathcal{N}}} \mathcal{R} {\mathrm{C}}_{1}^2 {\mathrm{S}}_{1}^2 Q_0^3 {\alpha}^{\prime} {\chi}^{\prime 3} \mathcal{V}^{\prime \prime}-2 e^{2 {\mathcal{M}}-2 {\mathcal{N}}} \mathcal{R} {\mathrm{S}}_{2}^2 Q_0^2 {\mathcal{M}}^{\prime} {\chi}^{\prime 2} \psi^{\prime} \mathcal{V}^{\prime \prime}  \phantom{\Big)} \nonumber \\
&\ - 4 e^{2 {\mathcal{M}}-2 {\mathcal{N}}} \mathcal{R} {\mathrm{S}}_{2}^2 Q_0^2 {\mathcal{N}}^{\prime} {\chi}^{\prime 2} \psi^{\prime} \mathcal{V}^{\prime \prime}+2 e^{2 {\mathcal{M}}-2 {\mathcal{N}}} \mathcal{R} {\mathrm{S}}_{2} Q_0^2 \alpha^{\prime} {\chi}^{\prime 2} \psi^{\prime} \mathcal{V}^{\prime \prime}+6 e^{2 {\mathcal{M}}-2 {\mathcal{N}}} \mathcal{R} {\mathrm{C}}_{2} {\mathrm{S}}_{2} Q_0^2 {\alpha}^{\prime} {\chi}^{\prime 2} \psi^{\prime} \mathcal{V}^{\prime \prime}-8 \mathcal{R} {\mathrm{C}}_{1}^4 {\mathcal{M}}^{\prime} \psi^{\prime 3} \mathcal{V}^{\prime \prime}  \phantom{\Big)} \nonumber \\
&\ - 8 e^{{\mathcal{M}}-{\mathcal{N}}} \mathcal{R} {\mathrm{C}}_{1}^2 {\mathrm{S}}_{2} Q_0 {\mathcal{M}}^{\prime} {\chi}^{\prime} \psi^{\prime 2} \mathcal{V}^{\prime \prime}-4 e^{{\mathcal{M}}-{\mathcal{N}}} \mathcal{R} {\mathrm{C}}_{1}^2 {\mathrm{S}}_{2} Q_0 {\mathcal{N}}^{\prime} {\chi}^{\prime} \psi^{\prime 2} \mathcal{V}^{\prime \prime}-4 e^{{\mathcal{M}}-{\mathcal{N}}} \mathcal{R} {\mathrm{C}}_{1}^2 Q_0 {\alpha}^{\prime} {\chi}^{\prime} \psi^{\prime 2} \mathcal{V}^{\prime \prime}+8 \mathcal{R} {\mathrm{C}}_{1}^3 {\mathrm{S}}_{1} {\alpha}^{\prime} \psi^{\prime 3} \mathcal{V}^{\prime \prime}  \phantom{\Big)} \nonumber \\
&\  + 12 e^{{\mathcal{M}}-{\mathcal{N}}} \mathcal{R} {\mathrm{C}}_{1}^2 {\mathrm{C}}_{2} Q_0 {\alpha}^{\prime} {\chi}^{\prime} \psi^{\prime 2} \mathcal{V}^{\prime \prime}+ 2 e^{2 {\mathcal{M}}-2 {\mathcal{N}}} \mathcal{R} {\mathrm{S}}_{2}^2 Q_0^2 {\chi}^{\prime 2} \psi^{\prime \prime} \mathcal{V}^{\prime \prime}+8 e^{{\mathcal{M}}-{\mathcal{N}}} \mathcal{R} {\mathrm{C}}_{1}^2 {\mathrm{S}}_{2} Q_0 {\chi}^{\prime} \psi^{\prime} \psi^{\prime \prime} \mathcal{V}^{\prime \prime}+8 \mathcal{R} {\mathrm{C}}_{1}^4 \psi^2 \psi^{\prime \prime} \mathcal{V}^{\prime \prime}  \phantom{\Big)} \nonumber \\
&\  + 8 e^{4 {\mathcal{M}}-4 {\mathcal{N}}} \mathcal{R} {\mathrm{S}}_{1}^4 Q_0^3 {\chi}^{\prime 2} {\chi}^{\prime \prime} \mathcal{V}^{\prime \prime}+16 e^{3 {\mathcal{M}}-3 {\mathcal{N}}} \mathcal{R} {\mathrm{C}}_{1} {\mathrm{S}}_{1}^3 Q_0^2 {\chi}^{\prime} \psi^{\prime} {\chi}^{\prime \prime} \mathcal{V}^{\prime \prime}+2 e^{2 {\mathcal{M}}-2 {\mathcal{N}}} \mathcal{R} {\mathrm{S}}_{2}^2 Q_0 \psi^{\prime 2} {\chi}^{\prime \prime} \mathcal{V}^{\prime \prime}=0, \label{eqP}
\\
	\frac{\delta \tensor*{S}{_{\text{\AE{}ST}}^{\text{SZ}}}}{\delta\tensor{g}{^{tt}}}\propto
&\ 
4 e^{2 {\mathcal{M}}}
+4 e^{2 {\mathcal{M}}} \mathcal{R}^2 \kappa \mathcal{F}_{20} Q_0^2
-4 e^{2 {\mathcal{M}}} \mathcal{R} \kappa^2 \mathcal{V}
+8 \KB{} \mathcal{R}^2 \kappa {\mathrm{C}}_{1}^4 {\mathcal{M}}^{\prime} {\mathcal{N}}^{\prime}
-4 \KB{} \mathcal{R}^2 \kappa {\mathrm{C}}_{1}^2 {\mathcal{N}}^{\prime 2}
+8 \mathcal{R} \mathcal{M}^{\prime} \mathcal{R}^{\prime}
-4 \mathcal{R}^{\prime 2}
-8 \mathcal{R} \mathcal{R}^{\prime \prime}
\phantom{\Big)} \nonumber \\
&\ 
-16 \KB{} \mathcal{R} \kappa {\mathrm{C}}_{1}^4 {\mathcal{N}}^{\prime} \mathcal{R}^{\prime}
+8 \KB{} \mathcal{R}^2 \kappa {\mathrm{C}}_{1}^3 {\mathrm{S}}_{1} {\mathcal{M}}^{\prime} \alpha^{\prime}
-8 \KB{} \mathcal{R}^2 \kappa {\mathrm{C}}_{1}^3 {\mathrm{S}}_{1} {\mathcal{N}}^{\prime} {\alpha}^{\prime}
-4 \KB{} \mathcal{R}^2 \kappa {\mathrm{S}}_{2} {\mathcal{N}}^{\prime} \alpha^{\prime}
-16 \KB{} \mathcal{R} \kappa {\mathrm{C}}_{1}^3 {\mathrm{S}}_{1} \mathcal{R}^{\prime} \alpha^{\prime}
\phantom{\Big)} \nonumber \\
&\ 
-8 \KB{} \mathcal{R}^2 \kappa {\mathrm{C}}_{1}^4 {\alpha}^{\prime 2}
-4 \KB{} \mathcal{R}^2 \kappa {\mathrm{S}}_{1}^2 \alpha^{\prime 2}
+8 e^{2 {\mathcal{M}}-{\mathcal{N}}} \mathcal{R}^2 \kappa {\mathrm{C}}_{1} \mathcal{F}_{20} Q_0^2 {\chi}^{\prime}
- 8 e^{2 {\mathcal{M}}-{\mathcal{N}}} \mathcal{R}^2 \kappa {\mathrm{C}}_{1}^3 \mathcal{F}_{20} Q_0^2 {\chi}^{\prime}
-8 b \mathcal{R} \kappa {\mathrm{C}}_{1}^4 \mathcal{R}^{\prime} \psi^{\prime}
\phantom{\Big)} \nonumber \\
&\ 
+8 e^{{\mathcal{M}}-{\mathcal{N}}} b \mathcal{R}^2 \kappa {\mathrm{C}}_{1}^3 {\mathrm{S}}_{1} Q_0 {\mathcal{N}}^{\prime} {{\chi}}^{\prime}
-2 e^{{\mathcal{M}}-{\mathcal{N}}} b \mathcal{R}^2 \kappa {\mathrm{S}}_{2} Q_0 {\mathcal{N}}^{\prime} \lambda^{\prime}
-4 e^{{\mathcal{M}}-{\mathcal{N}}} b \mathcal{R}^2 \kappa {\mathrm{S}}_{1}^2 Q_0 {\alpha}^{\prime} {\chi}^{\prime}
+2 b \mathcal{R}^2 \kappa {\mathrm{C}}_{1}^2 {\mathrm{C}}_{2} {\mathcal{N}}^{\prime} \psi^{\prime}
\phantom{\Big)} \nonumber \\
&\ 
-8 e^{{\mathcal{M}}-{\mathcal{N}}} b \mathcal{R} \kappa {\mathrm{C}}_{1}^3 {\mathrm{S}}_{1} Q_0 \mathcal{R}^{\prime} {\chi}^{\prime}
-4 e^{{\mathcal{M}}-{\mathcal{N}}} b \mathcal{R}^2 \kappa {\mathrm{C}}_{1}^2 Q_0 {\alpha}^{\prime} {\chi}^{\prime}
+4 b \mathcal{R}^2 \kappa {\mathrm{C}}_{1}^4 {\mathcal{M}}^{\prime} \psi^{\prime}
-2 b \mathcal{R}^2 \kappa {\mathrm{C}}_{1}^2 {\mathcal{N}}^{\prime} \psi^{\prime}
-2 b \mathcal{R}^2 \kappa {\mathrm{S}}_{2} \alpha^{\prime} \psi^{\prime}
\phantom{\Big)} \nonumber \\
&\ 
-16 e^{2 {\mathcal{M}}-2 {\mathcal{N}}} \mathcal{R}^2 \kappa {\mathrm{C}}_{1}^4 Q_0^2 {\chi}^{\prime 2}
+8 e^{2 {\mathcal{M}}-2 {\mathcal{N}}} \KB{} \mathcal{R}^2 \kappa {\mathrm{C}}_{1}^4 Q_0^2 {\chi}^{\prime 2}
-12 e^{2 {\mathcal{M}}-2 {\mathcal{N}}} \mathcal{R}^2 \kappa {\mathrm{C}}_{1}^2 \mathcal{F}_{20} Q_0^2 {\chi}^{\prime 2}  
-8 e^{\mathcal{M}} \mathcal{R}^2 \kappa {\mathrm{S}}_{1} \mathcal{F}_{20} Q_0 \psi^{\prime}
\phantom{\Big)} \nonumber \\
&\ 
+ 8 e^{2 {\mathcal{M}}-2 {\mathcal{N}}} \mathcal{R}^2 \kappa {\mathrm{C}}_{1}^4 \mathcal{F}_{20} Q_0^2 {\chi}^{\prime 2}
-8 e^{\mathcal{M}} \mathcal{R}^2 \kappa {\mathrm{C}}_{1}^2 {\mathrm{S}}_{1} \mathcal{F}_{20} Q_0 \psi^{\prime}
-32 e^{{\mathcal{M}}-{\mathcal{N}}} \mathcal{R}^2 \kappa {\mathrm{C}}_{1}^3 {\mathrm{S}}_{1} Q_0 {\chi}^{\prime} \psi^{\prime}
-4 e^{{\mathcal{M}}-{\mathcal{N}}} \mathcal{R}^2 \kappa {\mathrm{S}}_{2} \mathcal{F}_{20} Q_0 {\chi}^{\prime} \psi^{\prime}
\phantom{\Big)} \nonumber \\
&\ 
+16 e^{{\mathcal{M}}-{\mathcal{N}}} \mathcal{R}^2 \kappa {\mathrm{C}}_{1}^3 {\mathrm{S}}_{1} \mathcal{F}_{20} Q_0 {\chi}^{\prime} \psi^{\prime}
+16 e^{{\mathcal{M}}-{\mathcal{N}}} \KB{} \mathcal{R}^2 \kappa {\mathrm{C}}_{1}^3 {\mathrm{S}}_{1} Q_0 {\chi}^{\prime} \psi^{\prime}
-4 \mathcal{R}^2 \kappa {\mathrm{S}}_{2}^2 \psi^{\prime 2}
+2 \KB{} \mathcal{R}^2 \kappa {\mathrm{S}}_{2}^2 \psi^{\prime 2}
+2 \mathcal{R}^2 \kappa {\mathrm{S}}_{2}^2 \mathcal{F}_{20} \psi^{\prime 2}
\phantom{\Big)} \nonumber \\
&\ 
+ 4 \mathcal{R}^2 \kappa {\mathrm{S}}_{1}^2 \mathcal{F}_{20} \psi^{\prime 2}
-8 \KB{} \mathcal{R}^2 \kappa {\mathrm{C}}_{1}^4 {\mathcal{N}}^{\prime \prime}
-8 \KB{} \mathcal{R}^2 \kappa {\mathrm{C}}_{1}^3 {\mathrm{S}}_{1} {\alpha}^{\prime \prime}
-16 e^{{\mathcal{M}}-{\mathcal{N}}} \mathcal{R}^2 \kappa {\mathrm{C}}_{1}^3 {\mathrm{S}}_{1} Q_0 {\chi}^{\prime} \psi^{\prime} \mathcal{V}^{\prime}
+8 e^{{\mathcal{M}}-{\mathcal{N}}} \mathcal{R}^2 \kappa {\mathrm{S}}_{2} Q_0 {\chi}^{\prime} \psi^{\prime} \mathcal{V}^{\prime}
\phantom{\Big)} \nonumber \\
&\ 
+4 e^{2 {\mathcal{M}}-2 {\mathcal{N}}} b \mathcal{R}^2 \kappa {\mathrm{C}}_{1}^2 Q_0 \lambda^{\prime \prime}
-4 e^{2 {\mathcal{M}}-2 {\mathcal{N}}} \quad \mathcal{R}^2 \kappa {\mathrm{C}}_{1}^4 Q_0 {\chi}^{\prime \prime}
-4 b \mathcal{R}^2 \kappa {\mathrm{C}}_{1}^2 \psi^{\prime \prime}
-b \mathcal{R}^2 \kappa {\mathrm{S}}_{2}^2 \psi^{\prime \prime}
-8 e^{2 {\mathcal{M}}-2 {\mathcal{N}}} \mathcal{R}^2 \kappa {\mathrm{C}}_{1}^4 Q_0^2 \lambda^{\prime 2} \mathcal{V}^{\prime}
\phantom{\Big)} \nonumber \\
&\ 
-8 e^{2 {\mathcal{M}}-2 {\mathcal{N}}} \mathcal{R}^2 \kappa Q_0^2 {\chi}^{\prime 2} \mathcal{V}^{\prime}
+16 e^{2 {\mathcal{M}}-2 {\mathcal{N}}} \mathcal{R}^2 \kappa {\mathrm{C}}_{1}^2 Q_0^2 \lambda^{\prime 2} \mathcal{V}^{\prime}
-8 \mathcal{R}^2 \kappa {\mathrm{C}}_{1}^2 {\mathrm{S}}_{1}^2 \psi^{\prime 2} \mathcal{V}^{\prime}=0, \label{eqG00}
\\
\frac{\delta \tensor*{S}{_{\text{\AE{}ST}}^{\text{SZ}}}}{\delta\tensor{g}{^{tr}}}\propto
&\ 
-16 \KB{} \mathcal{R} {\mathrm{C}}_{1}^3 {\mathrm{S}}_{1} {\mathcal{M}}^{\prime} {\mathcal{N}}^{\prime}
+32 \KB{} {\mathrm{C}}_{1}^3 {\mathrm{S}}_{1} {\mathcal{N}}^{\prime} \mathcal{R}^{\prime}
-16 \KB{} \mathcal{R} {\mathrm{C}}_{1}^2 {\mathrm{S}}_{1}^2 {\mathcal{M}}^{\prime} {\alpha}^{\prime}
+16 \KB{} \mathcal{R} {\mathrm{C}}_{1}^2 {\mathrm{S}}_{1}^2 {\mathcal{N}}^{\prime} {\alpha}^{\prime}
+32 \KB{} {\mathrm{C}}_{1}^2 {\mathrm{S}}_{1}^2 \mathcal{R}^{\prime} {\alpha}^{\prime}
\phantom{\Big)} \nonumber \\
&\ 
+4 \KB{} \mathcal{R} {\mathrm{S}}_{2} {\alpha}^{\prime 2}
+2 \KB{} \mathcal{R} {\mathrm{S}}_{4} {\alpha}^{\prime 2}
-12 e^{2 {\mathcal{M}}-{\mathcal{N}}} \mathcal{R} {\mathrm{S}}_{1} \mathcal{F}_{20} Q_0^2 {\chi}^{\prime}
+4 e^{{\mathcal{M}}-{\mathcal{N}}} b \mathcal{R} {\mathrm{S}}_{2} Q_0 {\alpha}^{\prime} {\chi}^{\prime}
+8 e^{2 {\mathcal{M}}-2 {\mathcal{N}}} \mathcal{R} {\mathrm{S}}_{2} Q_0^2 {\chi}^{\prime 2}
\phantom{\Big)} \nonumber \\
&\ 
+4 e^{2 {\mathcal{M}}-{\mathcal{N}}} \mathcal{R} {\mathrm{S}}_{3} \mathcal{F}_{20} Q_0^2 {\chi}^{\prime}
-16 e^{{\mathcal{M}}-{\mathcal{N}}} b \mathcal{R} {\mathrm{C}}_{1}^2 {\mathrm{S}}_{1}^2 Q_0 {\mathcal{N}}^{\prime} {\chi}^{\prime}
+16 e^{{\mathcal{M}}-{\mathcal{N}}} b {\mathrm{C}}_{1}^2 {\mathrm{S}}_{1}^2 Q_0 \mathcal{R}^{\prime} {\chi}^{\prime}
-4 e^{2 {\mathcal{M}}-2 {\mathcal{N}}} \KB{} \mathcal{R} {\mathrm{S}}_{2} \mathcal{Q}_0^2 {\chi}^{\prime 2}
\phantom{\Big)} \nonumber \\
&\ 
+4 e^{2 {\mathcal{M}}-2 {\mathcal{N}}} \mathcal{R} {\mathrm{S}}_{4} Q_0^2 {\chi}^{\prime 2}
-2 e^{2 {\mathcal{M}}-2 {\mathcal{N}}} \KB{} \mathcal{R} {\mathrm{S}}_{4} Q_0^2 {\chi}^{\prime 2}
+4 e^{2 {\mathcal{M}}-2 {\mathcal{N}}} \mathcal{R} {\mathrm{S}}_{2} \mathcal{F}_{20} Q_0^2 {\chi}^{\prime 2}
-2 e^{2 {\mathcal{M}}-2 {\mathcal{N}}} \mathcal{R} {\mathrm{S}}_{4} \mathcal{F}_{20} Q_0^2 {\chi}^{\prime 2}
\phantom{\Big)} \nonumber \\
&\ 
+12 e^{{\mathcal{M}}} \mathcal{R} {\mathrm{C}}_{1} \mathcal{F}_{20} Q_0 \psi^{\prime}
+4 e^{{\mathcal{M}}} \mathcal{R} {\mathrm{C}}_{3} \mathcal{F}_{20} Q_0 \psi^{\prime}
-8 b \mathcal{R} {\mathrm{C}}_{1}^3 {\mathrm{S}}_{1} {\mathcal{N}}^{\prime} \psi^{\prime}
+16 b {\mathrm{C}}_{1}^3 {\mathrm{S}}_{1} \mathcal{R}^{\prime} \psi^{\prime}
-8 e^{{\mathcal{M}}-{\mathcal{N}}} \mathcal{R} Q_0 {\chi}^{\prime} \psi^{\prime}
\phantom{\Big)} \nonumber \\
&\ 
-8 b \mathcal{R} {\mathrm{C}}_{1}^3 {\mathrm{S}}_{1} {\mathcal{M}}^{\prime} \psi^{\prime}
-2 \KB{} \mathcal{R} {\mathrm{S}}_{4} \psi^{\prime 2}
-4 \mathcal{R} {\mathrm{S}}_{2} \mathcal{F}_{20} \psi^{\prime 2}
-2 \mathcal{R} {\mathrm{S}}_{4} \mathcal{F}_{20} \psi^{\prime 2}
+4 \KB{} \mathcal{R} {\mathrm{S}}_{2} {\mathcal{N}}^{\prime \prime}
+2 \KB{} \mathcal{R} {\mathrm{S}}_{4} {\mathcal{N}}^{\prime \prime}
\phantom{\Big)} \nonumber \\
&\ 
+4 e^{{\mathcal{M}}-{\mathcal{N}}} \KB{} \mathcal{R} Q_0 {\chi}^{\prime} \psi^{\prime}
+8 e^{{\mathcal{M}}-{\mathcal{N}}} \mathcal{R} {\mathrm{C}}_{4} Q_0 {\chi}^{\prime} \psi^{\prime}
-4 e^{{\mathcal{M}}-{\mathcal{N}}} \KB{} \mathcal{R} {\mathrm{C}}_{4} Q_0 {\chi}^{\prime} \psi^{\prime}
-12 e^{{\mathcal{M}}-{\mathcal{N}}} \mathcal{R} \mathcal{F}_{20} \mathcal{Q}_0 {\chi}^{\prime} \psi^{\prime}
\phantom{\Big)} \nonumber \\
&\ 
-4 e^{{\mathcal{M}}-{\mathcal{N}}} \mathcal{R} {\mathrm{C}}_{4} \mathcal{F}_{20} Q_0 {\chi}^{\prime} \psi^{\prime}
-8 \mathcal{R} {\mathrm{S}}_{2} \psi^2
+4 \KB{} \mathcal{R} {\mathrm{S}}_{2} \psi^{\prime 2}
+2 e^{2 {\mathcal{M}}-2 {\mathcal{N}}} \mathcal{R} {\mathrm{S}}_{4} Q_0^2{{\chi}^{\prime 2}}^{\prime \prime} \mathcal{V}^{\prime}
+e^{2 {\mathcal{M}}-2 {\mathcal{N}}} b \mathcal{R} {\mathrm{S}}_{4} Q_0 {\chi}^{\prime \prime}
\phantom{\Big)} \nonumber \\
&\ 
-4 e^{{\mathcal{M}}-{\mathcal{N}}} \mathcal{R} Q_0 {\chi}^{\prime} \psi^{\prime} \mathcal{V}^{\prime}
+4 e^{{\mathcal{M}}-{\mathcal{N}}} \mathcal{R} {\mathrm{C}}_{4} Q_0 {\chi}^{\prime} \psi^{\prime} \mathcal{V}^{\prime}
+4 \mathcal{R} {\mathrm{S}}_{2} \psi^{\prime 2} \mathcal{V}^{\prime}
+2 \mathcal{R} {\mathrm{S}}_{4} \psi^2 \mathcal{V}^{\prime}
-4 e^{2 {\mathcal{M}}-2 {\mathcal{N}}} \mathcal{R} {\mathrm{S}}_{2} Q_0^2 {\chi}^{\prime 2} \mathcal{V}^{\prime}
\phantom{\Big)} \nonumber \\
&\ 
-2 \KB{} \mathcal{R} {\alpha}^{\prime \prime}
+2 \KB{} \mathcal{R} {\mathrm{C}}_{4} {\alpha}^{\prime \prime}
-2 e^{2 {\mathcal{M}}-2 {\mathcal{N}}} b \mathcal{R} {\mathrm{S}}_{2} Q_0 {\chi}^{\prime \prime}
+2 b \mathcal{R} {\mathrm{S}}_{2} \psi^{\prime \prime}
+b \mathcal{R} {\mathrm{S}}_{4} \psi^{\prime \prime}
+4 \mathcal{R} {\mathrm{S}}_{4} \psi^{\prime 2}=0, \label{eqG01}
\\
\frac{\delta \tensor*{S}{_{\text{\AE{}ST}}^{\text{SZ}}}}{\delta\tensor{g}{^{rr}}}\propto
&\ 
-4 e^{2 {\mathcal{M}}}
-4 e^{2 {\mathcal{M}}} \mathcal{R}^2 \kappa \mathcal{F}_{20} Q_0^2
+4 e^{2 {\mathcal{M}}} \mathcal{R}^2 \kappa \mathcal{V}
+8 \KB{} \mathcal{R}^2 \kappa {\mathrm{C}}_{1}^2 {\mathrm{S}}_{1}^2 {\mathcal{M}}^{\prime} {\mathcal{N}}^{\prime}
+4 \KB{} \mathcal{R}^2 \kappa {\mathrm{C}}_{1}^2 {\mathcal{N}}^{\prime 2} 
-4 \KB{} \mathcal{R} \kappa {\mathrm{S}}_{2}^2 {\mathcal{N}}^{\prime} \mathcal{R}^{\prime}
\phantom{\Big)} \nonumber \\
&\ 
+8 \mathcal{R} {\mathcal{N}}^{\prime} \mathcal{R}^{\prime}
+4 \mathcal{R}^{\prime 2}
+8 \KB{} \mathcal{R}^2 \kappa {\mathrm{C}}_{1} {\mathrm{S}}_{1}^3 {\mathcal{M}}^{\prime} {\alpha}^{\prime}
+6 \KB{} \mathcal{R}^2 \kappa {\mathrm{S}}_{2} {\mathcal{N}}^{\prime} {\alpha}^{\prime}
+4 \KB{} \mathcal{R}^2 \kappa {\mathrm{S}}_{1}^2 {\alpha}^{\prime 2}
-16 \KB{} \mathcal{R} \kappa {\mathrm{C}}_{1} {\mathrm{S}}_{1}^3 \mathcal{R}^{\prime} {\alpha}^{\prime}
\phantom{\Big)} \nonumber \\
&\ 
-\KB{} \mathcal{R}^2 \kappa {\mathrm{S}}_{4} {\mathcal{N}}^{\prime} {\alpha}^{\prime}
-2 \KB{} \mathcal{R}^2 \kappa {\mathrm{S}}_{2}^2 {\alpha}^{\prime 2}
+8 e^{2 {\mathcal{M}}-{\mathcal{N}}} \mathcal{R}^2 \kappa {\mathrm{C}}_{1} \mathcal{F}_{20} Q_0^2 {\chi}^{\prime}
-8 e^{2 {\mathcal{M}}-{\mathcal{N}}} \mathcal{R}^2 \kappa {\mathrm{C}}_{1} {\mathrm{S}}_{1}^2 \mathcal{F}_{20} Q_0^2 {\chi}^{\prime}
\phantom{\Big)} \nonumber \\
&\ 
+e^{{\mathcal{M}}-{\mathcal{N}}} b \mathcal{R}^2 \kappa {\mathrm{S}}_{4} Q_0 {\mathcal{N}}^{\prime} {\chi}^{\prime} 
-8 e^{{\mathcal{M}}-{\mathcal{N}}} b \mathcal{R} \kappa {\mathrm{C}}_{1} {\mathrm{S}}_{1}^3 Q_0 \mathcal{R}^{\prime} {\chi}^{\prime}
-4 e^{2 {\mathcal{M}}-2 {\mathcal{N}}} \mathcal{R}^2 \kappa {\mathrm{S}}_{2}^2 Q_0^2 {\chi}^{\prime 2}
-8 e^{\mathcal{M}} \mathcal{R}^2 \kappa {\mathrm{S}}_{1} \mathcal{F}_{20} Q_0 \psi^{\prime}
\phantom{\Big)} \nonumber \\
&\ 
+2 e^{2 {\mathcal{M}}-2 {\mathcal{N}}} \kappa \mathcal{R}^2 \kappa {\mathrm{S}}_{2}^2 Q_0^2 {\chi}^{\prime 2}
+4 b \mathcal{R}^2 \kappa {\mathrm{C}}_{1}^2 {\mathrm{S}}_{1}^2 {\mathcal{M}}^{\prime} \psi^{\prime}
+4 b \mathcal{R}^2 \kappa {\mathrm{C}}_{1}^4 {\mathcal{N}}^{\prime} \psi^{\prime}
-8 b \mathcal{R} \kappa {\mathrm{C}}_{1}^2 {\mathrm{S}}_{1}^2 \mathcal{R}^{\prime} \psi^{\prime}
+2 b \mathcal{R}^2 \kappa {\mathrm{S}}_{2} {\alpha}^{\prime} \psi^{\prime}
\phantom{\Big)} \nonumber \\
&\ 
-4 e^{2 {\mathcal{M}}-2 {\mathcal{N}}} \mathcal{R}^2 \kappa {\mathrm{C}}_{1}^2 \mathcal{F}_{20} Q_0^2 {\chi}^{\prime 2}
+2 e^{2 {\mathcal{M}}-2 {\mathcal{N}}} \mathcal{R}^2 \kappa {\mathrm{S}}_{2}^2 \mathcal{F}_{20} Q_0^2 {\chi}^{\prime 2}
+4 e^{{\mathcal{M}}-{\mathcal{N}}} \mathcal{R}^2 \kappa {\mathrm{S}}_{2} \mathcal{F}_{20} Q_0 {\chi}^{\prime} \psi^{\prime}
-16 \mathcal{R}^2 \kappa {\mathrm{S}}_{1}^4 \psi^{\prime 2}
\phantom{\Big)} \nonumber \\
&\ 
-8 e^{\mathcal{M}} \mathcal{R}^2 \kappa {\mathrm{S}}_{1}^3 \mathcal{F}_{20} Q_0 \psi^{\prime}
+8 \KB{} \mathcal{R}^2 \kappa {\mathrm{S}}_{1}^4 \psi^{\prime 2}
+12 \mathcal{R}^2 \kappa {\mathrm{S}}_{1}^2 \mathcal{F}_{20} \psi^{\prime 2}
+8 \mathcal{R}^2 \kappa {\mathrm{S}}_{1}^4 \mathcal{F}_{20} \psi^{\prime 2}
-2 \KB{} \mathcal{R}^2 \kappa {\mathrm{S}}_{2}^2 {\mathcal{N}}^{\prime \prime}
\phantom{\Big)} \nonumber \\
&\ 
-32 e^{{\mathcal{M}}-{\mathcal{N}}} \mathcal{R}^2 \kappa {\mathrm{C}}_{1} {\mathrm{S}}_{1}^3 Q_0 {\chi}^{\prime} \psi^{\prime}
+16 e^{{\mathcal{M}}-{\mathcal{N}}} \KB{} \mathcal{R}^2 \kappa {\mathrm{C}}_{1} {\mathrm{S}}_{1}^3 Q_0 {\chi}^{\prime} \psi^{\prime}
+16 e^{{\mathcal{M}}-{\mathcal{N}}} \mathcal{R}^2 \kappa {\mathrm{C}}_{1} {\mathrm{S}}_{1}^3 \mathcal{F}_{20} Q_0 {\chi}^{\prime} \psi^{\prime}
\phantom{\Big)} \nonumber \\
&\ 
-8 \mathcal{R}^2 \kappa {\mathrm{C}}_{1} {\mathrm{S}}_{1}^3 {\alpha}^{\prime \prime}
+4 e^{2 {\mathcal{M}}-2 {\mathcal{N}}} b \mathcal{R}^2 \kappa {\mathrm{S}}_{1}^2 Q_0 {\chi}^{\prime \prime}
-e^{2 \mathcal{{\mathcal{M}} - 2 {\mathcal{N}}}} b \mathcal{R}^2 \kappa {\mathrm{S}}_{2}^2 Q_0 {\chi}^{\prime \prime}
-4 b \mathcal{R}^2 \kappa {\mathrm{S}}_{1}^2 \psi^{\prime \prime}
-4 b \mathcal{R}^2 \kappa {\mathrm{S}}_{1}^4 \psi^{\prime \prime}
\phantom{\Big)} \nonumber \\
&\ 
-8 e^{2 {\mathcal{M}}-2 {\mathcal{N}}} \mathcal{R}^2 \kappa {\mathrm{C}}_{1}^2 {\mathrm{S}}_{1}^2 Q_0^2 {\chi}^{\prime 2} \mathcal{V}^{\prime}
-16 e^{{\mathcal{M}}-{\mathcal{N}}} \mathcal{R}^2 \kappa {\mathrm{C}}_{1} {\mathrm{S}}_{1}^3 Q_0 {\chi}^{\prime} \psi^{\prime} \mathcal{V}^{\prime}
-16 \mathcal{R}^2 \kappa {\mathrm{S}}_{1}^2 \psi^{\prime 2} \mathcal{V}^{\prime}
-8 \mathcal{R}^2 \kappa {\mathrm{S}}_{1}^4 \psi^{\prime 2} \mathcal{V}^{\prime}
\phantom{\Big)} \nonumber \\
&\ 
-8 e^{{\mathcal{M}}-{\mathcal{N}}} \mathcal{R}^2 \kappa {\mathrm{S}}_{2} Q_0 {\chi}^{\prime} \psi^{\prime} \mathcal{V}^{\prime}
-8 \mathcal{R}^2 \kappa \psi^{\prime 2} \mathcal{V}^{\prime}=0, \label{eqG11}
\\
\frac{\delta \tensor*{S}{_{\text{\AE{}ST}}^{\text{SZ}}}}{\delta\tensor{g}{^{\theta\theta}}}\propto
&\  
e^{2 {\mathcal{M}}} \mathcal{R} \kappa \mathcal{F}_{20} Q_0^2
-e^{2 {\mathcal{M}}} \mathcal{R} \kappa \mathcal{V}
+\KB{} \mathcal{R} \kappa {\mathrm{C}}_{1}^2 {\mathcal{N}}^{\prime 2}
+\KB{} \mathcal{R} \kappa {\mathrm{S}}_{2} {\mathcal{N}}^{\prime} {\alpha}^{\prime}
+\KB{} \mathcal{R} \kappa {\mathrm{S}}_{1}^2 {\alpha}^{\prime 2}
-\mathcal{R} {\mathcal{N}}^{\prime 2}
+{\mathcal{M}}^{\prime} \mathcal{R}^{\prime}
-{\mathcal{N}}^{\prime} \mathcal{R}^{\prime}
\phantom{\Big)} \nonumber \\
&\ 
-2 e^{2 {\mathcal{M}}-{\mathcal{N}}} \mathcal{R} \kappa {\mathrm{C}}_{1} \mathcal{F}_{20} Q_0^2 {\chi}^{\prime}
+e^{{\mathcal{M}}-{\mathcal{N}}} b \mathcal{R} \kappa {\mathrm{C}}_{1} {\mathrm{S}}_{1} Q_0 {\mathcal{N}}^{\prime} {\chi}^{\prime}
+e^{{\mathcal{M}}-{\mathcal{N}}} b \mathcal{R} \kappa {\mathrm{S}}_{1}^2 Q_0 \alpha^{\prime} {\chi}^{\prime}
+e^{2 {\mathcal{M}}-2 {\mathcal{N}}} \mathcal{R} \kappa {\mathrm{C}}_{1}^2 \mathcal{F}_{20} Q_0^2 {\chi}^{\prime 2}
\phantom{\Big)} \nonumber \\
&\ 
-2 e^{\mathcal{M}} \mathcal{R} \kappa {\mathrm{S}}_{1} \mathcal{F}_{20} Q_0 \psi^{\prime}
+b \mathcal{R} \kappa {\mathrm{C}}_{1}^2 {\mathcal{N}}^{\prime} \psi^{\prime}
+b \mathcal{R} \kappa {\mathrm{C}}_{1} {\mathrm{S}}_{1} \alpha^{\prime} \psi^{\prime}
+e^{{\mathcal{M}}-{\mathcal{N}}} \mathcal{R} \kappa {\mathrm{S}}_{2} \mathcal{F}_{20} \mathcal{Q}_0 {\chi}^{\prime} \psi^{\prime}
+\mathcal{R} \kappa {\mathrm{S}}_{1}^2 \mathcal{F}_{20} \psi^{\prime 2}
-\mathcal{R} {\mathcal{N}}^{\prime \prime}
\phantom{\Big)} \nonumber \\
&\ 
+\mathcal{R} {\mathcal{M}}^{\prime} {\mathcal{N}}^{\prime}
-\mathcal{R}^{\prime \prime}=0. \label{eqG22}
\end{align}
\end{subequations}
Note that these equations are not all independent. Specifically,~\cref{eqA0} and~\cref{eqA1} are related by the normalization condition~\cref{UnitTimelike}, whilst~\cref{eqG00,eqG01,eqG11,eqG22} are related by the Bianchi identity.

\section{Conservation of \ae{}ther acceleration}\label{ConservedCurrent}
\paragraph*{Einstein-\ae{}ther case} In this appendix we prove the result in~\cref{ConservedCurrentEquation}, that in E\AE{} theory the \ae{}ther acceleration is generically a conserved current if \emph{both} the \ae{}ther and the spacetime geometry are static. The first step in this proof is to prepare certain results. From~\cref{UnitTimelike} it follows that~$\tensor{\nabla}{_\mu}\left(\tensor{A}{^\nu}\tensor{A}{_\nu}\right)=0$. This implies~$\tensor{A}{^\nu}\nabla_\mu\tensor{A}{_\nu}=0$, and hence
\begin{align}
  \tensor{F}{_{\mu\nu}}\tensor{A}{^\nu}
	&\equiv\tensor{\nabla}{_\mu}\left(\tensor{A}{_\nu}\tensor{A}{^\nu}\right)
	-\tensor{A}{^\nu}\tensor{\nabla}{_\nu}\tensor{A}{_\mu}
  =-\tensor{J}{_\mu}.
  \label{eq:MaxwellAcceleration}
\end{align}
Next, we start to use the two staticity conditions. Staticity of the spacetime implies the existence of a global timelike Killing vector~$\tensor{\xi}{^\mu}$. Staticity of the \ae{}ther and the condition in~\cref{UnitTimelike} implies that~$\tensor{A}{^\mu}$ describes precisely the field of four-velocities of static observers. Accordingly, we may write
\begin{equation}\label{Lapse}
	\tensor{A}{^\mu}=\frac{\tensor{\xi}{^\mu}}{N},
  \quad
	N\equiv\sqrt{-\tensor{\xi}{^\alpha}\tensor{\xi}{_\alpha}},
\end{equation}
where~$N$ is the lapse function. Together with the identification in~\cref{Lapse}, Killing’s identity~$\tensor{\nabla}{_\alpha}\tensor{\nabla}{^\alpha}\,\tensor{\xi}{^\mu}\equiv-\tensor*{R}{^\mu_\nu}\,\tensor{\xi}{^\nu}$ implies that the divergence of the \ae{}ther acceleration defined in~\cref{Q_identity} is related to the Ricci tensor by
\begin{align}
	\tensor{\nabla}{_\mu} \tensor{J}{^\mu}
	&= -\,\tensor*{R}{_{\mu\nu}}\,\tensor{A}{^\mu}\tensor{A}{^\nu}.
  \label{eq:divAFromRicci}
\end{align}
So far, the only field equation that has been exploited is~\cref{UnitTimelike}. Thus, assuming only that the model in question can be thought of as GR coupled (minimally) to some matter in such a way that doubly static solutions exist, then one can express~\cref{eq:divAFromRicci} in terms of the matter stress-energy tensor~$\tensor{T}{_{\mu\nu}}$ by using the Einstein equations~$\tensor*{R}{_{\mu\nu}}-\tfrac12R\,\tensor{g}{_{\mu\nu}}=\kappa\,\tensor{T}{_{\mu\nu}}$, so that
\begin{equation}
	  \tensor{\nabla}{_\mu} \tensor{J}{^\mu} =-\,\kappa\Bigl(\tensor{T}{_{\mu\nu}}\tensor{A}{^\mu}\tensor{A}{^\nu}+\frac{1}{2}\,T\Bigr),
  \label{eq:divAIdentity}
\end{equation}
where~$T\equiv\tensor*{T}{^\mu_\mu}$ is the trace of the stress-energy tensor. In practice, of course, both~\cref{original_action_EAE,original_action_AEST} satisfy these assumptions about the model, and so the Komar-type construction in~\cref{eq:divAIdentity} is valid for both E\AE{} and \AE{}ST theories. Lastly, static \ae{}ther is hypersurface orthogonal in static spacetime. This means that, in the kinematical decomposition of the congruence of~$\tensor{A}{^\mu}$, the expansion, shear and twist all vanish, which allows us to write~$\tensor{\nabla}{_\mu} \tensor{A}{_\nu} = -\,\tensor{A}{_\mu}\tensor{J}{_\nu}$ and~$\tensor{F}{_{\mu\nu}} = -\,\tensor{A}{_\mu}\tensor{J}{_\nu} +\tensor{A}{_\nu}\tensor{J}{_\mu}$. This leads to a further result (as quoted already in~\cref{Q_identity})
\begin{equation}
  \tensor{F}{_{\mu\nu}}\tensor{F}{^{\mu\nu}} = -2\,\tensor{J}{_\mu}\tensor{J}{^\mu},
  \label{eq:F2toA2}
\end{equation}
which can also be used in obtaining~\cref{Effective_Action}. We now apply~\cref{eq:MaxwellAcceleration,eq:divAIdentity,eq:F2toA2} to the specific case of E\AE{} theory. We will work in terms of the \ae{}ther coupling~$\KB{}$ as introduced in~\cref{original_action_EAE}, but of course the results also apply to the `effective' coupling~$\KBE{}$ defined in~\cref{modified_KB}. From~\cref{original_action_EAE} we deduce 
\begin{equation}
  \kappa\tensor{T}{_{\mu\nu}}
	= \KB{}\,\tensor{F}{_{\mu\alpha}}\tensor{F}{_{\nu}^{\alpha}}
	-\frac{\KB{}}{4}\,\tensor{g}{_{\mu\nu}}\,\tensor{F}{_{\alpha\beta}}\tensor{F}{^{\alpha\beta}}
	+\lambda\,\tensor{A}{_\mu}\tensor{A}{_\nu},
  \label{eq:aetherStressTensor}
\end{equation}
and by using~\cref{UnitTimelike,eq:MaxwellAcceleration} the relevant scalars formed from~\cref{eq:aetherStressTensor} are found to be
\begin{align}
	\kappa\tensor{T}{_{\mu\nu}}\tensor{A}{^\mu}\tensor{A}{^\nu} =\KB{}\,\tensor{J}{_\mu}\tensor{J}{^\mu} +\frac{\KB{}}{4}\,\tensor{F}{_{\mu\nu}}\tensor{F}{^{\mu\nu}} +\lambda,
\quad
\kappa T=-\lambda.
  \label{eq:TraceT}
\end{align}
When~\cref{eq:TraceT} is substituted into~\cref{eq:divAIdentity}, we conclude 
\begin{equation}\label{eq:divAIdentityEAE}
	\tensor{\nabla}{_\mu} \tensor{J}{^\mu}=-\,\KB{}\,\tensor{J}{_\mu}\tensor{J}{^\mu} -\frac{\KB{}}{4}\,\tensor{F}{_{\mu\nu}}\tensor{F}{^{\mu\nu}} -\frac{1}{2}\lambda.
\end{equation}
As mentioned in~\cref{section: Equations of motion}, the \ae{}ther field equation itself generally allows the Lagrange multiplier~$\lambda$ to be determined on the shell. In the case of the E\AE{} theory in~\cref{original_action_EAE}, an application of~\cref{eq:MaxwellAcceleration} then reduces this result to 
\begin{equation}\label{LambdaNew}
	\lambda=\KB{}\,\tensor{\nabla}{_\mu} \tensor{J}{^\mu} +\frac{\KB{}}{2}\,\tensor{F}{_{\mu\nu}}\tensor{F}{^{\mu\nu}}.
\end{equation}
After substituting~\cref{LambdaNew} into~\cref{eq:divAIdentityEAE} and making one last application of~\cref{eq:F2toA2}, we find
\begin{equation}\label{ConservedCurrentEquation2}
	\left(2+\KB{}\right)\tensor{\nabla}{_\mu} \tensor{J}{^\mu}=0,
\end{equation}
and~\cref{ConservedCurrentEquation2} implies~\cref{ConservedCurrentEquation}.

\paragraph*{\AE{}ther-scalar-tensor case} For the \AE{}ST action in~\cref{Action_general}, it is possible to replace the scalar-tensor kinetic interaction~$-\tensor{F}{^{\mu\nu}} \tensor{A}{_\nu}$ with~$\tensor{J}{_\mu}$. The energy momentum tensor is then easier to compute, and~\cref{eq:aetherStressTensor} is replaced by 
~\cref{GABDefinitions}
\begin{align}\label{eq:AESTStressTensor}
\kappa \tensor{T}{_{\mu\nu}}& = \KB{} \tensor{F}{_{\mu\alpha}} \tensor{F}{_{\nu}^{\alpha}} -\frac{\KB{}}{4}\tensor{g}{_{\mu\nu}} \tensor{F}{^{\alpha \beta}} \tensor{F}{_{\alpha \beta}} + \frac{b}{2} \left( \tensor{F}{^{\beta}_{\nu}} \tensor{A}{_\mu} \tensor{\nabla}{_\beta} \psi +\tensor{F}{^{\beta}_{\mu}} \tensor{A}{_\nu} \tensor{\nabla}{_\beta} \psi -\tensor{J}{_\mu} \tensor{\nabla}{_\nu} \psi-\tensor{J}{_\nu} \tensor{\nabla}{_\mu} \psi \right) + \frac{b}{2} \tensor{g}{_{\mu\nu}} \tensor{J}{^\alpha}\tensor{\nabla}{_\alpha} \psi 
\nonumber \\ & \ \ \ 
+ \tensor{\nabla}{_\mu} \psi \tensor{\nabla}{_\nu} \psi \frac{\mathrm{d} \mathcal{V}}{\mathrm{d} \mathcal{Y}} - \frac{1}{2} \tensor{g}{_{\mu\nu}} \mathcal{V}\left(\mathcal{Y}\right) + \lambda \tensor{A}{_\mu}\tensor{A}{_\nu} \, .
\end{align}
By using~\cref{UnitTimelike,eq:MaxwellAcceleration}, and~$\tensor{A}{^\mu} \tensor{J}{_\mu} =\tensor{A}{^\mu} \tensor{\nabla}{_\mu} \psi = 0$, the relevant scalars in~\cref{eq:TraceT} are replaced by 
\begin{align}\label{scalars_from_T}
	\kappa \tensor{T}{_{\mu\nu}}\tensor{A}{^\mu} \tensor{A}{^\nu} = \frac{\KB{}}{2}\tensor{J}{_\mu}\tensor{J}{^\mu} + \frac{b}{2}\tensor{J}{^\mu} \tensor{\nabla}{_\mu}\psi + \frac{1}{2} \mathcal{V}\left(\mathcal{Y}\right) + \lambda \, ,\quad \kappa T = \mathcal{Y} \frac{\mathrm{d} \mathcal{V}}{\mathrm{d}\mathcal{Y}} - 2\mathcal{V}\left(\mathcal{Y}\right) - \lambda \, .
\end{align}
When~\cref{scalars_from_T} is substituted into~\cref{eq:divAIdentity}, we conclude
\begin{equation}\label{divJ_inter}
    \tensor{\nabla}{_\mu} \tensor{J}{^\mu} = -\frac{\KB{}}{2} \tensor{J}{_\mu}\tensor{J}{^\mu} - \frac{b}{2} \tensor{J}{^\mu}\tensor{\nabla}{_\mu}\psi + \frac{1}{2} \mathcal{V}\left(\mathcal{Y}\right) - \frac{1}{2}  \mathcal{Y} \frac{\mathrm{d} \mathcal{V}}{\mathrm{d}\mathcal{Y}} - \frac{1}{2}\lambda \, .
\end{equation}
From~\cref{lag_exp}, and using the fact that~$\tensor{A}{^\mu} \tensor{\nabla}{_\mu} \psi= 0$, we have
\begin{equation}\label{LagForDivJ}
    \lambda = - b\tensor{J}{^\mu}\tensor{\nabla}{_\mu} \psi + \KB{} \tensor{\nabla}{_\mu}\tensor{J}{^\mu} - \KB{} \tensor{J}{^\mu}\tensor{J}{_\mu} \, .
\end{equation}
Finally, by combining~\cref{divJ_inter,LagForDivJ} we obain the analogue of~\cref{ConservedCurrentEquation2}
\begin{equation}\label{divJFinal}
    \left(2 + \KB{}\right)\tensor{\nabla}{_\mu} \tensor{J}{^\mu} =
    \mathcal{V}\left(\mathcal{Y}\right) - \mathcal{Y} \frac{\mathrm{d} \mathcal{V}}{\mathrm{d}\mathcal{Y}} \, .
\end{equation}
Therefore,~\cref{divJFinal} shows that~$\tensor{\nabla}{_\mu}\tensor{J}{^\mu}= 0$, if~$\mathcal{V}\left(\mathcal{Y}\right)\propto\mathcal{Y}$, consistent with the assumptions that lead to~\cref{NoMOND}.

\section{Details of exact solutions}\label{appendix: Wormhole solution: mathematical detail}

\paragraph*{New definitions} In this appendix, we shall use the following definitions for convenience:
\begin{equation}\label{NewDefinitions}
	{{\Upsilon}}\equiv \frac{\mathrm{d}\mathcal{N}}{\mathrm{d}r}, \quad \mathcal{P} \equiv \frac{\mathrm{d}\psi}{\mathrm{d}r}.
\end{equation}

\paragraph*{Ellis--Bronnikov drainhole}  Without loss of generality, we refine~\cref{sph_back} to a genuinely Schwarzschild-like coordinate system so that
\begin{equation}\label{SchwarzschildLike}
	\mathcal{R} = r.
\end{equation}
After substituting~\cref{EnforceTimelike,NoMOND,NoDust,SchwarzschildLike}, the only field equations that do not vanish are~\cref{eqP,eqG00,eqG11,eqG22}. These are respectively given by
\begin{subequations}\small
    \begin{align}
	    & -b  {{\Upsilon}}+a  r\frac{\mathrm{d}\mathcal{P}}{\mathrm{d}r}+a \mathcal{P}  \left( 2+r  {{\Upsilon}} - r  \frac{\mathrm{d}\mathcal{M}}{\mathrm{d}r} \right) - \frac{b}{2}  r   \left( {{\Upsilon}}^2 + \frac{\mathrm{d}\Upsilon}{\mathrm{d}r} - {{\Upsilon}} \frac{\mathrm{d}\mathcal{M}}{\mathrm{d}r} \right)=0, \label{eq1}\\
	    & -1+e^{2 {\mathcal{M}}}-\frac{r}{2}\left(2 b  \mathcal{P}+a  r \mathcal{P}^2+b  r\frac{\mathrm{d}\mathcal{P}}{\mathrm{d}r}+\KB{}  \left({{\Upsilon}}  (4+r {{\Upsilon}})+2   r \frac{\mathrm{d}{{\Upsilon}}}{\mathrm{d}r}\right)\right)  + r \frac{\mathrm{d}\mathcal{M}}{\mathrm{d}r}  \left(2+\frac{b}{2}   r \mathcal{P}+ \KB{}  r {{\Upsilon}}\right)=0, \label{eq2}\\
        & 1 - e^{2{\mathcal{M}}} + r  \left( 2{{\Upsilon}} + \frac{r}{2}   \left( -a  \mathcal{P}^2 + b  \mathcal{P}  {{\Upsilon}} + \KB{}   {{\Upsilon}}^2\right) \right)=0, \label{eq3}\\
        & {{\Upsilon}} - \frac{\mathrm{d}\mathcal{M}}{\mathrm{d}r} + r  \left( \frac{a}{2}  \mathcal{P}^2-\frac{b}{2}\mathcal{P}  {{\Upsilon}} + \frac{\mathrm{d}\Upsilon}{\mathrm{d}r} - {{\Upsilon}}   \left( \left(-1+\frac{\KB{}}{2}\right)  {{\Upsilon}}+\frac{\mathrm{d}\mathcal{M}}{\mathrm{d}r}\right)\right)=0.\label{eq4}
    \end{align}
\end{subequations}
In~\cref{eq1,eq2,eq3,eq4} we have used~\cref{NewDefinitions}. One can combine~\cref{eq2,eq3} to eliminate~$e^{2{\mathcal{M}}}$. The remaining equations can be rearranged to yield
\begin{subequations}\small
    \begin{align}
	    &\frac{\mathrm{d}\mathcal{M}}{\mathrm{d}r} =   \frac{a}{2}   r\mathcal{P}^2 -\frac{b}{2}   r\mathcal{P}  {{\Upsilon}} - {{\Upsilon}}  \left(1+\frac{\KB{}}{2}   r{{\Upsilon}} \right), \label{simplified_eqmu}\\
	    &\frac{\mathrm{d}{\Upsilon}}{\mathrm{d}r} = {{\Upsilon}}   \left(\frac{\mathrm{d}\mathcal{M}}{\mathrm{d}r} - {{\Upsilon}} -\frac{2}{r}\right), \label{simplified_eqV}\\
	    &\frac{\mathrm{d}\mathcal{P}}{\mathrm{d}r} = \mathcal{P}   \left( \frac{\mathrm{d}\mathcal{M}}{\mathrm{d}r} - {{\Upsilon}} -\frac{2}{r}\right). \label{simplified_eqP}
    \end{align}
\end{subequations}
From~\cref{simplified_eqV,simplified_eqP}, it is obvious that~$\mathcal{P} = q  {{\Upsilon}}$ for some constant~$q$, and this leads to our conjecture in~\cref{scalar_current,PsiBehaviour}. With reference to~\cref{modified_KB}, the system in~\cref{simplified_eqV,eq3} becomes
\begin{align}\label{final_differential_equation}
    &r\frac{\mathrm{d}{\Upsilon}}{\mathrm{d}r} = -2{{\Upsilon}} -2r {{\Upsilon}}^2 - \frac{\KBE{}}{2}  r^2  {{\Upsilon}}^3, \quad 
    e^{2{\mathcal{M}}} =  1+2r  {{\Upsilon}} + \frac{\KBE{}}{2}  r^2   {{\Upsilon}}^2. 
\end{align}
Notice that the final equations of motion in~\cref{final_differential_equation} are precisely the same as those of E\AE{} theory in Schwarzschild-like coordinates, see, for example,~\cite{BH_in_Ae,NS_Ae}. This is the result which was to be proven, but we now go beyond it so as to actually solve~\cref{final_differential_equation}. The choice of Schwarzschild-like coordinate in~\cref{SchwarzschildLike} is generally different from the radial gauge~\cref{RadialCoordinate} chosen to present the results in~\cref{section: wormhole solutions,section: A Cosmological Solution}.  We will now define a new radial coordinate~$\sigma$, a special case of the general~$r$ we have been using hitherto, which we require to be proportional to the affine parameter of radial null geodesics. More concretely, this choice of radial coordinate is motivated as follows. In general, we want our gauge condition in~\cref{RadialCoordinate} to always apply in the coordinates we happen to be using, and so it needs to be revised accordingly whenever we rescale the radial coordinate. Starting, therefore, from the general~$r$ coordinate, we want to move to the new radial coordinate~$\sigma$ in which this gauge condition is actually~$g_{tt} = -1/g_{\sigma \sigma}$. In terms of the original~$r$ coordinate, this implies~$-\mathcal{N} = \mathcal{M} + \ln(dr/d\sigma)$ which, taking derivatives with respect to~$\sigma$, becomes a differential equation for~$r(\sigma)$
\begin{equation}\label{radial_null_geodesics_final_equation}
	\frac{\mathrm{d}^2r}{\mathrm{d}\sigma^2}= -\frac{\mathrm{d}r}{\mathrm{d}\sigma} \left(\frac{\mathrm{d}\mathcal{N}}{\mathrm{d}\sigma}+\frac{\mathrm{d}\mathcal{M}}{\mathrm{d}\sigma}\right) \,.
\end{equation}
This is precisely the radial geodesic equation following from~\cref{sph_back}, which motivates this choice of radial coordinate. Accordingly, we move from~$r$ to~$\sigma$ as defined in~\cref{radial_null_geodesics_final_equation}. We then further have the identities
\begin{equation}\label{TrivialIdentities}
	\frac{\mathrm{d}\mathcal{N}}{\mathrm{d}\sigma} \equiv \frac{\mathrm{d}\mathcal{N}}{\mathrm{d}r}\frac{\mathrm{d}r}{\mathrm{d}\sigma}, \quad
	\frac{\mathrm{d}^2\mathcal{N}}{\mathrm{d}\sigma^2} \equiv -\left(\frac{\mathrm{d}\mathcal{N}}{\mathrm{d}\sigma}\right)^2 -\frac{\mathrm{d}\mathcal{N}}{\mathrm{d}\sigma}\frac{\mathrm{d}\mathcal{M}}{\mathrm{d}\sigma} + \frac{\mathrm{d}^2\mathcal{N}}{\mathrm{d}r^2}\left(\frac{\mathrm{d}r}{\mathrm{d}\sigma}\right)^2.
\end{equation}
By combining~\cref{final_differential_equation,radial_null_geodesics_final_equation,TrivialIdentities}, we find that~\cref{simplified_eqV,simplified_eqmu} become
\begin{equation}\label{our_solution_final_equation_for_nu}
	\frac{\mathrm{d}^2\mathcal{N}}{\mathrm{d}\sigma^2}= -\frac{2}{r}\frac{\mathrm{d}r}{\mathrm{d}\sigma}\frac{\mathrm{d}\mathcal{N}}{\mathrm{d}\sigma}- 2\left(\frac{\mathrm{d}\mathcal{N}}{\mathrm{d}\sigma}\right)^2,\quad 
	\frac{\mathrm{d}^2r}{\mathrm{d}\sigma^2} = \frac{\KBE{}}{2} r\left(\frac{\mathrm{d}\mathcal{N}}{\mathrm{d}\sigma}\right)^2.
\end{equation}
The solution to the first equation of~\cref{our_solution_final_equation_for_nu} is 
\begin{equation}\label{GeodesicRescalingSolution}
	r^2 =c_1^2e^{-2{\mathcal{N}}} \frac{\mathrm{d}\sigma}{\mathrm{d}\mathcal{N}},
\end{equation}
for some integration constant~$c_1$. Once again,~\cref{GeodesicRescalingSolution} is independent of the coupling parameters in the action and is a result of the conservation of the \ae{}ther acceleration in~\cref{ConservedCurrentEquation}. Now from~\cref{GeodesicRescalingSolution,our_solution_final_equation_for_nu}, we find 
\begin{equation}\label{MasterGeometryEquation}
	\frac{\mathrm{d}\mathcal{N}}{\mathrm{d}\sigma} = \frac{2c_3}{2\left(\sigma+c_2\right)^2 + \left(\KBE{}-2\right)  c_3^2},
\end{equation}
with further integration constants~$c_2$ and~$c_3$. For the case of the Ellis--Bronnikov drainhole, one can choose~$c_2=0$ and assume~\cref{UnphysicalCase2}. By re-labelling~$\sigma$ as~$r$ and integrating~\cref{MasterGeometryEquation} with the constraints of Minkowski geometry at spatial infinity, we then recover the line-element function in~\cref{EB_wormholeA}. The other function~\cref{EB_wormholeB} is obtained from~\cref{SchwarzschildLike,GeodesicRescalingSolution}, again with appropriate integration constants.

\paragraph*{First extended Eling--Jacobson solution} The Eling--Jacobson wormhole is obtained as the special case of~\cref{MasterGeometryEquation} where~$c_2 = -c_3   \sqrt{1-\KBE{}/2}$, and~\cref{NewAetherCouplingRange} is also assumed instead of~\cref{UnphysicalCase2}.

\paragraph*{Anti-Ellis--Bronnikov solution} Whilst the Eiling--Jacobson solution was a special case of the solution branch which led also to the Ellis--Bronnikov drainhole, we now start afresh by considering quite a different branch, in which~\cref{NewDefinitions} holds true. The condition in~\cref{NewCondition} translates to
\begin{equation}\label{NewerCondition}
	\Upsilon=0.
\end{equation}
With~\cref{NewerCondition,RadialCoordinate}, one is forced to have constant~$\mathcal{M} = -\mathcal{N}$, which we satisfy by the simple choice~$\mathcal{M} = \mathcal{N} =0$, which is already consistent with the first equality in~\cref{Ellis_wormholeA}. To proceed,~\cref{eqP} now reduces to
\begin{equation}\label{Conjugate}
	-\frac{2}{\mathcal{R}}\frac{\mathrm{d}\mathcal{R}}{\mathrm{d}r} = \frac{1}{\mathcal{P}}\frac{\mathrm{d}\mathcal{P}}{\mathrm{d}r}.
\end{equation}
Now~\cref{Conjugate} implies that
\begin{equation}\label{ConjugateImplication}
\mathcal{P} = {c_1}{\mathcal{R}^{-2}},
\end{equation}
for some constant~$c_1$. After substituting~\cref{ConjugateImplication}, it can be seen that~\cref{eqG00,eqG11,eqG22} all reduce to the same differential equation
\begin{equation}\label{FinalDifferential}
	\left(\frac{\mathrm{d}\mathcal{R}}{\mathrm{d}r}\right)^2 = 1 + \frac{a c_1^2}{2 \mathcal{R}^2}.
\end{equation}
A solution to~\cref{FinalDifferential} is 
\begin{equation}\label{FinalSolution}
    \mathcal{R}^2 = -\frac{a c_1^2}{2}+r^2 .
\end{equation}
Therefore, with the identification~$\ell^2={a c_1^2}/{2}$ in~\cref{FinalSolution} the second equality in~\cref{Ellis_wormholeA} is also recovered.

\paragraph*{Second extended Eling--Jacobson solution} By substituting~\cref{EnforceTimelike,NoMOND,SkordisCase} into~\cref{eqA1} we obtain
\begin{equation}\label{constantinos_solution_eqA1}
	e^{{\mathcal{M}}+{\mathcal{N}}}\mathcal{R}\frac{\mathrm{d}\chi}{\mathrm{d}t}\left( b  \frac{\mathrm{d}\mathcal{N}}{\mathrm{d}r} - 2a  \frac{\mathrm{d}\psi}{\mathrm{d}r} \right) = 0.
\end{equation}
For regions where the coordinates in~\cref{sph_back} are non-singular,~\cref{constantinos_solution_eqA1} evidently gives rise to~\cref{AdditionalCondition}. As a consequence, if~$\chi$ is non-constant we must have
\begin{equation}\label{PsiGradient}
	\frac{\mathrm{d}\psi}{\mathrm{d}r}=\frac{b}{2a}\frac{\mathrm{d}\mathcal{N}}{\mathrm{d}r}.
\end{equation}
We then substitute~\cref{PsiGradient} into~\cref{eqG00,eqG11,eqG22}, and combine these equations to yield
\begin{equation}\label{constantinos_final_equation_for_mu_and_R}
	2\frac{\mathrm{d}\mathcal{R}}{\mathrm{d}r}\frac{\mathrm{d}\mathcal{M}}{\mathrm{d}r} + \mathcal{R}  \left[ \frac{\mathrm{d}^2\mathcal{M}}{\mathrm{d}r^2}-2 \left(\frac{\mathrm{d}\mathcal{M}}{\mathrm{d}r}\right)^2\right] = 0.
\end{equation}
Notice that~\cref{constantinos_final_equation_for_mu_and_R,our_solution_final_equation_for_nu} have a similar form, except that~${\mathcal{N}}$ is replaced with~$-{\mathcal{M}}$ in~\cref{our_solution_final_equation_for_nu}, as expected from~\cref{RadialCoordinate}. Hence, the implication is that
\begin{equation}\label{ScalingSolution}
	\mathcal{R} = c_1 e^{{\mathcal{M}}}\sqrt{-\frac{\mathrm{d}r}{\mathrm{d}\mathcal{M}}},
\end{equation}
for some integration constant~$c_1$. Notice that~\cref{ScalingSolution} is independent of the coupling constant~$\KB{}$ and~$a$. In fact,~\cref{ScalingSolution} is a result of the conservation of the \ae{}ther current in~\cref{ConservedCurrentEquation}. By substituting~\cref{ScalingSolution} into~\cref{eqG11} we obtain
\begin{equation}\label{GradientM}
	\frac{\mathrm{d}\mathcal{M}}{\mathrm{d}r} =-\frac{c_1^2}{r^2}   \left(1-\frac{2 c_3}{r} \right)^{-1}, \quad
    c_3 \equiv c_1^2\sqrt{\frac{(2-\KB{})\lambda_s}{2\left(1+\lambda_s\right)} } \,.
\end{equation}
The solution to~\cref{GradientM}, in combination with~\cref{RadialCoordinate}, leads to the geometry in~\cref{constantinos_solution_metricA,constantinos_solution_metricB}. One can substitute~\cref{constantinos_solution_metricA,constantinos_solution_metricB,constantinos_solution_scalar} back to verify that all the other field equations are satisfied.

\paragraph*{Cosmological solution} When~\cref{EnforceTimelike,RadialCoordinate,LinearTime} are imposed, and~\cref{NoDust} is relaxed, we find that~\cref{eqG01} is satisfied by the first equality in~\cref{cosmological_solution_metric}. Meanwhile,~\cref{eqA1} reduces to 
\begin{equation}\label{RemainderEquation}
	\mathcal{Q}_0   \mathcal{R}  \frac{\mathrm{d}\mathcal{V}}{\mathrm{d}\mathcal{Y}}\frac{\mathrm{d}\psi}{\mathrm{d}r} = 0.
\end{equation}
From~\cref{RemainderEquation}, we are obliged to set~\cref{CCOrigin} which, when substituted into~\cref{eqG11,eqG22}, respectively gives
\begin{align}\label{TwoConditions}
	-1+\Lambda\kappa   \mathcal{R}^2 + \left(\frac{\mathrm{d}\mathcal{R}}{\mathrm{d}r}\right)^2=0, \quad \Lambda\kappa   \mathcal{R}+\frac{\mathrm{d}^2\mathcal{R}}{\mathrm{d}r^2}=0.
\end{align}
The equations in~\cref{TwoConditions} are satisfied by the second equality in~\cref{cosmological_solution_metric} which, when substituted into~\cref{eqG00}, gives rise to
\begin{equation}\label{cosmological_solution_eqG00}
	2\sqrt{\kappa \Lambda}   \cos\left(\sqrt{\kappa \Lambda }  r \right)   \left( \KB{}-2\right)\frac{\mathrm{d} \psi}{\mathrm{d}r}  + \sin \left( \sqrt{\kappa \Lambda}   r\right)   \left[ 2\kappa \Lambda + \left(\KB{}-2\right)   \frac{\mathrm{d}^2\psi}{\mathrm{d}r^2}\right]=0.
\end{equation}
It can then be shown that~\cref{cosmological_solution_scalar} solves~\cref{cosmological_solution_eqG00}.

\section{More like black holes than wormholes}\label{appendix: Nature of the horizon}
\paragraph*{Null singularity} In this appendix, we discuss the nature and extendibility of the solution in~\cref{JE_wormholeA,JE_wormholeB} beyond the hypersurface that appears to be a Killing horizon. We first assume the condition in~\cref{RadialCoordinate}, which allows~$r$ to be interpreted as the affine parameter of the radial null geodesics. Now let
\begin{equation}\label{TangentVector}
	\tensor{k}{^\mu}\equiv\frac{\mathrm{d}\tensor{x}{^\mu}}{\mathrm{d}r}, \quad \left[\tensor{x}{^\mu}\right]\equiv\left[  t, r, \theta, \phi\right],
\end{equation}
be the tangent vector to such geodesics for coordinates~$\tensor{x}{^\mu}$ in~\cref{sph_back}. Then from~\cref{sph_back,RadialCoordinate}, the particular curvature scalar considered in~\cite{BH_in_Ae} takes the form
\begin{align}\label{null_geodesics_scalar}
	\tensor{R}{_{\mu \nu}} \tensor{k}{^\mu}\tensor{k}{^\nu} =- \frac{2}{\mathcal{R}}\frac{\mathrm{d}^2\mathcal{R}}{\mathrm{d}r^2}.
\end{align}
If, as~$r\to\Rh{}$, the volume element behaves as~$\mathcal{R}\approx r e^{\mathcal{H}}  \left(1-\frac{\Rh{}}{r} \right)^\epsilon$, where~$\mathcal{H}\equiv\mathcal{H}(r)$ is regular at~$r = \Rh{}$ and~$\epsilon<0$, then the curvature scalar in~\cref{null_geodesics_scalar} takes the form
\begin{equation}\label{null_singularity_scalar}
	\tensor{R}{_{\mu \nu}} \tensor{k}{^\mu} \tensor{k}{^\nu} \approx -\frac{2\Rh{}^2   \epsilon (\epsilon - 1)}{r^2   \left( r - \Rh{}\right)^2} + \frac{4\left[ r + \Rh{}   (\epsilon - 1) \right]   }{r   \left(r - \Rh{}\right)}\frac{\mathrm{d}\mathcal{H}}{\mathrm{d}r} - 2\left(\frac{\mathrm{d}\mathcal{H}}{\mathrm{d}r}\right)^2 -2\frac{\mathrm{d}^2\mathcal{H}}{\mathrm{d}r^2}. 
\end{equation}
According to~\cref{null_singularity_scalar}, this scalar quantity always diverges at this hypersurface, i.e. it is a null singularity.

\paragraph*{No conformal rescaling} We will now show this singular behavior can be removed by some conformal transformations. However, for the solutions discussed in this paper, the cost of doing so is either shifting this surface to a null-infinity, or making the Ricci scalar diverges. Suppose, for some regular~$\mathcal{K}\equiv\mathcal{K}(r)$, the conformal factor we are looking for takes the form~$\Omega\equiv\Omega(r)$ so that 
\begin{align}\label{conformal_transformation_to_remove_null_singularity}
	\tensor{g}{_{\mu\nu}}\mapsto\Omega^2\tensor{g}{_{\mu\nu}},\quad \Omega\equiv e^{\mathcal{K}}   \left(1 - \frac{\Rh{}}{r} \right)^{\gamma}.
\end{align}
The condition from~\cref{conformal_transformation_to_remove_null_singularity} that the hypersurface is not shifted to null-infinity is~$2\gamma>-1$. Next, the cases in which the singularity in~\cref{null_singularity_scalar} can be removed are
\begin{equation}
	\left({\mathcal{K}}(r) = \Delta(r)-\ln(r)\right) \wedge \gamma \in \{\epsilon-1,-\epsilon\}  .\label{RegularCase1}
\end{equation}
The function~$\Delta(r)$ introduced in~\cref{RegularCase1} is the solution to the following differential equation, 
\begin{equation}
    0=
    2\left(r - \Rh{} \cdot \frac{\epsilon (\epsilon -1)}{\gamma} \right) \cdot \Delta'
    +2\left(r + \Rh{} \cdot {(\epsilon -1)} \right) \cdot {\mathcal{H}}'
    - r \left(r - \Rh{} \right) \cdot \left({\Delta'}^{2} - {\mathcal{H}'}^{2} \right) + r\left( r - \Rh{}\right) \cdot \left( \Delta'' + \mathcal{H}'' \right) \, . \label{diff_eq_for_Delta}
\end{equation}
From~\cref{diff_eq_for_Delta}, one immediately has~$\Delta = -\mathcal{H}$ for the case~$\gamma = -\epsilon$. The case when~$\gamma = \epsilon - 1$ is more complicated, but one can always find the series expansion of~$\Delta$ in the vicinity of the horizon ib terms of~$\mathcal{H}$ and its derivatives. Under such choices of~$\gamma$ and~$\mathcal{K}$, the scalar defined by~\cref{null_geodesics_scalar} is guaranteed to be~$0$ at~$r=\Rh{}$, hence the singularity is removed. Notice that the choice~$\gamma=\epsilon-1$ clearly violates the condition~$2\gamma>1$ since~$\epsilon<0$ by assumption. Which means, if one choose~$\gamma = \epsilon -1$, the cost of removing this singularity is shifting the surface at~$r=\Rh{}$ to a null-infinity. We now turn our attention to the Ricci scalar for the choice~$\gamma = -\epsilon$.\\

\paragraph*{General curvature scalars} We now study the behavior of other scalar quantities near this surface, noting how~\cref{sph_back,RadialCoordinate} imply for some regular, nonzero~$\mathcal{T}\equiv\mathcal{T}(r)$
\begin{equation}\label{asymptotic_behavior_near_the_horizon}
	e^{2\mathcal{N}} \approx \mathcal{T}  \left( 1 - \frac{\Rh{}}{r} \right)^\beta,\quad  \mathcal{R} \approx r   e^{\mathcal{H}}  \left(1-\frac{\Rh{}}{r} \right)^\epsilon . 
\end{equation}
From~\cref{asymptotic_behavior_near_the_horizon} we conclude that, after the conformal transformation described by~\cref{conformal_transformation_to_remove_null_singularity}, the scalar concomitants of the curvature tensor and the metric behave as (including the tidal stress experienced by radial massive observer)
\begin{equation}\label{curvature_scalars_after_conformal}
	\mathcal{S}\left(\tensor{R}{^\mu_{\nu\sigma\lambda}},\tensor{g}{_{\mu\nu}}\right) \propto \left( 1 - \frac{\Rh{}}{r}\right)^{\mu \zeta}, \quad \mu>0, \quad  \zeta \equiv \min\left(-2\epsilon -2\gamma, \beta - 2\gamma - 2 \right).
\end{equation}
For example,~\cref{curvature_scalars_after_conformal} implies if one choose~$\gamma = -\epsilon$, the Ricci scalar
\begin{equation}\label{curvature_scalars_after_conformal2}
	R \propto \left( 1 - \frac{\Rh{}}{r}\right)^{\zeta}, \quad  \zeta \equiv \min\left(0, \beta+2\epsilon  - 2 \right) =  \beta+2\epsilon  - 2.
\end{equation}
For the case under consideration, we have~$\beta>0$ and~$\epsilon<0$. Thus, for the choice~$\gamma = -\epsilon$,~$\zeta<0$ if and only if~$\beta +2\epsilon>2$. In such cases, the curvature scalar and~\cref{null_geodesics_scalar} are all regularized, and the null hypersurface is not a null-infinity. \\

\noindent For the solution discussed in~\cref{ElingJacobsonWormholeLineElement}, one always has~$\beta +2\epsilon=-2<2$. Hence, the cost of removing the singular behavior in the scalar~\cref{null_geodesics_scalar} for~$\gamma = -\epsilon$ for this solution is to make the Ricci scalar singular, which makes the hypersurface defined by~$r=\Rh{}$ remain a null singularity. We thus showed that, for spacetime described by~\cref{ElingJacobsonWormholeLineElement}, near the null hypersurface~$r=\Rh{}$, the following three conditions \textbf{cannot} be simultaneously satisfied via a conformal transformation of spacetime:
\begin{itemize}
    \item the surface can be reached by radial photon within finite amount of its affine parameter; 
    \item the scalar defined in~\cref{null_geodesics_scalar} is regular; 
    \item the curvature scalar is regular.
\end{itemize}
We therefore can safely conclude that the radial photon geodesics are non-extendible at the null hypersurface~$r=\Rh{}$.\\

\noindent For the physical spacetime, from~\cref{curvature_scalars_after_conformal} with~$\gamma = 0$, the null surface at~$r=\Rh{}$ is not a curvature singularity, if and only if
\begin{equation}\label{FinalCondition}
    \beta>2\, .
\end{equation}
\noindent In the solution of~\cref{JE_wormholeA}, the condition in~\cref{FinalCondition} is
\begin{equation}\label{FinalCondition2}
    \beta = \frac{2}{\sqrt{4-2\KBE{}}} >2,
\end{equation}
and~\cref{FinalCondition2} gives~$3/2<\KBE{}<2$. Within this range, the curvature scalar is regular at~$r=\Rh{}$.

\end{document}